\documentclass[11pt,a4paper]{article}
\pdfoutput=1
\usepackage{jheppub}
\usepackage{slashed}

\makeatletter
\def\@fpheader{\relax}
\makeatother

\usepackage{ulem}
\usepackage{tensor}
\usepackage[czech,english]{babel}
\usepackage{graphicx}
\usepackage{amsmath,amsfonts,amssymb}
\usepackage{url}
\usepackage{mathabx}
\DeclareMathOperator{\MyProd}{\scalebox{1.4}{$\mathrm{I\kern-0.2ex I}$}}

\usepackage{leftidx}
\usepackage{xcolor}
\numberwithin{equation}{section}
\usepackage{adjustbox}
\usepackage{appendix}
\usepackage{tikz}
\tikzstyle{process} = [rectangle, minimum width=3cm, minimum height=1cm, text centered, draw=black, fill=orange!30]
\tikzstyle{arrow} = [thick,->,>=stealth]

\preprint{LCTP-20-09}

\title{Gravitational Cardy Limit and AdS Black Hole Entropy}

\author[a]{Marina David}

\emailAdd{mmdavid@umich.edu}

\author[a]{Jun Nian}

\emailAdd{nian@umich.edu}

\affiliation[a]{Leinweber Center for Theoretical Physics, University of Michigan, Ann Arbor, MI 48109, U.S.A.}

\author[a, b]{and Leopoldo A. Pando Zayas}

\emailAdd{lpandoz@umich.edu}

\affiliation[b]{The Abdus Salam International Centre for Theoretical Physics, 34014 Trieste, Italy}

\abstract{We explore the gravitational implementation of the field theory Cardy-like limit recently used in the successful microstate countings of AdS black hole entropy in various dimensions. On the field theory side, the Cardy-like limit focuses on a particular scaling of conserved electric charges and angular momenta, and we first translate this scaling to the gravitational side by a limiting procedure on the black hole parameters. We note that the scaling naturally accompanies a near-horizon region for which these black hole solutions are greatly simplified. Applying the Kerr/CFT correspondence to the near-horizon region, we precisely reproduce the Bekenstein-Hawking entropy of asymptotically AdS$_{4, 5, 6, 7}$ BPS black holes. Our results explicitly  provide a microscopic and universal low energy description for AdS black holes across various dimensions.}

\keywords{}

\arxivnumber{}

\newcommand{\bea}{\begin{eqnarray}}
\newcommand{\eea}{\end{eqnarray}}

\newcommand{\be}{\begin{equation}}
\newcommand{\ee}{\end{equation}}

\begin{document}

\maketitle

%%%%%%%%%%%%%%%%%%%%%%%%%%%%%%%%%%%%%%%%%%%
%%%%%%%%%%%%%%%%%%%%%%%%%%%%%%%%%%%%%%%%%%%
\section{Introduction}\label{sec:Introduction}
%%%%%%%%%%%%%%%%%%%%%%%%%%%%%%%%%%%%%%%%%%%
%%%%%%%%%%%%%%%%%%%%%%%%%%%%%%%%%%%%%%%%%%%

The AdS/CFT correspondence \cite{Maldacena:1997re} posits  an equivalence between string theory in certain asymptotically AdS spacetimes and the  corresponding  dual conformal field theories.  This paradigm provides implicit resolutions to many of the puzzles presented by gravity as it  can now be reformulated in terms of a unitary field theory.  A necessary first step for this program involves the microscopic counting of the degrees of freedom responsible for the Bekenstein-Hawking entropy of asymptotically AdS  black holes.  In the context of string theory on AdS$_5\times S^5$, a  powerful early attempt to the microscopic computation of the black hole entropy was put forward in  \cite{Kinney:2005ej}; such an attempt has been recently improved in \cite{Cabo-Bizet:2018ehj, Choi:2018hmj, Benini:2018ywd} to provide a  precise matching between the Bekenstein-Hawking entropy of the black hole and the microstate counting in the dual ${\cal N}=4$ SYM theory. The key idea in these recent works has been to consider complex backgrounds or complex chemical potentials. This approach  has now been generalized to include the microscopic counting of the entropy of rotating, electrically charged asymptotically AdS black holes in  various  dimensions \cite{Choi:2019miv, Kantor:2019lfo, Nahmgoong:2019hko, Choi:2019zpz, Nian:2019pxj, Bobev:2019zmz, Benini:2019dyp,Hosseini:2019lkt,Crichigno:2020ouj}. These recent works, together with previous progress on magnetically charged asymptotically AdS black holes \cite{Benini:2015eyy, Zaffaroni:2019dhb}, provide a fairly complete and novel approach to microstate counting for asymptotically AdS black holes and mark the beginning of a new era in explorations of quantum gravity.

A so-called Cardy-like limit has played a prominent role among the recent literature on microstate counting  dual to the entropy of  rotating electrically charged asymptotically AdS black holes \cite{Choi:2018hmj, Choi:2019miv, Nahmgoong:2019hko, Choi:2019zpz, Nian:2019pxj}; the limit is  loosely defined as  
\be\label{eq:DefCardyLimit}
  |\omega_i| \ll 1\, ,
\ee
i.e., the angular velocities are very small compared to other parameters on the field theory side.  The  Cardy-like limit above has the advantage of greatly simplifying the analysis of the effective large$-N$ matrix model obtained from the  superconformal index or the partition function of the boundary conformal field theory.

Let us discuss this limit by focusing on the states or operators it singles out. One interpretation of the Cardy-like limit follows from selecting a particular sector in the dual field theory. More precisely, we consider the entropy function given schematically as $ S =  I + \sum_i \omega_i J_i + \sum_I \Delta_I Q_I - \Lambda (\sum_I \Delta_I - \sum_i \omega_i - 2 \pi i)$ in terms of conserved charges, chemical potentials and a Lagrange multiplier, $\Lambda$, and search for a self-consistent scaling of the form   $\omega_i \sim \epsilon$ and $\Delta_I \sim {\cal O}(1)$. Demanding that all terms contribute at the same order in $\epsilon$ leads to the following scalings of quantum numbers in Table~\ref{Table:Intro} that describes the singled-out sector.

\begin{table} 
\centering
\begin{tabular}{|c|c|c|c|c|c|}
\hline
Dimension of CFT &  $\omega$ & $\Delta$ & $J$ & $Q$ & Entropy Function\\
\hline \hline
$d=3$ &  $\epsilon$ & 1 &  $1/\epsilon^2$ &$1/\epsilon$ &  $1/\epsilon$
\\ \hline
$d=4$ &  $\epsilon$ & 1 & $1/\epsilon^3$ & $1/\epsilon^2$ &  $1/\epsilon^2 $
\\ \hline
$d=5$ & $\epsilon $ & 1 & $1/\epsilon^3$ & $1/\epsilon^2 $ &  $1/\epsilon^2$ 
\\ \hline 
$d=6$ & $\epsilon$  & 1 & $1/\epsilon^4 $ & $1/\epsilon^3$  &  $1/\epsilon^3$ 
\\ \hline\hline       
\end{tabular}
\caption{Scaling of conserved quantum numbers in various field theory dimensions. \label{Table:Intro}}
\end{table}

\vspace{0.5cm}

The self-consistency of these scalings is checked {\it a posteriori}. Since the charges (angular momenta and electric charges) are large, the intuition for a semiclassical regime is justified. It is precisely in this regime that the Legendre transform to obtain the entropy as the extremization of the entropy function is fully justified\footnote{ Our intuition closely follows the BMN paradigm where a  closed sector of large R-charge operators of a given scaling is singled out  \cite{Berenstein:2002jq}.}.

In this paper we study the Cardy-like limit on the gravity side, which we refer to as the gravitational Cardy limit. To define this limit, we start with the field theory Cardy-like limit \eqref{eq:DefCardyLimit} and find its counterpart on the gravity side using some relations between field theory and gravity parameters. More explicitly, the gravitational Cardy limit has specific expressions for various examples considered in this paper given by \eqref{eq:AdS5CardyLimit}, \eqref{eq:AdS4CardyLimit}, \eqref{eq:SimpleAdS7CardyLimit}, \eqref{eq:GeneralAdS7CardyLimit} and \eqref{eq:AdS6CardyLimit}, which impose some special limits on the parameters of the corresponding black hole solutions.  For all the asymptotically AdS$_{4,5,6,7}$ solutions studied in this paper, the gravitational Cardy limit can be universally written as 
\begin{align}
	|a_i g| \to 1,
\end{align}
where $a_i$ roughly characterize angular momenta in units of the inverse radius of AdS, $g$.  It turns out that when combined with a particular near-horizon limit for rotating asymptotically AdS BPS black holes, the gravitational Cardy limit provides further insight by yielding a near-horizon AdS$_3$ geometry, independent of the dimension of the black hole. Hence, the black hole entropy can be obtained from this near-horizon AdS$_3$ {\it a la} the Kerr/CFT correspondence.

More specifically, the asymptotically flat extremal black hole entropy can be obtained using the Kerr/CFT correspondence near the horizon region \cite{Guica:2008mu}, and this approach has been generalized to asymptotically AdS extremal black holes \cite{Lu:2008jk, Chow:2008dp}. The main idea is to apply the limit introduced in \cite{Bardeen:1999px, Lu:2008jk, Chow:2008dp} to probe  a particular near-horizon region of extremal black holes.  In this limit the near-horizon geometry is simplified to be some circles fibered over AdS$_2$, from which one can find a Virasoro algebra and, consequently, compute the corresponding central charge and Frolov-Thorne temperature. Using the Cardy formula, the black hole entropy can be successfully reproduced for extremal black holes. The conceptual advantage of this approach is that it microscopically explains the black hole entropy based on general near-horizon symmetries and does not require intimate knowledge of the full higher-dimensional dual CFT. The situation is akin to microscopically computing the entropy of the AdS black holes without full knowledge of the UV complete theory, which in this case is the higher-dimensional boundary CFT.  We thus present steps toward certain universality of asymptotically AdS black holes. This is similar to Strominger's discussion of the entropy of asymptotically flat black holes, using only the near-horizon geometry \cite{Strominger:1997eq} without recourse to the full string theory description originally used in \cite{Strominger:1996sh}.

\begin{figure}
\begin{center}
  \includegraphics[width=13cm]{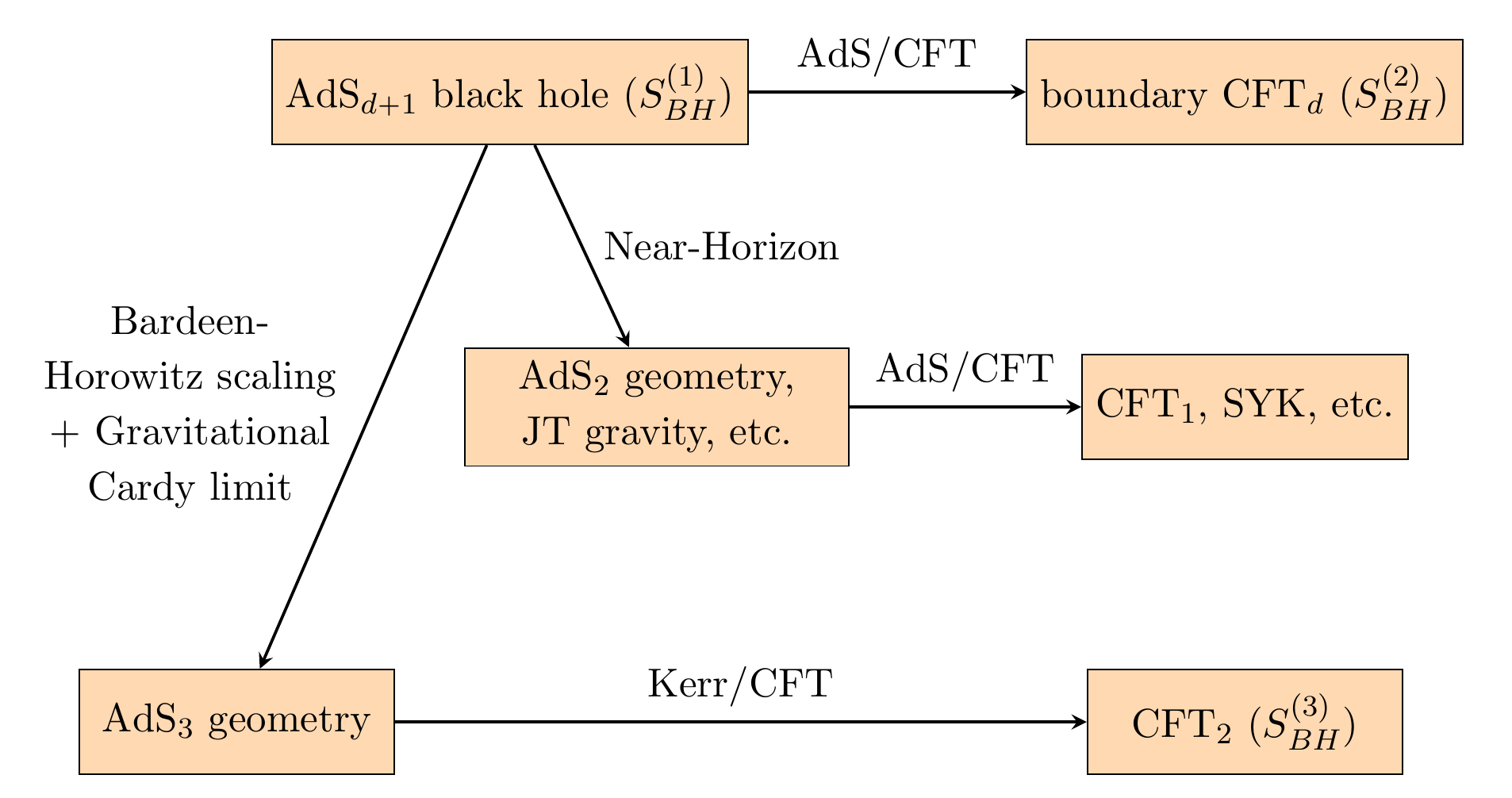}
	\caption{The Gravitational Cardy limit plays a role in the broader picture of the study of AdS black holes, giving us a near-horizon AdS$_3$ geometry, similar to that of the extremal vanishing horizon limit. The entropy can be computed in three separate regimes ($S_{BH}^{(i)}$), and are all found to be equal.} \label{fig:M1}
\end{center}
\end{figure}

In this paper, we apply the same near-horizon limit introduced in \cite{Bardeen:1999px, Lu:2008jk, Chow:2008dp} to the BPS, i.e., to  supersymmetric and extremal, asymptotically AdS black holes in various dimensions. As a new ingredient, we take the gravitational Cardy limit, and the near-horizon geometry will be further reduced to only one $U(1)$ fibered  over  AdS$_2$, or equivalently, an AdS$_3$ geometry. Hence, the gravitational Cardy limit provides the minimal amount of information for computing the black hole entropy using the Kerr/CFT correspondence. We will demonstrate this statement using several examples of asymptotically AdS black holes in various dimensions.

A pictorial way to summarize the situation is presented in Fig.~\ref{fig:M1} where we indicate how to reproduce the Bekenstein-Hawking entropy of asymptotically AdS$_{d+1}$ BPS black holes from two points of view: (i) the boundary CFT$_d$ and (ii) the near-horizon Virasoro algebra arising as the asymptotic symmetry algebra.

The near-horizon Virasoro-based results are simpler. The most important point, however, is that our explicit computations demonstrate the existence of certain universality for the class of rotating electrically charged asymptotically AdS black holes in dimensions 4, 5, 6 and 7. This universality manifests itself as an underlying AdS$_3$/CFT$_2$ correspondence living in the near-horizon region.

To further gain intuition into the gravitational Cardy limit we discuss in this paper, it is instructive to compare our approach with a somewhat related Extremal Vanishing Horizon (EVH) limit discussed in \cite{Goldstein:2019gpz} (see also \cite{SheikhJabbaria:2011gc, deBoer:2011zt, Johnstone:2013eg, Sadeghian:2015laa}).  For asymptotically AdS$_{d+1}$  BPS black holes, the authors of  \cite{Goldstein:2019gpz}  related the boundary CFT$_{d}$ and the corresponding near-horizon computations  in the EVH limit. In this case, there arises an  effective 2d CFT obtained in the near-horizon region.  However, the (near-)EVH limit is strictly restricted to the (nearly) vanishing horizon case, i.e., $A_{\text{BH}}\to 0$, $T\to0$, $A_{\text{BH}}/T\to$ constant.  In contrast,  the gravitational Cardy limit works for generic asymptotically AdS BPS black holes without restricting the horizon area. This is because the gravitational Cardy limit also rescales the conserved charges.

This paper is organized as follows. In Section~\ref{sec:AdS5} through Section~\ref{sec:AdS6} we discuss the asymptotically AdS$_5$, AdS$_4$, AdS$_7$ and AdS$_6$ cases, respectively. For each case, we first review the black hole solution, and then define the corresponding gravitational Cardy limit. After taking the gravitational Cardy limit under the Bardeen-Horowitz near-horizon scaling, we can see a structure of Virasoro algebra and compute the black hole entropy using the microscopic Cardy formula, which matches the results from the boundary CFT$_d$ and the Bekenstein-Hawking formula for AdS$_{d+1}$. Some possible extensions for  future research are discussed in Section~\ref{sec:Discussions}. In Appendix~\ref{app:NearHorizonEOM}, we present some details of verifying the equations of motion for the near-horizon geometries.

%%%%%%%%%%%%%%%%%%%%%%%%%%%%%%%%%%%%%%%%%%%
%%%%%%%%%%%%%%%%%%%%%%%%%%%%%%%%%%%%%%%%%%%
\section{Asymptotically AdS$_5$ Black Holes}\label{sec:AdS5}
%%%%%%%%%%%%%%%%%%%%%%%%%%%%%%%%%%%%%%%%%%%
%%%%%%%%%%%%%%%%%%%%%%%%%%%%%%%%%%%%%%%%%%%

In this section, we consider the asymptotically AdS$_5$ black holes and the corresponding gravitational Cardy limit. We will demonstrate that the black hole entropy can be computed in various ways as shown in Fig.~\ref{fig:M1}, and that  the other thermodynamic quantities scale  in  the gravitational Cardy limit precisely as in the field theory approach following Table~\ref{Table:Intro}.

\subsection{AdS$_5$ Black Hole Solution}

In this subsection, we first review the non-extremal asymptotically AdS$_5$ black hole solution found in \cite{Chong:2005hr} with degenerate electric charges $Q_1 = Q_2 = Q_3 = Q$ and two angular momenta $J_{1, 2}$, and then take the BPS limit to obtain its  supersymmetric version.

The non-extremal asymptotically AdS$_5$ black hole background was found in \cite{Chong:2005hr} as a solution to the equations of motion of the 5d minimal gauged supergravity in the Boyer-Lindquist coordinates $x^\mu = (t,\, r,\, \theta,\, \phi,\, \psi)$. The metric and the gauge field of the black hole solution are given by
\begin{align}
  ds^2 & = - \frac{\Delta_\theta \left[(1 + g^2 r^2) \rho^2 dt + 2 q \nu \right]\, dt}{\Xi_a \Xi_b \rho^2} + \frac{2 q \nu \omega}{\rho^2} + \frac{f}{\rho^4} \left(\frac{\Delta_\theta dt}{\Xi_a \Xi_b} - \omega \right)^2 + \frac{\rho^2 dr^2}{\Delta_r} + \frac{\rho^2 d\theta^2}{\Delta_\theta} \nonumber\\
  {} & \quad + \frac{r^2 + a^2}{\Xi_a}\, \textrm{sin}^2 \theta\, d\phi^2 + \frac{r^2 + b^2}{\Xi_b}\, \textrm{cos}^2 \theta\, d\psi^2\, ,\label{eq:AdS5Metric}\\
  A & = \frac{\sqrt{3}\, q}{\rho^2} \left(\frac{\Delta_\theta\, dt}{\Xi_a \Xi_b} - \omega\right) + \alpha_{5}\, dt\, ,
\end{align}
where
\begin{align}
\begin{split}
  \nu & \equiv b\, \textrm{sin}^2 \theta\, d\phi + a\, \textrm{cos}^2 \theta\, d\psi\, ,\\
  \omega & \equiv a\, \textrm{sin}^2 \theta\, \frac{d\phi}{\Xi_a} + b\, \textrm{cos}^2 \theta\, \frac{d\psi}{\Xi_b}\, ,\\
  \Delta_\theta & \equiv 1 - a^2 g^2\, \textrm{cos}^2 \theta - b^2 g^2\, \textrm{sin}^2 \theta\, ,\\
  \Delta_r & \equiv \frac{(r^2 + a^2) (r^2 + b^2) (1 + g^2 r^2) + q^2 + 2 a b q}{r^2} - 2 m\, , \label{eq:Delta-r}\\
  \rho^2 & \equiv r^2 + a^2\, \textrm{cos}^2 \theta + b^2\, \textrm{sin}^2 \theta\, ,\\
  \Xi_a & \equiv 1 - a^2 g^2\, ,\\
  \Xi_b & \equiv 1 - b^2 g^2\, ,\\
  f & \equiv 2 m \rho^2 - q^2 + 2 a b q g^2 \rho^2\, ,
\end{split}
\end{align}
and $\alpha_{5}\, dt$ is a pure gauge term with $\alpha_5$ a constant. These black hole solutions are characterized by four independent parameters $(a, b, m, q)$. The thermodynamical quantities, including the mass $E$, the temperature $T$ and the entropy $S$, can all be expressed in terms of these independent parameters. The other physical quantities, such as the electric charge $Q$, the electric potential $\Delta$, the angular momenta $J_{1, 2}$ and the angular velocities $\Omega_{1, 2}$ can similarly be written in terms of the four independent parameters. For example,  the gravitational angular velocities $\Omega_{1, 2}$ and the temperature $T$ are given by
\begin{align}
\begin{split}\label{eq:AdS5thermo}
  \Omega_1 & = \frac{a (r_+^2 + b^2) (1 + g^2 r_+^2) + b q}{(r_+^2 + a^2) (r_+^2 + b^2) + a b q}\, ,\\
  \Omega_2 & = \frac{b (r_+^2 + a^2) (1 + g^2 r_+^2) + a q}{(r_+^2 + a^2) (r_+^2 + b^2) + a b q}\, ,\\
  T & = \frac{r_+^4 \Big[1 + g^2 (2 r_+^2 + a^2 + b^2) \Big] - (a b + q)^2}{2 \pi r_+ \Big[ (r_+^2 + a^2) (r_+^2 + b^2) + a b q \Big]}\, ,
\end{split}
\end{align}
where $r_+$ denotes the position of the outer horizon given by the largest root of $\Delta_r$ in \eqref{eq:Delta-r}.

As carefully discussed in \cite{Cabo-Bizet:2018ehj}, it is crucial to make the following important distinctions of these solutions, in the broader context when complex potentials are allowed. The extremal black hole solution is characterized by the appearance of a double root in  $\Delta_r = 0$, while the BPS black hole solution is obtained by solving the supersymmetry equations.

The BPS limit is a special limit in the parameter space, such that the backgrounds in this limit are both extremal and supersymmetric. For the class of AdS$_5$ black hole solutions \eqref{eq:AdS5Metric}, the BPS limit corresponds to the following condition
\be\label{eq:AdS5 BPScond-1}
  q = \frac{m}{1 + a g + b g}\, .
\ee
Moreover, to prevent unphysical naked closed timelike curves (CTC), it is shown in \cite{Chong:2005hr} that the BPS solutions should further satisfy the constraint
\be\label{eq:AdS5 BPScond-2}
  m = \frac{1}{g} (a + b) (1 + a g) (1 + b g) (1 + a g + b g)\, .
\ee
Hence, in the BPS limit only two of the four parameters $(a, b, m, q)$ are independent, which can be chosen to be $(a, b)$. The special case $a = b$ corresponds to the supersymmetric AdS$_5$ black hole solutions found by Gutowski and Reall \cite{Gutowski:2004ez}. In the BPS limit, the outer horizon $r_+$ coincides with the inner horizon at $r_0$
\be
  r_0^2 = \frac{a + b + a b g}{g}\, ,
\ee
and the black hole entropy $S_*$, the electric charge $Q_*$ and the angular momenta $J_{1, 2}^*$ have the following expressions in terms of $(a, b)$
\begin{align}
\begin{split}\label{eq:AdS5 BPSthermo}
  S_* & = \frac{\pi^2 (a + b) \sqrt{a + b + a b g}}{2 g^{3/2} (1 - a g) (1 - b g)}\, ,\\
  Q_* & = \frac{\pi (a + b)}{4 g (1 - a g) (1 - b g)}\, ,\\
  J_1^* & = \frac{\pi (a + b) (2 a + b + a b g)}{4 g (1 - a g)^2 (1 - b g)}\, ,\\
  J_2^* & = \frac{\pi (a + b) (a + 2 b + a b g)}{4 g (1 - a g) (1 - b g)^2}\, ,
\end{split}
\end{align}
where the entropy $S_*$ is computed from the Bekenstein-Hawking entropy formula
\be
  S_{BH} = \frac{A}{4 G_N}\, ,
\ee
a quarter of the horizon area in units of Planck length.  Using the expressions \eqref{eq:AdS5 BPSthermo}, we can also rewrite the black hole entropy as a function of $Q$ and $J_{1, 2}$
\be\label{eq:AdS5 S area to bdy CFT}
  S_{BH} = 2 \pi \sqrt{\frac{3 Q^2}{g^2} - \frac{\pi}{4 g^3} (J_1 + J_2)}\, .
\ee

The AdS/CFT dictionary helps translate the parameters of the AdS$_5$ black holes to quantities in $\mathcal{N}=4$ SYM
\be
 \frac{1}{2} N^2 = \frac{\pi}{4 G_N} \ell_5^3,
\ee
with $\ell_5 = g^{-1}$ denoting the AdS$_5$ radius. We can rewrite the expression \eqref{eq:AdS5 S area to bdy CFT} of the AdS$_5$ black hole entropy (in the unit $G=1$)
\be\label{eq:AdS5 S area to bdy CFT New}
  S_{BH} = 2 \pi \sqrt{\frac{3 Q^2}{g^2} - \frac{N^2}{2} (J_1 + J_2)}\, .
\ee
This expression has recently been extracted  directly from the boundary CFT in \cite{Cabo-Bizet:2018ehj, Choi:2018hmj, Benini:2018ywd} with further clarifying field theory work presented in \cite{Honda:2019cio, ArabiArdehali:2019tdm, Kim:2019yrz, Cabo-Bizet:2019osg, Amariti:2019mgp, Lezcano:2019pae, Lanir:2019abx, Larsen:2019oll,Cabo-Bizet:2019eaf,ArabiArdehali:2019orz}. We show below that this boundary CFT result can also be obtained from a particular near-horizon Cardy formula.

\subsection{Gravitational Cardy Limit}

The Cardy-like limit for the $\mathcal{N}=4$ SYM index was defined in \cite{Choi:2018hmj}. This limit has been discussed in the context of ${\cal N}=4 $ SYM also in \cite{Honda:2019cio, ArabiArdehali:2019tdm, Cabo-Bizet:2019eaf}. In the more general context of ${\cal N}=1$ superconformal theories, it has been discussed in \cite{Cabo-Bizet:2019osg, Kim:2019yrz, Amariti:2019mgp}. A key ingredient in the limit is the regime
\be\label{eq:CardyLimitAdS5}
   |\omega_i| \ll 1\, ,\quad \Delta_I \sim \mathcal{O} (1)\, ,\qquad (\textrm{$i = 1,\, 2$};\, \textrm{$I = 1,\, 2,\, 3$})\, .
\ee
Using the relation found in \cite{Cabo-Bizet:2018ehj, Larsen:2019oll}
\be\label{eq:Rel of temp derivative}
  \textrm{Re} (\omega_i) = \frac{\partial \Omega_i}{\partial T}\bigg|_{T=0}\, ,\quad \textrm{Re} (\Delta_I) = \frac{\partial \Phi_I}{\partial T}\bigg|_{T=0}\, ,
\ee
we can express the Cardy-like limit \eqref{eq:CardyLimitAdS5} in terms of quantities in the dual gravity theory, such that
\be
  \bigg|\left(\frac{\partial \Omega_i}{\partial T}\right)_{T=0} \bigg| \ll 1\, ,\quad \frac{\partial \Phi_I}{\partial T}\bigg|_{T=0} \sim \mathcal{O} (1)\, ,
\ee
with $i = 1,\, 2$ and $I = 1,\, 2,\, 3$. Using the expressions of the thermodynamic quantities \eqref{eq:AdS5thermo}, we obtain for the asymptotically AdS$_5$ BPS black holes,
\begin{align}
\begin{split}\label{eq:dOmega dT BPS}
  \frac{\partial \Omega_1}{\partial T} \bigg|_{\textrm{BPS}} & = \lim_{T \to 0} \frac{\Omega_1 - \Omega_1^*}{T} = \frac{2 \pi (-1 + a g)}{3 g} \sqrt{\frac{1 + a g + b g}{a b}}\, ,\\
  \frac{\partial \Omega_2}{\partial T} \bigg|_{\textrm{BPS}} & = \lim_{T \to 0} \frac{\Omega_2 - \Omega_2^*}{T} = \frac{2 \pi (-1 + b g)}{3 g} \sqrt{\frac{1 + a g + b g}{a b}}\, ,
\end{split}
\end{align}
where $\Omega^*_{1, 2}$ are the values of $\Omega_{1, 2}$ in the BPS limit. From the expressions of $\frac{\partial \Omega_i}{\partial T} \big|_{\textrm{BPS}}$ ($i = 1, 2$), we conclude that for asymptotically AdS$_5$ BPS black holes, the gravitational Cardy limit corresponds to the special limit of the parameters on the gravity side
\be
  a \to \frac{1}{g}\, ,\quad b \to \frac{1}{g}\, .
\ee
For later convenience, we parameterize $a$ and $b$ as
\be\label{eq:AdS5CardyLimit}
  a = \frac{1}{g} - \epsilon\, ,\quad b = \frac{1}{g} - \epsilon\, .
\ee
For this case, $\epsilon$ has the dimension of length. Taking the gravitational Cardy limit \eqref{eq:AdS5CardyLimit} for the parameters into account, the BPS thermodynamic quantities \eqref{eq:AdS5 BPSthermo} become
\begin{align}
\begin{split}\label{eq:AdS5 BPSthermo Cardy limit}
  S_* & = \frac{\sqrt{3} \pi^2}{g^5 \epsilon^2} + \mathcal{O} (\epsilon^{-1})\, ,\\
  Q_* & = \frac{\pi}{2 g^4 \epsilon^2} + \mathcal{O} (\epsilon^{-1})\, ,\\
  J_1^* & = \frac{2 \pi}{g^6 \epsilon^3} + \mathcal{O} (\epsilon^{-2})\, ,\\
  J_2^* & = \frac{2 \pi}{g^6 \epsilon^3} + \mathcal{O} (\epsilon^{-2})\, ,
\end{split}
\end{align}
which are precisely the scalings of the field theory results \cite{Choi:2018hmj, ZaffaroniSlides}.

%%%%%%%%%%%%%%%%%%%%%%%%%%%%%%%%%%%%%%%%%%%%%%%%%%%%%%%%%%

\subsection{Black Hole Solution in the Near-Horizon + Gravitational Cardy Limit}

In the previous subsection, we have obtained the gravitational Cardy limit for the parameters on the gravity side. In this subsection, we discuss how the near-horizon metric changes in this limit as well as clarify other ingredients.  

The asymptotically AdS$_5$ metric \eqref{eq:AdS5Metric} can be written in the following equivalent form, which is more convenient for the discussions in this subsection,
\be\label{eq:AdS5MetricNew}
  ds^2 = - \frac{\Delta_r \Delta_\theta r^2\, \textrm{sin}^2 (2 \theta)}{4 \Xi_a^2 \Xi_b^2 B_\phi B_\psi} dt^2 + \rho^2 \left(\frac{dr^2}{\Delta_r} + \frac{d\theta^2}{\Delta_\theta} \right) + B_\psi (d\psi + v_1 d\phi + v_2 dt)^2 + B_\phi (d\phi + v_3 dt)^2\, ,
\ee
where
\begin{align}
\begin{split}
  B_\phi \equiv \frac{g_{33}\, g_{44} - g_{34}^2}{g_{44}}\, , & \qquad B_\psi \equiv g_{44}\, ,\\
  v_1 \equiv \frac{g_{34}}{g_{44}}\, ,\quad v_2 \equiv \frac{g_{04}}{g_{44}}\, , & \quad v_3 \equiv \frac{g_{04}\, g_{34} - g_{03}\, g_{44}}{g_{34}^2 - g_{33}\, g_{44}}\, ,
\end{split}
\end{align}
with the non-vanishing components of the metric \eqref{eq:AdS5Metric} explicitly in the coordinates $(t, r, \theta, \varphi, \psi)$
\begin{align}
\begin{split}
  g_{00} & = - \frac{\Delta_\theta (1 + g^2 r^2)}{\Xi_a \Xi_b} + \frac{\Delta_\theta^2 (2 m \rho^2 - q^2 + 2 a b q g^2 \rho^2)}{\rho^4 \Xi_a^2 \Xi_b^2}\, ,\\
  g_{03} & = g_{30} = - \frac{\Delta_\theta \Big[a (2 m \rho^2 - q^2) + b q \rho^2 (1 + a^2 g^2) \Big]\, \textrm{sin}^2 \theta}{\rho^4 \Xi_a^2 \Xi_b}\, ,\\
  g_{04} & = g_{40} = - \frac{\Delta_\theta \Big[b (2 m \rho^2 - q^2) + a q \rho^2 (1 + b^2 g^2) \Big]\, \textrm{cos}^2 \theta}{\rho^4 \Xi_b^2 \Xi_a}\, ,\\
  g_{11} & = \frac{\rho^2}{\Delta_r}\, ,\qquad\qquad g_{22} = \frac{\rho^2}{\Delta_\theta}\, ,\\
  g_{33} & = \frac{(r^2 + a^2)\, \textrm{sin}^2 \theta}{\Xi_a} + \frac{a \Big[a (2 m \rho^2 - q^2) + 2 b q \rho^2 \Big]\, \textrm{sin}^4 \theta}{\rho^4 \Xi_a^2}\, ,\\
  g_{44} & = \frac{(r^2 + b^2)\, \textrm{cos}^2 \theta}{\Xi_b} + \frac{b \Big[b (2 m \rho^2 - q^2) + 2 a q \rho^2 \Big]\, \textrm{cos}^4 \theta}{\rho^4 \Xi_b^2}\, ,\\
  g_{34} & = g_{43} = \frac{\Big[a b (2 m \rho^2 - q^2) + (a^2 + b^2) q \rho^2 \Big]\, \textrm{sin}^2 \theta\, \textrm{cos}^2 \theta}{\rho^4 \Xi_a \Xi_b}\, .
\end{split}
\end{align}

A central element in our approach is a near-horizon limit following the prescription of Bardeen and Horowitz \cite{Bardeen:1999px} to zoom into a near-horizon region, and at the same time we move to a rotating frame by  implementing the following coordinate change
\be\label{eq:AdS5scaling}
  r \to r_0 + \lambda\, \widetilde{r}\, ,\quad t \to \frac{\widetilde{t}}{\lambda}\, ,\quad \phi \to \widetilde{\phi} + g \frac{\widetilde{t}}{\lambda}\, ,\quad \psi \to \widetilde{\psi} + g \frac{\widetilde{t}}{\lambda}\, .
\ee
Taking $\lambda \to 0$ brings us to a particular near-horizon region of the AdS$_5$ BPS black holes given by the following metric in the coordinates $(\widetilde{t}, \widetilde{r}, \theta, \widetilde{\phi}, \widetilde{\psi})$
\begin{align}
  ds^2 & = - \frac{2 g (1 + 5 a g)}{a (1 + a g)^2}\, \widetilde{r}^2\, d\widetilde{t}^2 + \frac{a}{2 g (1 + 5 a g)} \frac{d\widetilde{r}^2}{\widetilde{r}^2} + \Lambda_{\textrm{AdS$_5$}} (\theta) \left[d\widetilde{\phi} + \frac{3 g (1 - a g)}{(1 + a g) \sqrt{a \left(a + \frac{2}{g} \right)}} \, \widetilde{r}\, d\widetilde{t} \right]^2 \nonumber\\
  {} & \quad + \frac{a \Big(4 - a g + 3 a g\, \textrm{cos} (2 \theta) \Big)\, \textrm{cos}^2 \theta}{2 g (1 - a g)^2} \left[d\widetilde{\psi} + \frac{6 a g\, \textrm{sin}^2 \theta}{4 - a g + 3 a g\, \textrm{cos} (2 \theta)} d\widetilde{\phi} + V(\theta)\, \widetilde{r}\, d\widetilde{t} \right]^2 \nonumber\\
  {} & \quad + \frac{2 a}{g (1 - a g)}\, d\theta^2\, ,\label{eq:AdS5CP NHmetric}
\end{align}
where
\begin{align}
  \Lambda_{\textrm{AdS$_5$}} (\theta) & \equiv \frac{4 a (2 + a g)\, \textrm{sin}^2 \theta}{g (1 - a g) \Big(4 - a g + 3 a g\, \textrm{cos} (2 \theta) \Big)}\, ,\\
  V (\theta) & \equiv \frac{6 g^2 (1 - a g) \sqrt{a \left(a + \frac{2}{g} \right)}}{a (1 + a g) \left(4 - a g + 3 a g\, \textrm{cos} (2 \theta) \right)}\, ,
\end{align}
and for simplicity, we have set $a = b$, in consistency with the gravitational Cardy limit \eqref{eq:AdS5CardyLimit} that will be imposed later. For some special values of $\theta$, the metric \eqref{eq:AdS5CP NHmetric} has the topology of two $U(1)$ circles fibered over  the AdS$_2$ parametrized by $(\tilde{t},\tilde{r})$, as pointed out in \cite{Lu:2008jk, Chow:2008dp}.

After a further change of coordinates
\be
  \tau \equiv \frac{2 g (1 + 5 a g)}{a (1 + a g)}\, \widetilde{t}\, ,
\ee
we can bring the metric \eqref{eq:AdS5CP NHmetric} into the form
\begin{align}
  ds^2 & = \frac{a}{2 g (1 + 5 a g)} \left[- \widetilde{r}^2\, d\tau^2 + \frac{d\widetilde{r}^2}{\widetilde{r}^2}\right] + \Lambda_{\textrm{AdS$_5$}} (\theta) \left[d\widetilde{\phi} + \frac{3 a (1 - a g)}{2 (1 + 5 a g) \sqrt{a \left(a + \frac{2}{g} \right)}} \, \widetilde{r}\, d\tau \right]^2 \nonumber\\
  {} & \quad + \frac{a \Big(4 - a g + 3 a g\, \textrm{cos} (2 \theta) \Big)\, \textrm{cos}^2 \theta}{2 g (1 - a g)^2} \left[d\widetilde{\psi} + \frac{6 a g\, \textrm{sin}^2 \theta}{4 - a g + 3 a g\, \textrm{cos} (2 \theta)} d\widetilde{\phi} + \widetilde{V} (\theta)\, \widetilde{r}\, d\tau \right]^2 \nonumber\\
  {} & \quad + \frac{2 a}{g (1 - a g)}\, d\theta^2\, ,\label{eq:AdS5 NHmetricNew}
\end{align}
where
\be
  \widetilde{V} (\theta) \equiv \frac{3 g (1 - a g) \sqrt{a \left(a + \frac{2}{g} \right)}}{(1 + 5 a g) \left(4 - a g + 3 a g\, \textrm{cos} (2 \theta) \right)}\, .
\ee
In both U(1) fibrations, the coefficients in front of $\widetilde{r}\, d\tau$ are proportional to $\partial_T \Omega$ \eqref{eq:dOmega dT BPS} with $a=b$. Hence, according to the relation \eqref{eq:Rel of temp derivative}, $\omega_i$ from $\mathcal{N}=4$ SYM indeed play the role of angular velocities in the metric \eqref{eq:AdS5 NHmetricNew}, and the Cardy-like limit from the field theory side means the angular velocities slow down on some $U(1)$ circles in the near-horizon metric \eqref{eq:AdS5 NHmetricNew}.

In Appendix~\ref{app:AdS5}, we verify explicitly that the resulting background is a solution of the 5d minimal gauged supergravity equations of motion. This statement holds for arbitrary values of $a = b$.  Up to this point, our approach is completely rigorous and verifying the equations of motion explicitly provides a powerful seal of approval. However, to flesh out the scaling properties of the solution, in what follows we implement the gravitational Cardy limit in the space of parameters which further simplifies the geometry.

We apply the gravitational Cardy limit \eqref{eq:AdS5CardyLimit} to the metric \eqref{eq:AdS5 NHmetricNew} and keep the leading orders in $\epsilon$, which leads to
\begin{align}
  ds^2 & = \frac{1}{12 g^2} \left[- \widetilde{r}^2\, d\tau^2 + \frac{d\widetilde{r}^2}{\widetilde{r}^2}\right] - \frac{2}{g^3 \epsilon}\, d\theta^2 - \frac{4\, \textrm{sin}^2 (\theta)\, \epsilon}{g^3 (1 + \textrm{cos} (2 \theta))} \left[\frac{1}{\epsilon} d\widetilde{\phi} - \frac{g}{4 \sqrt{3}}\, \widetilde{r}\, d\tau \right]^2 \nonumber\\
  {} & \quad + \frac{3\, \textrm{cos}^4 (\theta)}{g^4} \left[\frac{1}{\epsilon} d\widetilde{\psi} + \frac{2\, \textrm{sin}^2 (\theta)}{\epsilon \left(1 + \textrm{cos} (2 \theta) \right)} d\widetilde{\phi} - \frac{g\, \textrm{sec}^2 (\theta)}{4 \sqrt{3}} \, \widetilde{r}\, d\tau \right]^2\, .\label{eq:AdS5 NHmetricNew 2}
\end{align}
From this metric, we can see that in the gravitational Cardy limit $\epsilon \to 0$ only one U(1) circle remains non-trivially fibered over AdS$_2$. We have only assumed that $\epsilon$ is small without strictly taking the limit $\epsilon \to 0$, and the near-horizon metric will approximate to AdS$_3$, as $\epsilon$ becomes smaller. However, since the two initial U(1) fibrations give the same result of the black hole entropy according to the Cardy formula and the extreme black hole/CFT correspondence \cite{Chow:2008dp}, the remaining U(1) is enough to compute the AdS$_5$ black hole entropy. We will demonstrate this point in the next subsection.  To summarize, the gravitational Cardy limit simplifies the near-horizon geometry but keeps the minimal amount of information for computing the black hole entropy.

Let us finish by warning the potentially puzzled reader.  The analysis above, surrounding equation \eqref{eq:AdS5 NHmetricNew 2}, is local and has the sole intention of clarifying the geometry of the gravitational Cardy limit. If bothered by this last limiting procedure it is possible to step back and derive all the quantities from the safer background  obtained in equation \eqref{eq:AdS5 NHmetricNew}. However, without this gravitational Cardy limit the connection to the field theory approach would be very tenuous.

%%%%%%%%%%%%%%%%%%%%%%%%%%%%%%%%%%%%%%%%%%%%%%%%%%%%%%%%%%
\subsection{Black Hole Entropy from Cardy Formula}\label{sec:Review Cardy}

In the previous subsection, we showed that a warped AdS$_3$ geometry appears in the near-horizon region of asymptotically AdS$_5$ BPS black holes in the gravitational Cardy limit. This circumstance permits the use of ideas presented in \cite{Guica:2008mu}, which lead to the identification of a Virasoro algebra as the asymptotic symmetries in the near-horizon geometry and, subsequently, to a microscopic description of the black hole entropy via the Cardy formula.

Let us briefly review how the Virasoro algebra emerges as the algebra of asymptotic symmetries of the near-horizon region of the extremal Kerr black hole \cite{Guica:2008mu} (see also \cite{Bredberg:2011hp}). Recall that the asymptotic symmetry group is the group of all allowed diffeomorphisms modulo trivial ones where allowed diffeomorphisms are defined as those that preserve certain boundary conditions of the asymptotic metric. The starting element in determining the algebra of asymptotic symmetries is, therefore,  to consider diffeomorphims generated by vectors of the form
\be
  \zeta_\epsilon = \epsilon (\phi) \frac{\partial}{\partial \phi} - r\, \epsilon' (\phi) \frac{\partial}{\partial r}\, ,
\ee
where $\epsilon (\phi)$ is a function periodic in $\phi$. For simplicity we can choose to be $\epsilon (\phi) = - e^{- i n \phi}$, and consequently obtain the mode expansion of $\zeta_\epsilon$ as
\be
  \zeta_{(n)} = - e^{- i n \phi} \frac{\partial}{\partial \phi} - i n r e^{- i n \phi} \frac{\partial}{\partial r}\, ,
\ee
which satisfies a centreless Virasoro algebra
\be
  i [\zeta_{(m)},\, \zeta_{(n)}] = (m - n) \zeta_{(m+n)}\, .
\ee
The charge associated with the diffeomorphis $\zeta_\epsilon$ is given by an integral over the boundary of a spatial slice $\partial \Sigma$
\be
  Q_\zeta = \frac{1}{8 \pi G} \int_{\partial \Sigma} k_\zeta\, ,
\ee
where $k_\zeta$ is a 2-form defined for a general perturbation $h_{\mu\nu}$ around the background metric $g_{\mu\nu}$
\begin{align}
  k_\zeta [h,\, g] & \equiv - \frac{1}{4} \epsilon_{\alpha\beta\mu\nu} \Big[\zeta^\nu D^\mu h - \zeta^\nu D_\sigma h^{\mu\sigma} + \zeta_\sigma D^\nu h^{\mu\sigma} + \frac{1}{2} h D^\nu \zeta^\mu - h^{\nu\sigma} D_\sigma \zeta^\mu \nonumber\\
  {} & \qquad\qquad\quad + \frac{1}{2} h^{\sigma\nu} (D^\mu \zeta_\sigma + D_\sigma \zeta^\mu)\Big] dx^\alpha \wedge dx^\beta\, ,
\end{align}
with $h \equiv h_{\alpha\beta} g^{\alpha\beta}$. The Dirac bracket of the charges is
\be\label{eq:DiracBracket}
  \{Q_{\zeta_{(m)}},\, Q_{\zeta_{(n)}} \} = Q_{[\zeta_{(m)},\, \zeta_{(n)}]} + \frac{1}{8 \pi G} \int_{\partial \Sigma} k_\zeta [\mathcal{L}_\zeta g,\, g]\, ,
\ee
where $\mathcal{L}_\zeta$ denotes the Lie derivative with respect to $\zeta$
\be
  \mathcal{L}_\zeta g_{\mu\nu} \equiv \zeta^\rho \partial_\rho g_{\mu\nu} + g_{\rho\nu} \partial_\mu \zeta^\rho + g_{\mu\rho} \partial_\nu \zeta^\rho\, .
\ee
The mode expansion of the Dirac bracket \eqref{eq:DiracBracket} leads to a Virasora algebra
\be
  [L_m,\, L_n] = (m - n) L_{m+n} + \frac{1}{12} c_L\, (m^3 + \alpha m)\, \delta_{m+n,\, 0}\, ,
\ee
where $c_L$ can be obtained from the integral
\be\label{eq:AdS5 VirasoroIntegral}
  \frac{1}{8 \pi G} \int_{\partial \Sigma} k_{\zeta_{(m)}} [\mathcal{L}_{\zeta_{(n)}} g,\, g] = - \frac{i}{12} c_L\, (m^3 + \alpha m)\, \delta_{m+n,\, 0}\, ,
\ee
and $\alpha$ is an irrelevant constant.

To compute the black hole entropy using the Cardy formula, we still need the Frolov-Thorne temperature, which can be obtained in the following way. The quantum fields on the background \eqref{eq:AdS5Metric} can be expanded in the modes $e^{- i \omega t + i m \phi}$. After taking the scaling \eqref{eq:AdS5scaling}, these modes become
\be
  e^{- i \omega t + i m \phi} = e^{- i \omega \frac{\widetilde{t}}{\lambda} + i m \left(\widetilde{\phi} + g \frac{\widetilde{t}}{\lambda} \right)} = e^{- i \left(\frac{\omega}{\lambda} - \frac{m g}{\lambda} \right) \widetilde{t} + i m \widetilde{\phi}} \equiv e^{- i n_R \widetilde{t} + i n_L \widetilde{\phi}}\, ,
\ee
from which we can read off the left-moving and the right-moving mode numbers
\be\label{eq:ModeNum}
  n_L \equiv m\, ,\quad n_R \equiv \frac{\omega - m g}{\lambda}\, .
\ee
The Boltzmann factor is
\be\label{eq:BoltzmannFactor}
  e^{- \frac{\omega - m \Omega}{T_H}} = e^{- \frac{n_L}{T_L} - \frac{n_R}{T_R}}\, ,
\ee
where $T_H$ is the Hawking temperature, and $T_{L, R}$ are the left-moving and the right-moving Frolov-Thorne temperatures. Combining \eqref{eq:ModeNum} and \eqref{eq:BoltzmannFactor}, we obtain the near-extremal Frolov-Thorne temperatures
\be
  T_L = \frac{T_H}{g - \Omega}\, ,\quad T_R = \frac{T_H}{\lambda}\, .
\ee
The values for the extremal AdS$_5$ black holes can be obtained by taking the extremal limit ($T_H \to 0$).

In order to apply the technique described above, we need to first transform the AdS$_2$ Poincar\'e coordinates $(\widetilde{r}, \tau)$ in the metric \eqref{eq:AdS5 NHmetricNew} to global coordinates $(\hat{r}, \hat{t})$
\be\label{eq:PoincareToGlobal 1}
  g\, \widetilde{r} = \hat{r} + \sqrt{1 + \hat{r}^2}\, \textrm{cos} (\hat{t})\, ,\quad g^{-1}\, \tau = \frac{\sqrt{1 + \hat{r}^2}\, \textrm{sin} (\hat{t})}{\hat{r} + \sqrt{1 + \hat{r}^2}\, \textrm{cos} (\hat{t})}\, ,
\ee
which leads to
\begin{align}
\begin{split}\label{eq:PoincareToGlobal 2}
  - \widetilde{r}^2\, d\tau^2 + \frac{d\widetilde{r}^2}{\widetilde{r}^2} & = - (1 + \hat{r}^2)\, d\hat{t}^2 + \frac{d\hat{r}^2}{1 + \hat{r}^2}\, ,\\
  \widetilde{r}\, d\tau & = \hat{r}\, d\hat{t} + d\gamma\, ,
\end{split}
\end{align}
where
\be\label{eq:PoincareToGlobal 3}
  \gamma \equiv \textrm{log} \left(\frac{1 + \sqrt{1 + \hat{r}^2}\, \textrm{sin} (\hat{t})}{\textrm{cos} (\hat{t}) + \hat{r}\, \textrm{sin} (\hat{t})} \right)\, .
\ee
Consequently, the near-horizon metric \eqref{eq:AdS5 NHmetricNew} of the AdS$_5$ BPS black holes can be written as
\begin{align}
  ds^2 & = \frac{a}{2 g (1 + 5 a g)} \left[- (1 + \hat{r}^2)\, d\hat{t}^2 + \frac{d\hat{r}^2}{1 + \hat{r}^2} \right] + \Lambda_{\textrm{AdS$_5$}} (\theta) \left[d\hat{\phi} + \frac{3 a (1 - a g)}{2 (1 + 5 a g) \sqrt{a \left(a + \frac{2}{g} \right)}} \, \hat{r}\, d\hat{t} \right]^2 \nonumber\\
  {} & \quad + \frac{a \Big(4 - a g + 3 a g\, \textrm{cos} (2 \theta) \Big)\, \textrm{cos}^2 \theta}{2 g (1 - a g)^2} \left[d\hat{\psi} + \frac{6 a g\, \textrm{sin}^2 \theta}{4 - a g + 3 a g\, \textrm{cos} (2 \theta)} d\hat{\phi} + \widetilde{V} (\theta)\, \hat{r}\, d\hat{t} \right]^2 \nonumber\\
  {} & \quad + \frac{2 a}{g (1 - a g)}\, d\theta^2\, ,\label{eq:AdS5 NHmetricNew3}
\end{align}
where
\be
  \hat{\phi} \equiv \widetilde{\phi} + \frac{3 a (1 - a g) \gamma}{2 (1 + 5 a g) \sqrt{a \left(a + \frac{2}{g} \right)}} \, ,\quad \hat{\psi} \equiv \widetilde{\psi} + \frac{3 a (1 - a g) \gamma}{2 (1 + 5 a g) \sqrt{a \left(a + \frac{2}{g} \right)}}\, .
\ee

Applying the formalism reviewed in this subsection, we can compute the central charge and the extremal Frolov-Thorne temperature in the near-horizon region of the asymptotically AdS$_5$ BPS black hole solutions \eqref{eq:AdS5 NHmetricNew3}

\begin{align}
%  c_L & = \frac{9 \pi a^2 \sqrt{2 a + a^2 g}}{g^{3/2} (1 - a g) (1 + 5 a g) \sqrt{a \left(a + \frac{2}{g} \right)}}\, ,\label{eq:AdS5 cL}\\
  c_L & = \frac{9 \pi a^2}{g (1 - a g) (1 + 5 a g)}\, ,\\
  T_L & = \frac{1 + 5 a g}{3 a (1 - a g) \pi} \sqrt{a \left(a + \frac{2}{g} \right)}\, .\label{eq:AdS5 TL}
\end{align}
The BPS black hole entropy in this case is then given by the Cardy formula
\be\label{eq:AdS5 S Cardy}
  S_{BH} = \frac{\pi^2}{3} c_L T_L = \frac{\pi^2 a \sqrt{2 a + a^2 g}}{g^{3/2} (1 - a g)^2}\, ,
\ee
which is the same as the result from gravity \eqref{eq:AdS5 BPSthermo} with $a = b$. In fact, we can also apply the formalism discussed in this subsection to the near-horizon metric in the gravitational Cardy limit \eqref{eq:AdS5 NHmetricNew 2}, which can be recast into the global coordinates
\begin{align}
  ds^2 & = \frac{1}{12 g^2} \left[- (1 + \hat{r}^2)\, d\hat{t}^2 + \frac{d\hat{r}^2}{1 + \hat{r}^2}\right] - \frac{2}{g^3 \epsilon}\, d\theta^2 - \frac{4\, \textrm{sin}^2 (\theta)\, \epsilon}{g^3 (1 + \textrm{cos} (2 \theta))} \left[\frac{1}{\epsilon} d\hat{\phi} - \frac{g}{4 \sqrt{3}}\, \hat{r}\, d\hat{t} \right]^2 \nonumber\\
  {} & \quad + \frac{3\, \textrm{cos}^4 (\theta)}{g^4} \left[\frac{1}{\epsilon} d\hat{\psi} + \frac{2\, \textrm{sin}^2 (\theta)}{\epsilon \left(1 + \textrm{cos} (2 \theta) \right)} d\hat{\phi} - \frac{g\, \textrm{sec}^2 (\theta)}{4 \sqrt{3}} \, \hat{r}\, d\hat{t} \right]^2 \, .
\end{align}
The corresponding central charge and the extremal Frolov-Thorne temperature are

\be
  c_L = \frac{3 \pi}{2 g^4 \epsilon}\, ,\quad T_L = \frac{2 \sqrt{3}}{\pi g \epsilon}\, .
\ee
The black hole entropy can obtained from the Cardy formula
\be
  S_{BH} = \frac{\pi^2}{3} c_L T_L = \frac{\sqrt{3} \pi^2}{g^5 \epsilon^2}\, ,
\ee
which exactly matches the gravity result in the gravitational Cardy limit \eqref{eq:AdS5 BPSthermo Cardy limit}.

%%%%%%%%%%%%%%%%%%%%%%%%%%%%%%%%%%%%%%%%5
\subsection{Comparison with Results from Boundary CFT}

The asymptotically AdS$_5$ BPS black hole entropy can also be obtained from the boundary $\mathcal{N}=4$ SYM by extremizing an entropy function \cite{Cabo-Bizet:2018ehj, Choi:2018hmj, Benini:2018ywd} originally motivated in \cite{Hosseini:2017mds}  and more recently studied in \cite{Hosseini:2019iad}. One can first compute the free energy in the large-$N$ limit using the partition function via localization or the 4d superconformal index. The entropy function is then defined as the Legendre transform of the free energy in the large-$N$ limit
\be\label{eq:AdS5EntropyFct}
  S (\Delta_I,\, \omega) = \frac{N^2}{2} \frac{\Delta_1 \Delta_2 \Delta_3}{\omega_1 \omega_2} + \sum_{I=1}^3 Q_I \Delta_I + \sum_{i=1}^2 J_i \omega_i - \Lambda \left(\sum_I \Delta_I - \sum_i \omega - 2 \pi i \right)\, .
\ee
In the Cardy-like limit \eqref{eq:CardyLimitAdS5}
\be
  \omega \sim \epsilon\, ,\quad \Delta_I \sim \mathcal{O} (1)\, ,
\ee
we can read off from the entropy function \eqref{eq:AdS5EntropyFct}
\be
  S \sim \frac{1}{\epsilon^2}\, ,\quad J \sim \frac{1}{\epsilon^3}\, ,\quad Q_I \sim \frac{1}{\epsilon^2}\, ,
\ee
which have been summarized in Table~\ref{Table:Intro}.

The electric charges $Q_I$ and the angular momenta $J_i$ are real, while the chemical potentials $\Delta_I$ and the angular velocities $\omega_i$ can be complex, and so can the entropy function $S$. By requiring that the black hole entropy $S_{BH}$ be real after extremizing the entropy function $S$, we obtain one more constraint on $Q_I$ and $J_i$. More precisely, the asymptotically AdS$_5$ black hole entropy is given by \cite{Cabo-Bizet:2018ehj, Choi:2018hmj, Benini:2018ywd}
\be
  S_{BH} = 2 \pi \sqrt{Q_1 Q_2 + Q_2 Q_3 + Q_3 Q_1 - \frac{N^2}{2} (J_1 + J_2)}\, ,
\ee
subject to the constraint
\begin{align}
  & \Bigg(Q_1 + Q_2 + Q_3 + \frac{N^2}{2} \Bigg)\,  \Bigg(Q_1 Q_2 + Q_2 Q_3 + Q_3 Q_1 - \frac{N^2}{2} (J_1 + J_2) \Bigg)\nonumber\\
  & - \Bigg(Q_1 Q_2 Q_3 + \frac{N^2}{2} J_1 J_2\Bigg) = 0\, ,\label{eq:AdS5 constr}
\end{align}
which is a consequence of the reality condition on the black hole entropy $S_{BH}$.

For the AdS$_5$ black hole solutions in \cite{Chong:2005hr}, the electric charges are degenerate, i.e. $Q_1 = Q_2 = Q_3 = Q$. Hence, for this class of black hole solutions in the BPS limit, the black hole entropy becomes
\be
  S_{BH} = 2 \pi \sqrt{3 Q^2 - \frac{N^2}{2} (J_1 + J_2)}\, .
\ee
This is exactly the same as the result from the horizon area \eqref{eq:AdS5 S area to bdy CFT New} in the unit $g=1$, and the one from the Cardy formula \eqref{eq:AdS5 S Cardy}. The constraint \eqref{eq:AdS5 constr} for this degenerate case becomes
\be
  \left(3 Q + \frac{N^2}{2} \right) \left(3 Q^2 - \frac{N^2}{2} (J_1 + J_2) \right) = Q^3 + \frac{N^2}{2} J_1 J_2\, ,
\ee
which is also consistent with the thermodynamic quantities from the gravity side \eqref{eq:AdS5 BPSthermo}.

%%%%%%%%%%%%%%%%%%%%%%%%%%%%%%%%%%%%%%%%%%%
%%%%%%%%%%%%%%%%%%%%%%%%%%%%%%%%%%%%%%%%%%%
\section{Asymptotically AdS$_4$ Black Holes}\label{sec:AdS4}
%%%%%%%%%%%%%%%%%%%%%%%%%%%%%%%%%%%%%%%%%%%
%%%%%%%%%%%%%%%%%%%%%%%%%%%%%%%%%%%%%%%%%%%

In this section, we consider the asymptotically AdS$_4$ black holes and the corresponding gravitational Cardy limit. Similar to the AdS$_5$ case, we demonstrate that the AdS$_4$ black hole entropy can be computed in various ways as shown in Fig.~\ref{fig:M1}, and the other thermodynamic quantities scale correspondingly (see Table~\ref{Table:Intro}) in the gravitational Cardy limit.

\subsection{AdS$_4$ Black Hole Solution}

The non-extremal rotating, electrically  charged asymptotically AdS$_4$ black hole solution with gauge group $U(1) \times U(1)$ in 4d $\mathcal{N}=4$ gauged supergravity was constructed in \cite{Chong:2004na}. The solution is characterized by four parameters $(a, m, \delta_1, \delta_2)$. The metric, the scalars and the gauge fields are given by
\be
  ds^2 = - \frac{\Delta_r}{W} \left(dt - \frac{a\, \textrm{sin}^2 \theta}{\Xi} d\phi \right)^2 + W \left(\frac{dr^2}{\Delta_r} + \frac{d\theta^2}{\Delta_\theta} \right) + \frac{\Delta_\theta\, \textrm{sin}^2 \theta}{W} \left[a\, dt - \frac{r_1 r_2 + a^2}{\Xi} d\phi \right]^2\, ,\label{eq:AdS4Metric1}
\ee
\begin{align}
\begin{split}
  e^{\varphi_1} & = \frac{r_1^2 + a^2\, \textrm{cos}^2 \theta}{W}\, ,\qquad \chi_1 = \frac{a (r_2 - r_1)\, \textrm{cos}\, \theta}{r_1^2 + a^2\, \textrm{cos}^2 \theta}\, ,\\
  A_1 & = \frac{2 \sqrt{2} m\, \textrm{sinh} (\delta_1)\, \textrm{cosh} (\delta_1)\, r_2}{W} \left(dt - \frac{a\, \textrm{sin}^2 \theta}{\Xi} d\phi \right) + \alpha_{41}\, dt\, ,\\
  A_2 & = \frac{2 \sqrt{2} m\, \textrm{sinh} (\delta_2)\, \textrm{cosh} (\delta_2)\, r_1}{W} \left(dt - \frac{a\, \textrm{sin}^2 \theta}{\Xi} d\phi \right) + \alpha_{42}\, dt\, ,
\end{split}
\end{align}
where
\begin{align}
\begin{split}
  r_i & \equiv r + 2 m\, \textrm{sinh}^2 (\delta_i)\, ,\quad (i = 1, 2)\\
  \Delta_r & \equiv r^2 + a^2 - 2 m r + g^2 r_1 r_2 (r_1 r_2 + a^2)\, ,\\
  \Delta_\theta & \equiv 1 - g^2 a^2\, \textrm{cos}^2 \theta\, ,\\
  W & \equiv r_1 r_2 + a^2\, \textrm{cos}^2 \theta\, ,\\
  \Xi & \equiv 1 - a^2 g^2\, ,
\end{split}
\end{align}
and $g \equiv \ell_4^{-1}$ is the inverse of the AdS$_4$ radius. Note that we have added pure gauge terms to the two gauge fields where $\alpha_{41}$ and $\alpha_{42}$ are constant. The metric \eqref{eq:AdS4Metric1} can also be written in the following equivalent expression, which is more convenient for later discussions,
\be\label{eq:AdS4Metric2}
  ds^2 = - \frac{\Delta_r \Delta_\theta}{B \Xi^2} dt^2 + B\, \textrm{sin}^2 \theta (d\phi + f\, dt)^2 + W \left(\frac{dr^2}{\Delta_r} + \frac{d\theta^2}{\Delta_\theta} \right)\, ,
\ee
with
\begin{align}
\begin{split}
  B & \equiv \frac{(a^2 + r_1 r_2)^2 \Delta_\theta - a^2\, \textrm{sin}^2 (\theta)\, \Delta_r}{W \Xi^2}\, ,\\
  f & \equiv \frac{a \Xi \left(\Delta_r - \Delta_\theta (a^2 + r_1 r_2) \right)}{\Delta_\theta (a^2 + r_1 r_2)^2 - a^2 \Delta_r\, \textrm{sin}^2 \theta}\, .
\end{split}
\end{align}

The non-extremal asymptotically AdS$_4$ black holes with four degenerate electric charges ($Q_1 = Q_2$, $Q_3 = Q_4$) and one angular momentum $J$ have been found in \cite{Cvetic:2005zi}, which are characterized by four parameters $(a, m, \delta_1, \delta_2)$. The BPS limit imposes a condition
\be\label{eq:AdS4BPSCond}
  e^{2 \delta_1 + 2 \delta_2} = 1 + \frac{2}{a g}\, .
\ee
For the black hole solution to have a regular horizon, we impose an additional constraint
\be\label{eq:AdS4RegCond}
  (m g)^2 = \frac{\textrm{cosh}^2 (\delta_1 + \delta_2)}{e^{\delta_1 + \delta_2}\, \textrm{sinh}^3 (\delta_1 + \delta_2)\, \textrm{sinh} (2 \delta_1)\, \textrm{sinh} (2 \delta_2)}\, .
\ee
The two conditions \eqref{eq:AdS4BPSCond} \eqref{eq:AdS4RegCond} in \cite{Cvetic:2005zi} have typos, which have been corrected in \cite{Chow:2013gba, Choi:2018fdc}, see also \cite{Cassani:2019mms}. With these constraints, there are only two independent parameters for asymptotically AdS$_4$ BPS black holes, which we choose to be $(\delta_1, \delta_2)$ for convenience.  In the BPS limit, the position of the outer horizon is
\be\label{eq:AdS4 r+}
  r_+ = \frac{2 m\, \textrm{sinh} (\delta_1)\, \textrm{sinh} (\delta_2)}{\textrm{cosh} (\delta_1 + \delta_2)}\, ,
\ee
which coincides with the inner horizon.

The physical quantities of non-extremal  AdS$_4$ black holes can also be solved as functions of $(a, m, \delta_1, \delta_2)$. In particular, the gravitational angular velocity $\Omega$ and the temperature $T$ are given by
\be
  \Omega = \frac{a (1 + g^2 r_1 r_2)}{r_1 r_2 + a^2}\, ,\quad T = \frac{\Delta'_r}{4 \pi (r_1 r_2 + a^2)}\, .
\ee
Moreover, the other thermodynamic quantities of asymptotically AdS$_4$ black holes are \cite{Cvetic:2005zi}
\begin{align}
\begin{split}\label{eq:AdS4thermo}
  S & = \frac{\pi (r_1 r_2 + a^2)}{\Xi}\, ,\\
  J & = \frac{m a}{2 \Xi^2} \left(\textrm{cosh} (2 \delta_1) + \textrm{cosh} (2 \delta_2) \right)\, ,\\
  Q_1 = Q_2 & = \frac{m}{4 \Xi}\, \textrm{sinh} (2 \delta_1)\, ,\\
  Q_3 = Q_4 & = \frac{m}{4 \Xi}\, \textrm{sinh} (2 \delta_2)\, .
\end{split}
\end{align}

\subsection{Gravitational Cardy Limit}

The Cardy-like limit for the 3d ABJM theory was defined in \cite{Choi:2019zpz, Nian:2019pxj},
\be\label{eq:CardyLimitAdS4}
  |\omega| \ll 1\, ,\quad \Delta_I \sim \mathcal{O} (1)\, ,\qquad (\textrm{$I = 1,\, \cdots,\, 4$})\, .
\ee
Using the relations found in \cite{Choi:2018fdc}
\be
  \omega = - \lim_{T \to 0} \frac{\Omega - \Omega^*}{T}\, ,\quad \Delta_I = - \lim_{T\to 0} \frac{\Phi_I - \Phi_I^*}{T}\, ,
\ee
with $\Omega^* = g$ and $\Phi_I^* = 1$ denoting the BPS values of $\Omega$ and $\Phi_I$, we can find the gravitational counterpart of the Cardy-like limit \eqref{eq:CardyLimitAdS4}
\be
  \bigg| \left(\frac{\partial \Omega}{\partial T}\right)_{T=0} \bigg| \ll 1\, ,\quad \frac{\partial \Phi_I}{\partial T} \bigg|_{T=0} \sim \mathcal{O} (1)\, .
\ee
Hence, we obtain for the near-extremal AdS$_4$ black holes
\be
  \frac{\partial \Omega}{\partial T} \bigg|_{\textrm{BPS}} = \lim_{T \to 0} \frac{\Omega - \Omega_*}{T} = - \frac{\pi e^{\frac{5}{2} (\delta_1 + \delta_2)} \left(\textrm{coth} (\delta_1 + \delta_2) - 2 \right) \sqrt{\textrm{sinh} (2 \delta_1)\, \textrm{sinh} (2 \delta_2)}}{\left(\textrm{coth} (\delta_1 + \delta_2) + 1 \right) \sqrt{\textrm{sinh} (\delta_1 + \delta_2)}\, \textrm{cosh} (\delta_1 - \delta_2)}\, .
\ee
This expression has several roots
\be
  \delta_1 = 0\, ,\quad \delta_2 = 0\, ,\quad \delta_1 + \delta_2 = \textrm{arccoth} (2)\, .
\ee
However, $\delta_1 = 0$ and $\delta_2 = 0$ are unphysical, because according to \eqref{eq:AdS4 r+}, $\delta_1 = 0$ or $\delta_2 = 0$ will cause $r_+ \to 0$. Hence, we conclude that the gravitational Cardy limit for asymptotically AdS$_4$ BPS black holes is
\be
  \delta_1 + \delta_2 \to \textrm{arccoth} (2)\, .
\ee
 Equivalently, this can be written in terms of the other parameters as
\begin{align}
	a g \to 1.
\end{align}
We introduce a small parameter $\epsilon$ to denote small deviations from this limit, i.e.,
\be\label{eq:AdS4CardyLimit}
  \delta_1 + \delta_2 = \textrm{arccoth} (2) + \epsilon\, .
\ee
For this case $\epsilon$ is dimensionless. Imposing first the BPS constraint \eqref{eq:AdS4BPSCond} and the regularity condition \eqref{eq:AdS4RegCond} near the horizon, and then taking the gravitational Cardy limit \eqref{eq:AdS4CardyLimit}, we obtain the thermodynamic quantities \eqref{eq:AdS4thermo} to the leading order in $\epsilon$
\begin{align}
\begin{split}\label{eq:AdS4BPSthermoCardy}
  S_* & = \frac{\pi}{3 g^2 \epsilon} + \mathcal{O} (1)\, ,\\
  J_* & = \frac{\textrm{cosh} \left(2 \delta_1 - \frac{1}{2} \textrm{log} (3) \right)}{9 g^2 \epsilon^2 \sqrt{2\, \textrm{sinh} (4 \delta_1) - 5\, \textrm{sinh}^2 (2 \delta_1)}} + \mathcal{O} (\epsilon^{-1})\, ,\\
  Q_1^* = Q_2^* & = \frac{1}{4 g \epsilon \sqrt{6\, \textrm{tanh} (\delta_1) + 6\, \textrm{coth} (\delta_1) - 15}} + \mathcal{O} (1)\, ,\\
  Q_3^* = Q_4^* & = \frac{\sqrt{2\, \textrm{tanh} (\delta_1) + 2\, \textrm{coth} (\delta_1) - 5}}{12 \sqrt{3}\, g \epsilon} + \mathcal{O} (1)\, ,
\end{split}
\end{align}
which are consistent with \cite{ZaffaroniSlides, Choi:2019zpz} and the Cardy-like limit on the field theory side \eqref{eq:CardyLimitAdS4}
\be
  \omega_* \sim \epsilon\, ,\quad \Delta_{I*} \sim \mathcal{O} (1)\, .
\ee

\subsection{Black Hole Solution in the Near-Horizon + Gravitational Cardy Limit}

In the previous subsection, we have obtained the gravitational Cardy limit for the parameters on the gravity side. In this subsection, we discuss how the near-horizon metric changes when taking the gravitational Cardy limit. In Appendix~\ref{app:AdS4}, we verify explicitly that the resulting background is a solution of the 4d gauged supergravity equations of motion. In the following, we implement the gravitational Cardy limit in the space of parameters, which further simplifies the geometry.

For the asymptotically AdS$_4$ black hole metric \eqref{eq:AdS4Metric2}, we perform a near-horizon scaling similar to the AdS$_5$ case \eqref{eq:AdS5scaling}
\be\label{eq:AdS4Scaling}
  r \to r_* + \lambda\, \widetilde{r}\, ,\quad t \to \frac{\widetilde{t}}{\lambda}\, ,\quad \phi \to \widetilde{\phi} - g \big[\textrm{coth} (\delta_1 + \delta_2) - 2 \big] \frac{\widetilde{t}}{\lambda}\, .
\ee
Furthermore, we take the gravitational Cardy limit \eqref{eq:AdS4CardyLimit} and keep only the leading orders in $\epsilon$. The metric \eqref{eq:AdS4Metric2} finally becomes
\begin{align}
  ds^2 & = - \frac{g^2 \left(9 - 18\, e^{4 \delta_1} + e^{8 \delta_1} \right) \left(3 + \textrm{cos} (2 \theta) \right)}{3 \left(9 - 10\, e^{4 \delta_1} + e^{8 \delta_1} \right)}\, \widetilde{r}^2 d\widetilde{t}^2 + \frac{3 \left(3 + \textrm{cos} (2 \theta) \right) \left(e^{4 \delta_1} - 1 \right) \left(e^{4 \delta_1} - 9 \right)}{16 g^2 \left(9 - 18 e^{4 \delta_1} + e^{8 \delta_1} \right)} \frac{d\widetilde{r}^2}{\widetilde{r}^2} \nonumber\\
  {} & \quad + \frac{3 + \textrm{cos} (2 \theta)}{2 g^2\, \textrm{sin}^2 \theta}\, d\theta^2 + \frac{2\, \textrm{sin}^4 \theta}{9 g^2 \left(3 + \textrm{cos} (2 \theta) \right)} \bigg[\frac{d\widetilde{\phi}}{\epsilon} + \frac{\sqrt{3} g^2\, \textrm{cosh} \left(2\, \delta_1 - \frac{1}{2}\, \textrm{log} (3) \right)}{\sqrt{\textrm{sinh} (2 \delta_1)\, \textrm{sinh} \left(\textrm{log} (3) - 2 \delta_1 \right)}} \widetilde{r}\, d\widetilde{t}  \bigg]^2 \nonumber\\
  {} & \quad + \mathcal{O} (\epsilon)\, .\label{eq:AdS4Metric2NearHorizon}
\end{align}
Defining
\be
  \tau \equiv \frac{4 g^2 \left(9 - 5\, \textrm{cosh} (4\, \delta_1) + 4\, \textrm{sinh} (4\, \delta_1) \right)^2}{3 \left(5 - 5\, \textrm{cosh} (4\, \delta_1) + 4\, \textrm{sinh} (4\, \delta_1) \right)}\, \widetilde{t}\, ,
\ee
we can rewrite the metric \eqref{eq:AdS4Metric2NearHorizon} as
\begin{align}
  ds^2 & = \frac{3 \left(3 + \textrm{cos} (2 \theta) \right) \left(e^{4 \delta_1} - 1 \right) \left(e^{4 \delta_1} - 9 \right)}{16 g^2 \left(9 - 18 e^{4 \delta_1} + e^{8 \delta_1} \right)} \bigg[- \widetilde{r}^2 d\tau^2 + \frac{d\widetilde{r}^2}{\widetilde{r}^2} \bigg] + \frac{3 + \textrm{cos} (2 \theta)}{2 g^2\, \textrm{sin}^2 \theta}\, d\theta^2 \nonumber\\
  {} & \quad + \frac{2\, \textrm{sin}^4 \theta}{9 g^2 \left(3 + \textrm{cos} (2 \theta) \right)} \bigg[\frac{d\widetilde{\phi}}{\epsilon} + V (\delta_1) \widetilde{r}\, d\tau \bigg]^2 + \mathcal{O} (\epsilon)\, ,\label{eq:AdS4Metric2NearHorizon3}
\end{align}
where
\be
  V(\delta_1) \equiv \frac{9\, \textrm{cosh} \left(2\, \delta_1 - \frac{1}{2}\, \textrm{log} (3) \right)\, \left(5 - 5\, \textrm{cosh} (4\, \delta_1) + 4\, \textrm{sinh} (4\, \delta_1) \right)}{2 \sqrt{10 - 6\, \textrm{cosh} \left(4\, \delta_1 - \textrm{log} (3) \right)} \left(9 - 5\, \textrm{cosh} (4\, \delta_1) + 4\, \textrm{sinh} (4\, \delta_1) \right)}\, .
\ee

\subsection{Black Hole Entropy from Cardy Formula}

For the asymptotically AdS$_4$ black holes discussed in this section, we apply the Cardy formula to the near-horizon metric only after taking the gravitational Cardy limit. More explicitly, we first rewrite the metric \eqref{eq:AdS4Metric2NearHorizon3} from the Poincar\'e coordinates $(\widetilde{r},\, \tau)$ to the global coordinates $(\hat{r},\, \hat{t})$ using the relations \eqref{eq:PoincareToGlobal 1} - \eqref{eq:PoincareToGlobal 3}. Consequently, the near-horizon metric in the gravitational Cardy limit \eqref{eq:AdS4Metric2NearHorizon3} becomes
\begin{align}
  ds^2 & = \frac{3 \left(3 + \textrm{cos} (2 \theta) \right) \left(e^{4 \delta_1} - 1 \right) \left(e^{4 \delta_1} - 9 \right)}{16 g^2 \left(9 - 18 e^{4 \delta_1} + e^{8 \delta_1} \right)} \bigg[- (1 + \hat{r}^2) d\hat{t}^2 + \frac{d \hat{r}^2}{1 + \hat{r}^2} \bigg] + \frac{3 + \textrm{cos} (2 \theta)}{2 g^2\, \textrm{sin}^2 \theta}\, d\theta^2 \nonumber\\
  {} & \quad + \frac{2\, \textrm{sin}^4 \theta}{9 g^2 \left(3 + \textrm{cos} (2 \theta) \right)} \bigg[\frac{d\hat{\psi}}{\epsilon} + V(\delta_1) \hat{r}\, d\hat{t} \bigg]^2 + \mathcal{O} (\epsilon)\, ,\label{eq:AdS4Metric2NearHorizonGlobal}
\end{align}
where $\hat{t}$, $\hat{r}$ and $\gamma$ are defined in \eqref{eq:PoincareToGlobal 1} and \eqref{eq:PoincareToGlobal 3}, while
\be
  \hat{\psi} \equiv \hat{\phi} + V(\delta_1) \gamma \epsilon\, .
\ee

Applying the same formalism in Subsection~\ref{sec:Review Cardy}, we obtain the central charge and the extremal Frolov-Thorne temperature in the near-horizon region of the asymptotically AdS$_4$ BPS black holes,
\begin{align}
\begin{split}
  c_L & = \frac{3 \sqrt{\frac{3}{2}}\, e^{2 \delta_1} (3 + e^{4 \delta_1}) \sqrt{5 + 4\, \textrm{sinh} (4\, \delta_1) - 5\, \textrm{cosh} (4\, \delta_1)}}{g^2 (18\, e^{4\, \delta_1} - e^{8\, \delta_1} - 9)}\, ,\\
  T_L & = \frac{9 + 4\, \textrm{sinh} (4\, \delta_1) - 5\, \textrm{cosh} (4\, \delta_1)}{18 \pi \epsilon\, \textrm{sinh} (\delta_1)\, \textrm{cosh} (\delta_1)\, \textrm{cosh} \left(2\, \delta_1 - \frac{1}{2}\, \textrm{log} (3) \right) \sqrt{2\, \textrm{tanh} (\delta_1) + 2\, \textrm{coth} (\delta_1) - 5}}\, .
\end{split}
\end{align}
Using the Cardy formula, we can compute the black hole entropy of the asymptotically AdS$_4$ BPS black holes:
\be\label{eq:AdS4CardyFormula}
  S_{BH} = \frac{\pi^2}{3} c_L T_L = \frac{\pi}{3 g^2 \epsilon}\, ,
\ee
which is the same as the black hole entropy in the gravitational Cardy limit \eqref{eq:AdS4BPSthermoCardy} from the gravity side.

\subsection{Comparison with Results from Boundary CFT}

The asymptotically AdS$_4$ BPS black hole entropy can also be obtained from the boundary 3d ABJM theory by extremizing an entropy function \cite{Choi:2019zpz, Nian:2019pxj}, which has also been studied in \cite{Hosseini:2019iad}. One can first compute the free energy in the large-$N$ limit using the 3d superconformal index or the partition function via localization. The entropy function is then defined as a Legendre transform of the free energy in the large-$N$ limit
\be\label{eq:AdS4EntropyFct}
  S (\Delta_I,\, \omega) = \frac{2 \sqrt{2}\, i\, k^{\frac{1}{2}} N^{\frac{3}{2}}}{3} \frac{\sqrt{\Delta_1 \Delta_2 \Delta_3 \Delta_4}}{\omega} - 2 \omega J - \sum_I \Delta_I Q_I - \Lambda \left(\sum_I \Delta_I - 2 \omega + 2 \pi i \right)\, .
\ee
In the Cardy-like limit \eqref{eq:CardyLimitAdS4}
\be
  \omega \sim \epsilon\, ,\quad \Delta_I \sim \mathcal{O} (1)\, ,
\ee
we can read off from the entropy function \eqref{eq:AdS4EntropyFct}
\be
  S \sim \frac{1}{\epsilon}\, ,\quad J \sim \frac{1}{\epsilon^2}\, ,\quad Q_I \sim \frac{1}{\epsilon}\, ,
\ee
which have been summarized in Table~\ref{Table:Intro}.

Similar to the AdS$_5$ case, for AdS$_4$ the electric charges $Q_I$ and the angular momentum $J$ are real, while the chemical potentials $\Delta_I$ and the angular velocity $\omega$ can be complex, and so can the entropy function $S$. By requiring that the black hole entropy $S_{BH}$ to be real after extremizing the entropy function $S$, we obtain one more constraint on $Q_I$ and $J$. More precisely, for the degenerate case with $Q_1 = Q_3$, $Q_2 = Q_4$ and one angular momentum $J$, the asymptotically AdS$_4$ black hole entropy is given by \cite{Choi:2018fdc, Choi:2019zpz, Nian:2019pxj}
\be\label{eq:AdS4 S_BH from ABJM}
  S_{BH} = \frac{2 \pi}{3} \sqrt{\frac{9 Q_1 Q_2 (Q_1 + Q_2) - 2 k J N^3}{Q_1 + Q_2}}\, ,
\ee
subject to the constraint
\be
  2 k J^2 N^3 + 2 k J N^3 (Q_1 + Q_2) - 9 Q_1 Q_2 (Q_1 + Q_2)^2 = 0\, ,
\ee
which is a consequence of the reality condition on the black hole entropy $S_{BH}$.

If we identify the field theory parameters with the ones on the gravity side in the following way \cite{Nian:2017hac, Choi:2018fdc}
\be
  \frac{1}{G} = \frac{2 \sqrt{2}}{3} g^2 k^{\frac{1}{2}} N^{\frac{3}{2}}\, ,\quad Q_{BH} = \frac{g}{2} Q\, ,\quad J_{BH} = J\, ,
\ee
we can rewrite the black hole entropy \eqref{eq:AdS4 S_BH from ABJM} and the angular momentum as
\begin{align}
  S_{BH} & = \frac{\pi}{g^2 G} \frac{J_{BH}}{ \left( \frac{2}{g} Q_{BH, 1} + \frac{2}{g} Q_{BH, 2} \right)}\, ,\label{eq:S_BH-3}\\
  J_{BH} & = \frac{1}{2} \left(\frac{2}{g} Q_{BH, 1} + \frac{2}{g} Q_{BH, 2} \right) \left(- 1 + \sqrt{1 + 16 g^4 G^2 \frac{2 Q_{BH, 1}}{g} \frac{2 Q_{BH, 2}}{g}} \right)\, ,\label{eq:Sol_J-2}
\end{align}
which are consistent with both the thermodynamic quantities on the gravity side \eqref{eq:AdS4thermo} \eqref{eq:AdS4BPSthermoCardy} and the black hole entropy in the gravitational Cardy limit from the Cardy formula \eqref{eq:AdS4CardyFormula}.

%%%%%%%%%%%%%%%%%%%%%%%%%%%%%%%%%%%%%%%%%%%
%%%%%%%%%%%%%%%%%%%%%%%%%%%%%%%%%%%%%%%%%%%
\section{Asymptotically AdS$_7$ Black Holes}\label{sec:AdS7}
%%%%%%%%%%%%%%%%%%%%%%%%%%%%%%%%%%%%%%%%%%%
%%%%%%%%%%%%%%%%%%%%%%%%%%%%%%%%%%%%%%%%%%%

In this section, we consider asymptotically AdS$_7$ black holes and the corresponding gravitational Cardy limit. Similar to the previous sections, we demonstrate that the AdS$_7$ black hole entropy can be computed in various ways as shown in Fig.~\ref{fig:M1}, and the other thermodynamic quantities scale correspondingly in gravitational Cardy limit. For completeness, we discuss two asymptotically AdS$_7$ black hole solutions in the literature: a special case with all equal charges and all equal angular momenta in Subsection~\ref{sec:Special AdS7} and a more general case with two equal charges and three independent angular momenta in Subsection~\ref{sec:General AdS7}.

\subsection{A Special Case}\label{sec:Special AdS7}

In this subsection, we consider the gravitational Cardy limit of a special class of non-extremal asymptotically AdS$_7$ black holes discussed in \cite{Kantor:2019lfo}.

\subsubsection{AdS$_7$ Black Hole Solution}

The solutions $\mathcal{M}_7 \times S^4$ to 11d gauged supergravity, with $\mathcal{M}_7$ denoting an asymptotically AdS$_7$ black hole, have the isometry group $SO(2, 6) \times SO(5)$. Hence, this class of solutions has three angular momenta from the Cartan of the maximal compact subgroup $SO(6) \subset SO(2, 6)$ and two electric charges from the Cartan of $SO(5)$. The most generic solution has not been constructed in the literature so far. Instead, the solutions with two charges and three equal angular momenta were found in \cite{Chong:2004dy}, while the ones with two equal charges and three angular momenta were found in \cite{Chow:2007ts}. As the intersection of these two classes, the solution with two equal charges $Q_1 = Q_2$ and three equal angular momenta $J_1 = J_2 = J_3$ has been considered in \cite{Kantor:2019lfo}.

For this special solution, the metric of the asymptotically AdS$_7$ black hole part is given by
\be\label{eq:SimpleAdS7Metric}
  ds_7^2 = H^{\frac{2}{5}} \left[- \frac{Y}{f_1 \Xi_-^2} dt^2 + \frac{\Xi\, \rho^6}{Y} dr^2 + \frac{f_1}{H^2\, \Xi^2\, \rho^4} \left(\sigma - \frac{2 f_2}{f_1} dt \right)^2 + \frac{\rho^2}{\Xi}\, ds^2_{\mathbb{C} \mathbb{P}^2} \right]\, ,
\ee
where
\begin{align}
  \sigma & \equiv d\chi + \frac{1}{2} l_3\, \textrm{sin}^2 \xi\, ,\\
  ds^2_{\mathbb{C} \mathbb{P}^2} & = d\xi^2 + \frac{1}{4} \sin^2 \xi\, (l_1^2 + l_2^2) + \frac{1}{4} l_3^2\, \sin^2 \xi\, \cos^2 \xi\, ,
\end{align}
with $(l_1, l_2, l_3)$ denoting the left-invariant 1-forms of $SU(2)$, which can be explicitly chosen to be \cite{Nian:2013qwa}
\begin{align}
\begin{split}
	& l_1 = \sin\psi\, d\theta - \cos\psi \sin\theta\, d\phi\, , \\
	& l_2 = \cos\psi\, d\theta + \sin\psi \sin\theta\, d\phi\, , \\
	& l_3 = - (d\psi + \cos\theta\, d\phi)\, ,
\end{split}
\end{align}
\be
  0 \leq \theta \leq \pi\, ,\quad 0 \leq \phi \leq 2 \pi\, ,\quad 0 \leq \psi \leq 4 \pi\, .
\ee
Moreover, the asymptotically AdS$_7$ black hole solution also contains two 1-forms, one 3-form and two scalars, which are given by
\begin{align}
\begin{split}
  A^1_{(1)} & = A^2_{(1)} =A_{(1)}=\frac{ m\, \textrm{sinh} (\delta)\, \textrm{cosh}(\delta)}{\rho^4\, \Xi H} ( dt - a \sigma) + \frac{\alpha_{70}}{\Xi_-} dt\, ,\\
  A_{(3)} & = \frac{ m a\, \textrm{sinh}^2 (\delta)}{\rho^2 \Xi\, \Xi_-}\, \sigma \wedge \, d\sigma + \alpha_{71}\, dt\wedge d\theta \wedge d\psi +\alpha_{72}\, dt\wedge d\xi \wedge d\phi +\alpha_{73}\, dt\wedge d\xi \wedge d\psi\, ,\\
  X_1 & = X_2 = H^{-1/5}\, ,
\end{split}
\end{align}
where we have added some pure gauge terms to the potentials, and
\begin{align}
\begin{split}
  \rho^2 & \equiv \Xi\, r^2\, ,\\
  H & \equiv 1 + \frac{2 m\, \textrm{sinh}^2 (\delta)}{\rho^4}\, ,\\
  f_1 & \equiv \Xi\, \rho^6 H^2 - \frac{\big[2 \Xi_+ m a\, \textrm{sinh}^2 (\delta)\big]^2}{\rho^4} + 2 m a^2 \big[\Xi_+^2 + \textrm{cosh}^2 (\delta)\, (1 - \Xi_+^2) \big]\, ,\\
  f_2 & \equiv - \frac{g\, \Xi_+ \rho^6\, H^2}{2} + m a\, \textrm{cosh}^2 (\delta)\, ,\\
  Y & \equiv g^2 \rho^8 H^2 + \Xi\, \rho^6 - 2 m \rho^2 \left[a^2 g^2\, \textrm{cosh}^2 (\delta) + \Xi \right] + 2 m a^2 \left[\Xi_+^2 + \textrm{cosh}^2 (\delta)\, (1 - \Xi_+^2) \right]\, ,\\
  \Xi_\pm & \equiv 1 \pm a g\, ,\quad \Xi \equiv 1 - a^2 g^2\, ,
\end{split}
\end{align}
with $g = \ell_7^{-1}$ denoting the inverse of the AdS$_7$ radius. The thermodynamic quantities of the black hole have the following expressions
\begin{align}
\begin{split}
  T & = \frac{\partial_r Y}{4 \pi g\, \rho^3 \sqrt{\Xi f_1}}\, ,\\
  S & = \frac{\pi^3 \rho^2 \sqrt{f_1}}{4 G_N \Xi^3}\, ,\\
  \Omega & = - \frac{1}{g} \left(g + \frac{2 f_2\, \Xi_-}{f_1} \right)\, ,\\
  \Phi & = \frac{4 m\, \textrm{sinh} (\delta)\, \textrm{cosh} (\delta)}{\rho^4\, \Xi H} \left(\Xi_- - \frac{2 a f_2 \Xi_-}{f_1} \right)\, ,\\
  E & = \frac{m \pi^2}{32 G_N g \Xi^4} \Big[12 (a g + 1)^2 \left(a g (a g + 2) - 1 \right) - 4\, \textrm{cosh}^2 (\delta) \left(3 a^4 g^4 + 12 a^3 g^3 + 11 a^2 g^2 - 8 \right) \Big]\, ,\\
  J & = - \frac{m a \pi^2}{16 G_N \Xi^4} \Big[4 a g (a g + 1)^2 - 4\, \textrm{cosh}^2 (\delta) \left(a^3 g^3 + 2 a^2 g^2 + a g - 1 \right) \Big]\, ,\\
  Q & = \frac{m \pi^2\, \textrm{sinh} (\delta)\, \textrm{cosh} (\delta)}{4 G_N g \Xi^3}\, .
\end{split}
\end{align}
The expression of the temperature $T$ has three roots $r_\pm$ and $r_0$, all of which coincide at the extremality. As we can see, the solution depends only on three parameters $(a, m, \delta)$. As shown in \cite{Kantor:2019lfo}, the BPS condition and the absence of naked closed timelike curves (CTCs) require that
\be
   e^{2 \delta} = 1 - \frac{2}{3 a g}\, ,\quad m = \frac{128\, e^{2 \delta} (3\, e^{2 \delta} - 1)^3}{729\, g^4 (e^{2 \delta} + 1)^2 (e^{2 \delta} - 1)^6}\, .
\ee
Hence, there is only one independent parameter in the BPS limit, which we choose to be $\delta$. In addition, all three roots of $T$, i.e. $r_\pm$ and $r_0$, coincide in the BPS limit, and its value is
\be\label{eq:SimpleAdS7Horizon}
  r_*^2 = \frac{16}{3 g^2 (3\, e^{2 \delta} - 5) (e^{2 \delta} + 1)}\, .
\ee
The thermodynamic quantities in the BPS limit become
\begin{align}
\begin{split}\label{eq:SimpleAdS7BPSthermo}
  T_* & = 0\, ,\\
  S_* & = \frac{32 \pi^3 \sqrt{9\, e^{2 \delta} - 7}}{3 \sqrt{3}\, G_N g^5 (3\, e^{2 \delta} - 5)^3\, (e^{2 \delta} + 1)^{3/2}}\, ,\\
  \Omega_* & = 1\, ,\\
  \Phi_* & = 2\, ,\\
  E_* & = \frac{16 \pi^2 (18\, e^{6 \delta} - 21\, e^{4 \delta} + 7)}{3\, G_N g^5 (3\, e^{2 \delta} - 5)^4 (e^{2 \delta} + 1)^2}\, ,\\
  J_* & = \frac{16 \pi^2 (9\, e^{4 \delta} + 18\, e^{2 \delta} - 23)}{9\, G_N g^5 (3\, e^{2 \delta} - 5)^4 (e^{2 \delta} + 1)^2}\, ,\\
  Q_* & = \frac{8 \pi^2 (3\, e^{6 \delta} - 5\, e^{4 \delta} - 3\, e^{2 \delta} + 5)}{G_N g^5 (3\, e^{2 \delta} - 5)^4 (e^{2 \delta} + 1)^2}\, .
\end{split}
\end{align}

\subsubsection{Gravitational Cardy Limit}\label{sec:CardyLimitSpecialAdS7}

The Cardy-like limit for the 6d $\mathcal{N} = (2, 0)$ theory was defined in \cite{Nahmgoong:2019hko}, which for the special solution with three equal angular momenta is
\be\label{eq:CardyLimitAdS7}
  |\omega| \ll 1\, ,\quad \Delta \sim \mathcal{O} (1)\, .
\ee
Using the following relations found in \cite{Kantor:2019lfo}
\be\label{eq:SimpleAdS7Dict}
  \omega = \frac{1}{T} (\Omega - \Omega_*)\, ,\quad \phi = \frac{1}{T} (\Phi - \Phi_*)\, ,
\ee
we obtain the corresponding limit as
\be
  \bigg| \left(\frac{\partial \Omega}{\partial T}\right)_{T=0} \bigg| \ll 1\, ,\quad \frac{\partial \Phi}{\partial T} \bigg|_{T=0} \sim \mathcal{O} (1)\, .
\ee
Using the relation \eqref{eq:SimpleAdS7Dict}, we can express $\left(\frac{\partial \Omega}{\partial T} \right)_*$ in terms of the paramter $\delta$. The explicit form is not very elucidating, but we do find a root to the equation $\left(\frac{\partial \Omega}{\partial T} \right)_* = 0$, which is
\be
  \delta_* = \frac{1}{2}\, \textrm{log} \left(\frac{5}{3} \right)\, .
\ee
Hence, the gravitational Cardy limit for the class of asymptotically AdS$_7$ BPS black holes \eqref{eq:SimpleAdS7Metric} is
\be
  \delta \to \frac{1}{2}\, \textrm{log} \left(\frac{5}{3} \right)\, .
\ee
Note that this is equivalent to
\begin{align}
	 a g \to -1,
\end{align}
as in the other black hole solutions. We can introduce a small parameter to denote the deviation from this limit, i.e.,
\be\label{eq:SimpleAdS7CardyLimit}
  \delta = \frac{1}{2}\, \textrm{log} \left(\frac{5}{3} \right) + \epsilon\, .
\ee
For this case $\epsilon$ is dimensionless. Expanding in $\epsilon$, we find the BPS thermodynamic quantities \eqref{eq:SimpleAdS7BPSthermo} in the gravitational Cardy limit \eqref{eq:SimpleAdS7CardyLimit} to the leading order
\begin{align}
\begin{split}\label{eq:SimpleAdS7BPSthermoCardy}
  T_* & = 0\, ,\\
  S_* & = \frac{\pi^3}{250\, G_N g^5 \epsilon^3} + \mathcal{O} (\epsilon^{-2})\, ,\\
  \Omega_* & = 1\, ,\\
  \Phi_* & = 2\, ,\\
  E_* & = \frac{3 \pi^2}{1250\, G_N g^5 \epsilon^4} + \mathcal{O} (\epsilon^{-3})\, ,\\
  J_* & = \frac{\pi^2}{1250\, G_N g^5 \epsilon^4} + \mathcal{O} (\epsilon^{-3})\, ,\\
  Q_* & = \frac{\pi^2}{500\, G_N g^5 \epsilon^3} + \mathcal{O} (\epsilon^{-2})\, ,
\end{split}
\end{align}
which are consistent with \cite{Nahmgoong:2019hko, ZaffaroniSlides} and the Cardy-like limit on the field theory side \eqref{eq:CardyLimitAdS7}
\be
  \omega_* \sim \epsilon\, ,\quad \Delta_* \sim \mathcal{O} (1)\, .
\ee

%%%%%%%%%%%%%%%%%%%%%%%%%%%%%%%%%%
\subsubsection{Black Hole Solution in the Near-Horizon + Gravitational Cardy Limit}

In the previous subsection, we have obtained the gravitational Cardy limit for the parameters on the gravity side. In this subsection, we discuss how the near-horizon metric changes when taking the gravitational Cardy limit. In Appendix~\ref{app:AdS7}, we verify explicitly that the resulting background is a solution of the 7d gauged supergravity equations of motion. In the following, we implement the gravitational Cardy limit in the space of parameter which further simplifies the geometry.

We can apply the following scaling near the horizon $r_*$ \eqref{eq:SimpleAdS7Horizon} to the BPS AdS$_7$ black hole metric \eqref{eq:SimpleAdS7Metric}
\be\label{eq:SimpleAdS7scaling} 
  r \to r_* + \lambda\, \widetilde{r}\, ,\quad t \to \frac{\widetilde{t}}{\lambda}\, ,\quad \chi \to \widetilde{\chi} - \frac{6 g\, \textrm{sinh} (\delta)}{\textrm{cosh} (\delta) + 2 \textrm{sinh} (\delta)} \frac{\widetilde{t}}{\lambda}\, ,
\ee
with $\lambda \to 0$. In addition, taking the gravitational Cardy limit \eqref{eq:SimpleAdS7CardyLimit}, we obtain the near-horizon metric to the leading order in $\epsilon$
%\begin{align}
%  ds^2 & = - 10\, g^2\, 2^{2/5}\, \epsilon\, \widetilde{r}%^2\, d\widetilde{t}^2 + \frac{1}{16\, g^2\, 2^{3/5}} \frac{d%%\widetilde{r}^2}{\widetilde{r}^2} + \frac{2^{2/5}}{25\, g^2 %%\epsilon^2} \left(d\widetilde{\chi} + \frac{1}{2} l_3\, %\textrm{sin}^2 \xi - 5 \sqrt{5}\, g^2 \epsilon^{3/2}\, %\widetilde{r}\, d\widetilde{t} \right)^2 \nonumber\\
%  {} & \quad + \frac{2^{2/5}}{5 g^2 \epsilon}\, ds^2_{\mathbb{C} \mathbb{P}^2}\, .\label{eq:SimpleAdS7MetricCardy}
%\end{align}

\begin{align}
  ds^2 & = - 10\, g^2\, 2^{2/5}\, \epsilon\, \widetilde{r}^2\, d\widetilde{t}^2 + \frac{1}{16\, g^2\, 2^{3/5}} \frac{d\widetilde{r}^2}{\widetilde{r}^2} \nonumber\\
  {} & \quad + \frac{2^{2/5}}{25\, g^2 \epsilon^2} \left(d\widetilde{\chi} + \frac{1}{2} l_3\, \textrm{sin}^2 \xi - 5 \sqrt{5}\, g^2 \epsilon^{3/2}\, \widetilde{r}\, d\widetilde{t} \right)^2 + \frac{2^{2/5}}{5 g^2 \epsilon}\, ds^2_{\mathbb{C} \mathbb{P}^2}\, .\label{eq:SimpleAdS7MetricCardy}
\end{align}
Defining
\be
  \tau \equiv 8 \sqrt{5}\, g^2 \sqrt{\epsilon}\, \widetilde{t}\, ,
\ee
we can rewrite the near-horizon metric in the gravitational Cardy limit \eqref{eq:SimpleAdS7MetricCardy} as follows
\be
  ds^2 = \frac{1}{16\, g^2\, 2^{3/5}} \bigg[- \widetilde{r}^2\, d\tau^2 + \frac{d\widetilde{r}^2}{\widetilde{r}^2}\bigg] + \frac{2^{2/5}}{25\, g^2 \epsilon^2} \left(d\widetilde{\chi} + \frac{1}{2} l_3\, \textrm{sin}^2 \xi - \frac{5 \epsilon}{8}\, \widetilde{r}\, d\tau \right)^2 + \frac{2^{2/5}}{5 g^2 \epsilon}\, ds^2_{\mathbb{C} \mathbb{P}^2}\, .\label{eq:SimpleAdS7MetricCardy 2}
\ee

\subsubsection{Black Hole Entropy from Cardy Formula}

For the asymptotically AdS$_7$ black holes discussed in this section, we apply the Cardy formula to the near-horizon metric only after taking the gravitational Cardy limit. More explicitly, we first rewrite the metric \eqref{eq:SimpleAdS7MetricCardy 2} from the Poincar\'e coordinates $(\widetilde{r},\, \tau)$ to the global coordinates $(\hat{r},\, \hat{t})$ using the relations \eqref{eq:PoincareToGlobal 1} - \eqref{eq:PoincareToGlobal 3}. Consequently, the near-horizon metric in the gravitational Cardy limit \eqref{eq:SimpleAdS7MetricCardy 2} becomes
\be
  ds^2 = \frac{1}{16\, g^2\, 2^{3/5}} \bigg[- (1 + \hat{r}^2)\, d\hat{t}^2 + \frac{d\hat{r}^2}{1 + \hat{r}^2}\bigg] + \frac{2^{2/5}}{25\, g^2 \epsilon^2} \left(d\hat{\chi} + \frac{1}{2} l_3\, \textrm{sin}^2 \xi - \frac{5 \epsilon}{8}\, \hat{r}\, d\hat{t} \right)^2 + \frac{2^{2/5}}{5 g^2 \epsilon}\, ds^2_{\mathbb{C} \mathbb{P}^2}\, ,
\ee
where $\hat{t}$, $\hat{r}$ and $\gamma$ are defined in \eqref{eq:PoincareToGlobal 1} and \eqref{eq:PoincareToGlobal 3}, while
\be
  \hat{\chi} \equiv \widetilde{\chi} - \frac{5 \epsilon}{8}\, \gamma\, .
\ee

Applying the same formalism in Subsection~\ref{sec:Review Cardy}, we obtain the central charge and the extremal Frolov-Thorne temperature in the near-horizon region of the asymptotically AdS$_7$ BPS black holes as follows
\be
  c_L = \frac{3 \pi^2}{200\, g^5 \epsilon^2}\, ,\quad T_L = \frac{4}{5 \pi \epsilon}\, .
\ee
Using the Cardy formula, we can compute the black hole entropy of the asymptotically AdS$_7$ BPS black holes
\be\label{eq:SimpleAdS7CardyFormula}
  S_{BH} = \frac{\pi^2}{3} c_L T_L = \frac{\pi^3}{250\, g^5 \epsilon^3}\, ,
\ee
which is the same as the black hole entropy in the gravitational Cardy limit \eqref{eq:SimpleAdS7BPSthermoCardy} from the gravity side in the unit $G_N = 1$.

\subsubsection{Comparison with Results from Boundary CFT}\label{sec:General AdS7 S_BH from 6d (2,0)}

The asymptotically AdS$_7$ BPS black hole entropy can also be obtained from the boundary 6d $(2, 0)$ theory by extremizing an entropy function \cite{Choi:2018hmj, Kantor:2019lfo, Nahmgoong:2019hko} originally motivated in \cite{Hosseini:2018dob} and more recently studied in \cite{Hosseini:2019iad}. We can first compute the free energy in the large-$N$ limit using the background field method on $S^5$, the partition function via localization or the 6d superconformal index. The entropy function is then defined as a Legendre transform of the free energy in the large-$N$ limit
\be\label{eq:AdS7EntropyFct}
  S (\Delta_I,\, \omega_i) = - \frac{N^3}{24} \frac{\Delta_1^2 \Delta_2^2}{\omega_1 \omega_2 \omega_3} + \sum_{I=1}^2 Q_I \Delta_I + \sum_{i=1}^3 J_i \omega_i - \Lambda \left(\sum_{I=1}^2 \Delta_I - \sum_{i=1}^3 \omega_i - 2 \pi i \right)\, .
\ee
In the Cardy-like limit \eqref{eq:CardyLimitAdS7}
\be
  \omega \sim \epsilon\, ,\quad \Delta_I \sim \mathcal{O} (1)\, ,
\ee
we can read off from the entropy function \eqref{eq:AdS7EntropyFct}
\be
  S \sim \frac{1}{\epsilon^3}\, ,\quad J \sim \frac{1}{\epsilon^4}\, ,\quad Q_I \sim \frac{1}{\epsilon^3}\, ,
\ee
which have been summarized in Table~\ref{Table:Intro}.

Similar to AdS$_{4, 5}$, for AdS$_7$ the electric charges $Q_I$ and the angular momenta $J_i$ are real, while the chemical potentials $\Delta_I$ and the angular velocities $\omega_i$ can be complex, and so can the entropy function $S$. By requiring that the black hole entropy $S_{BH}$ to be real after extremizing the entropy function $S$, we obtain one more constraint on $Q_I$ and $J_i$. More precisely, the most general case with two charges $(Q_1,\, Q_2)$ and three angular momenta $(J_1,\, J_2,\, J_3)$ was discussed in \cite{Choi:2018hmj, Nahmgoong:2019hko}, while the degenerate case with $Q_1 = Q_2$ and $J_1 = J_2 = J_3$ was discussed in \cite{Kantor:2019lfo}. For the most general case, the asymptotically AdS$_7$ black hole entropy is \cite{Choi:2018hmj, Nahmgoong:2019hko}
\be\label{eq:AdS7 S_BH from 6d}
  S_{BH} = 2 \pi \sqrt{\frac{3 (Q_1^2 Q_2 + Q_1 Q_2^2) - N^3 (J_1 J_2 + J_2 J_3 + J_3 J_1)}{3 (Q_1 + Q_2) - N^3}}\, ,
\ee
subject to the constraint
\begin{align}
  {} & \frac{3 (Q_1^2 Q_2 + Q_1 Q_2^2) - N^3 (J_1 J_2 + J_2 J_3 + J_3 J_1)}{3 (Q_1 + Q_2) - N^3}\nonumber\\
  = & \left[\frac{N^3}{3} (J_1 + J_2 + J_3) + \frac{Q_1^2 + Q_2^2}{2} + 2 Q_1 Q_2 \right] \nonumber 
  \\ & \times \left[1 - \sqrt{1 - \frac{\frac{2}{3} N^3 J_1 J_2 J_3 + Q_1^2 Q_2^2}{\left(\frac{N^3}{3} (J_1 + J_2 + J_3) + \frac{Q_1^2 + Q_2^2}{2} + 2 Q_1 Q_2 \right)^2}} \right]\, ,\label{eq:AdS7 constraint from 6d}
\end{align}
which is a consequence of the reality condition on the black hole entropy $S_{BH}$.

We apply the general result to the special case $Q_1 = Q_2 = Q$ and $J_1 = J_2 = J_3 = J$,
\be
  S_{BH} = 2 \pi \sqrt{\frac{6 Q^3 - 3 N^3 J^2}{6 Q - N^3}}\, ,
\ee
with the constraint
\be
  \frac{6 Q^3 - 3 N^3 J^2}{6 Q - N^3} = \left(N^3 J + 3 Q^2 \right)\cdot \left[1 - \sqrt{1 - \frac{\frac{2}{3} N^3 J^3 + Q^4}{N^3 J + 3 Q^2}} \right]\, ,
\ee
which are consistent with both the thermodynamic quantities on the gravity side \eqref{eq:SimpleAdS7BPSthermo} \eqref{eq:SimpleAdS7BPSthermoCardy} and the black hole entropy in the gravitational Cardy limit from the Cardy formula \eqref{eq:SimpleAdS7CardyFormula} under the AdS$_7$/CFT$_6$ dictionary of parameters \cite{Choi:2018hmj, Nahmgoong:2019hko}
\be
  G_N = \frac{3 \pi^2}{16\, g^5 N^3}\, .
\ee

\subsection{More General Case}\label{sec:General AdS7}

In the previous section, we have discussed a special solution of asymptotically AdS$_7$ black holes with two equal charges $Q_1 = Q_2$ and three equal angular momenta $J_1 = J_2 = J_3$. In this subsection, we consider a more general solution with two equal charges $Q_1 = Q_2$ and three independent angular momenta $(J_1,\, J_2,\, J_3)$, which was first introduced in \cite{Chow:2007ts}.

\subsubsection{AdS$_7$ Black Hole Solution}\label{sec:GeneralAdS7Review}

The metric for this class of asymptotically AdS$_7$ black holes is
\begin{align}
  ds^2 & = H^{2/5} \Bigg[ \frac{(r^2 + y^2) (r^2 + z^2)}{R} dr^2 + \frac{(r^2 + y^2) (y^2 - z^2)}{Y} dy^2 + \frac{(r^2 + z^2) (z^2 - y^2)}{Z} dz^2 \nonumber\\
  {} & \qquad\qquad - \frac{R}{H^2 (r^2 + y^2) (r^2 + z^2)} \mathcal{A}^2 \nonumber\\
  {} & \qquad\qquad + \frac{Y}{(r^2 + y^2) (y^2 - z^2)} \left(dt' + (z^2 - r^2) d\psi_1 - r^2 z^2 d\psi_2 - \frac{q}{H (r^2 + y^2) (r^2 + z^2)} \mathcal{A} \right)^2 \nonumber\\
  {} & \qquad\qquad + \frac{Z}{(r^2 + z^2) (z^2 - y^2)} \left(dt' + (y^2 - r^2) d\psi_1 - r^2 y^2 d\psi_2 - \frac{q}{H (r^2 + y^2) (r^2 + z^2)} \mathcal{A} \right)^2 \nonumber\\
  {} & \qquad\qquad + \frac{a_1^2\, a_2^2\, a_3^2}{r^2 y^2 z^2} \Bigg(dt' + (y^2 + z^2 - r^2) d\psi_1 + (y^2 z^2 - r^2 y^2 - r^2 z^2) d\psi_2 - r^2 y^2 z^2 d\psi_3 \nonumber\\
  {} & \qquad\qquad\qquad - \frac{q}{H (r^2 + y^2) (r^2 + z^2)} \left(1 + \frac{g y^2 z^2}{a_1 a_2 a_3} \right) \mathcal{A} \Bigg)^2 \Bigg]\, ,\label{eq:GeneralAdS7Metric}
\end{align}
while the 1-form, the 2-form, the 3-form and the scalar are
\begin{align}
\begin{split}
  A_{(1)} & = \frac{2 m\, \textrm{sinh} (\delta)\, \textrm{cosh} (\delta)}{H (r^2 + y^2) (r^2 + z^2)} \mathcal{A}\, ,\\
  A_{(2)} & = \frac{q}{H (r^2 + y^2) (r^2 + z^2)} \mathcal{A} \wedge \\
  {} & \quad \bigg[dt' + \sum_i a_i^2 (g^2 dt' + d\psi_1+ \sum_{i < j} a_i^2 a_j^2 (g^2 d\psi_1 + d\psi_2) + a_1^2\, a_2^2\, a_3^2 (g^2 d\psi_2 + d\psi_3) \\
  {} & \qquad - g^2 (y^2 + z^2) dt' - g^2 y^2 z^2 d\psi_1 + a_1\, a_2\, a_3 \Big(d\psi_1 + (y^2 + z^2) d\psi_2 + y^2 z^2 d\psi_3 \Big) \bigg]\, ,\\
  A_{(3)} & = q a_1 a_2 a_3 \Big[d\psi_1 + (y^2 + z^2) d\psi_2 + y^2 z^2 d\psi_3 \Big] \\
  {} & \quad \wedge \Bigg(\frac{1}{(r^2 + y^2) z} dz \wedge (d\psi_1 + y^2 d\psi_2) + \frac{1}{(r^2 + z^2) y} dy \wedge (d\psi_1 + z^2 d\psi_2) \Bigg) \\
  {} & \quad - q g \mathcal{A} \wedge \Bigg(\frac{z}{r^2 + y^2} dz \wedge (d\psi_1 + y^2 d\psi_2) + \frac{y}{r^2 + z^2} dy \wedge (d\psi_1 + z^2 d\psi_2) \Bigg)\, ,\\
  X & = H^{-1/5}\, ,
\end{split}
\end{align}
where
\begin{align}
\begin{split}
  R & \equiv \frac{1 + g^2 r^2}{r^2} \prod_{i=1}^3 (r^2 + a_i^2) + q g^2 (2 r^2 + a_1^2 + a_2^2 + a_3^2) - \frac{2 q g a_1 a_2 a_3}{r^2} + \frac{q^2 g^2}{r^2} - 2 m\, ,\\
  Y & \equiv \frac{1 - g^2 y^2}{y^2} \prod_{i=1}^3 (a_i^2 - y^2)\, ,\\
  Z & \equiv \frac{1 - g^2 z^2}{z^2} \prod_{i=1}^3 (a_i^2 - z^2)\, ,\\
  \mathcal{A} & \equiv dt' + (y^2 + z^2) d\psi_1 + y^2 z^2 d\psi_2\, ,\\
  H & \equiv 1 + \frac{q}{(r^2 + y^2) (r^2 + z^2)}\, ,\\
  q & \equiv 2 m\, \textrm{sinh}^2 (\delta)\, .
\end{split}
\end{align}
It can be shown that after the change of coordinates
\begin{align}
\begin{split}\label{eq:GeneralAdS7ChangeCoord}
  t & = t' + (a_1^2 + a_2^2 + a_3^2) \psi_1 + (a_1^2\, a_2^2 + a_2^2\, a_3^2 + a_3^2\, a_1^2) \psi_2 + a_1^2\, a_2^2\, a_3^2\, \psi_3\, ,\\
  \frac{\phi_i}{a_i} & = g^2 t' + \psi_1 + \sum_{j \neq i} a_j^2 (g^2 \psi_1 + \psi_2) + \prod_{j \neq i} a_j^2 (g^2 \psi_2 + \psi_3)\, ,
\end{split}
\end{align}
the metric \eqref{eq:GeneralAdS7Metric} can be written in an equivalent form
\begin{align}
  ds^2 & = H^{2/5} \Bigg[ \frac{(r^2 + y^2) (r^2 + z^2)}{R} dr^2 + \frac{(r^2 + y^2) (y^2 - z^2)}{Y} dy^2 + \frac{(r^2 + z^2) (z^2 - y^2)}{Z} dz^2 \nonumber\\
  {} & \qquad - \frac{r^2 y^2 z^2 R Y Z}{H^2 \prod_{i < j} (a_i^2 - a_j^2)^2 B_1 B_2 B_3} dt^2 + B_3 (d\phi_3 + v_{32} d\phi_2 + v_{31} d\phi_1 + v_{30} dt)^2 \nonumber\\
  {} & \qquad + B_2 (d\phi_2 + v_{21} d\phi_1 + v_{20} dt)^2 + B_1 (d\phi_1 + v_{10} dt)^2 \Bigg]\, ,\label{eq:GeneralAdS7MetricNew}
\end{align}
where $B_1$, $B_2$, $B_3$, $v_{10}$, $v_{20}$, $v_{21}$, $v_{30}$, $v_{31}$ and $v_{32}$ can be determined by comparing \eqref{eq:GeneralAdS7MetricNew} with \eqref{eq:GeneralAdS7Metric}. We can see that in the gravitational Cardy limit $B_1$ and $B_2$ are subleading compared to $B_3$. Hence, qualitatively the term $B_3 (d\phi_3 + v_{32} d\phi_2 + v_{31} d\phi_1 + v_{30} dt)^2$ in the metric forms the only $U(1)$ circle fibered over AdS$_2$ in the gravitational Cardy limit of the near-horizon solution, similar to the other cases in this paper. However, because the explicit expressions of these coefficients are lengthy and not very illuminating, we do not list them here.

The thermodynamic quantities can be expressed as
\begin{align}
\begin{split}
  E & = \frac{\pi^2}{8 \Xi_1 \Xi_2 \Xi_3} \left[\sum_i \frac{2 m}{\Xi_i} - m + \frac{5 q}{2} + \frac{q}{2} \sum_i \left(\sum_{j \neq i} \frac{2 \Xi_j}{\Xi_i} - \Xi_i - \frac{2 (1 + 2 a_1 a_2 a_3 g^3)}{\Xi_i} \right) \right]\, ,\\
  T & = \frac{(1 + g^2 r_+^2) r_+^2 \sum_i \prod_{j \neq i} (r_+^2 + a_j^2) - \prod_i (r_+^2 + a_i^2) + 2 q (g^2 r_+^4 + g a_1 a_2 a_3) - q^2 g^2}{2 \pi r_+ \left[(r_+^2 + a_1^2) (r_+^2 + a_2^2) (r_+^2 + a_3^2) + q (r_+^2 - a_1 a_2 a_3 g) \right]}\, ,\\
  S & = \frac{\pi^3 \left[(r_+^2 + a_1^2) (r_+^2 + a_2^2) (r_+^2 + a_3^2) + q (r_+^2 - a_1 a_2 a_3 g) \right]}{4 \Xi_1 \Xi_2 \Xi_3 r_+}\, ,\\
  \Omega_i & = \frac{a_i \left[ (1 + g^2 r_+^2) \prod_{j \neq i} (r_+^2 + a_j^2) + q g^2 r_+^2\right] - q \prod_{j \neq i} a_j g}{(r_+^2 + a_1^2) (r_+^2 + a_2^2) (r_+^2 + a_3^2) + q (r_+^2 - a_1 a_2 a_3 g)}\, ,\\
  J_i & = \frac{\pi^2 m \left[ a_i\, \textrm{cosh}^2 (\delta) - g\, \textrm{sinh}^2 (\delta) \left(\prod_{j \neq i} a_j + a_i \sum_{j \neq i} a_j^2 g + a_1 a_2 a_3 a_i g^2 \right) \right]}{4 \Xi_1 \Xi_2 \Xi_3 \Xi_i}\, ,\\
  \Phi & = \frac{2 m\, \textrm{sinh} (\delta)\, \textrm{cosh} (\delta)\, r_+^2}{(r_+^2 + a_1^2) (r_+^2 + a_2^2) (r_+^2 + a_3^2) + q (r_+^2 - a_1 a_2 a_3 g)}\, ,\\
  Q & = \frac{\pi^2 m\, \textrm{sinh} (\delta)\, \textrm{cosh} (\delta)}{\Xi_1 \Xi_2 \Xi_3}\, ,
\end{split}
\end{align}
where $r_+$ denotes the position of the outer horizon, and
\be
  \Xi_i \equiv 1 - a_i^2 g^2\, ,\quad \Xi_{i \pm} \equiv 1 \pm a_i g\, ,\quad (i = 1,\, 2,\, 3)\, .
\ee

This class of asymptotically AdS$_7$ black hole solutions is characterized by five parameters $(m,\, \delta,\, a_1,\, a_2,\, a_3)$. The BPS limit for this class of solutions is
\be
  e^{2 \delta} = 1 - \frac{2}{(a_1 + a_2 + a_3) g}\, ,
\ee
while the naked closed timelike curves (CTCs) can be avoided by requiring an additional condition
\be
  q = - \frac{\Xi_{1-} \Xi_{2-} \Xi_{3-} (a_1 + a_2) (a_2 + a_3) (a_3 + a_1)}{(1 - a_1 g - a_2 g - a_3 g)^2 g}\, .
\ee
Hence, only three parameters are independent, which we can choose to be $(a_1,\, a_2,\, a_3)$. In the BPS limit, the thermodynamic quantities can be simplified as follows
\begin{align}
\begin{split}\label{eq:GeneralAdS7BPSthermo}
  E & = - \frac{\pi^2 \prod_{k < l} (a_k + a_l) \Big[\sum_i \Xi_i + \sum_{i < j} \Xi_i \Xi_j - (1 + a_1 a_2 a_3 g^3) (2 + \sum_i a_i g + \sum_{i < j} a_i a_j g^2) \Big]}{8\, \Xi_{1+}^2 \Xi_{2+}^2 \Xi_{3+}^2 (1 - a_1 g - a_2 g - a_3 g)^2 g}\, ,\\
  T & = 0\, ,\qquad \Omega_i = - g\, ,\qquad \Phi = 1\, ,\\
  S & = - \frac{\pi^3 (a_1 + a_2) (a_2 + a_3) (a_3 + a_1) (a_1 a_2 + a_2 a_3 + a_3 a_1 - a_1 a_2 a_3 g)}{4 \Xi_{1+} \Xi_{2+} \Xi_{3+} (1 - a_1 g - a_2 g - a_3 g)^2 g r_0}\, ,\\
  J_i & = - \frac{\pi^2 (a_1 + a_2) (a_2 + a_3) (a_3 + a_1) \Big[a_i - (a_i^2 + 2 a_i \sum_{j \neq i} a_j + \prod_{j \neq i} a_j) g + a_1 a_2 a_3 g^2\Big]}{8 \Xi_{1+} \Xi_{2+} \Xi_{3+} \Xi_{i+} (1 - a_1 g - a_2 g - a_3 g)^2 g}\, ,\\
  Q & = - \frac{\pi^2 (a_1 + a_2) (a_2 + a_3) (a_3 + a_1)}{2 \Xi_{1+} \Xi_{2+} \Xi_{3+} (1 - a_1 g - a_2 g - a_3 g) g}\, ,
\end{split}
\end{align}
where
\be
  r_0 = \sqrt{\frac{a_1 a_2 + a_2 a_3 + a_3 a_1 - a_1 a_2 a_3 g}{1 - a_1 g - a_2 g - a_3 g}}\, .
\ee

\subsubsection{Gravitational Cardy Limit}

Similar to Subsection~\ref{sec:CardyLimitSpecialAdS7}, for the more general AdS$_7$ solution with three independent angular momenta, we can translate the Cardy limit for the 6d $\mathcal{N} = (2, 0)$ theory defined in \cite{Nahmgoong:2019hko}
\be\label{eq:CardyLimitGeneralAdS7}
  |\omega_i| \ll 1\, ,\quad \Delta \sim \mathcal{O} (1)\, ,\quad (i = 1,\, 2,\, 3)
\ee
into the gravitational Cardy-like limit for this class of asymptotically AdS$_7$ black holes
\be\label{eq:GeneralAdS7CardyLimitDef}
  \bigg| \left(\frac{\partial \Omega_i}{\partial T}\right)_{T=0} \bigg| \ll 1\, ,\quad \frac{\partial \Phi}{\partial T} \bigg|_{T=0} \sim \mathcal{O} (1)\, .
\ee
A choice of the parameters $(a_1,\, a_2,\, a_3)$ that satisfies the limit \eqref{eq:GeneralAdS7CardyLimitDef} is
\be
  a_1 = a_2 = a_3 = - \frac{1}{g}\, .
\ee
As in the other black hole solutions, this can be summarized as
\begin{align}
	a_{i} g \to -1.
\end{align}
We can introduce a small parameter $\epsilon$ to denote the deviation from this limit, i.e.,
\be\label{eq:GeneralAdS7CardyLimit}
  a_i = - \frac{1}{g} + \epsilon\, ,\quad (i = 1,\, 2,\, 3)\, ,
\ee
or in a more refined way
\be\label{eq:GeneralAdS7CardyLimitRefined}
  a_1 = - \frac{1}{g} + \epsilon\, ,\quad a_2 = - \frac{1}{g} + \epsilon + \eta_1\, ,\quad a_3 = - \frac{1}{g} + \epsilon + \eta_2\, ,\quad (\eta_1,\, \eta_2 \ll \epsilon)\, .
\ee
Expanding in $\epsilon$, after expanding in $\eta_1$ and $\eta_2$, we find the BPS thermodynamic quantities \eqref{eq:GeneralAdS7BPSthermo} in the gravitational Cardy limit \eqref{eq:GeneralAdS7CardyLimit} to the leading order
\begin{align}
\begin{split}\label{eq:General AdS7 BPSthermo Cardy limit}
  S_* & = \frac{\pi^3}{2 g^8 \epsilon^3} + \mathcal{O} (\epsilon^{-2})\, ,\\
  J_i^* & = - \frac{\pi^2}{2 g^9 \epsilon^4} + \mathcal{O} (\epsilon^{-3})\, ,\\
  Q_* & = \frac{\pi^2}{g^7 \epsilon^3} + \mathcal{O} (\epsilon^{-2})\, ,
\end{split}
\end{align}
which are consistent with \cite{Nahmgoong:2019hko, ZaffaroniSlides} and the Cardy-like limit on the field theory side \eqref{eq:CardyLimitGeneralAdS7}
\be
  \omega_{i*} \sim \epsilon\, ,\quad \Delta_{*} \sim \mathcal{O} (1)\, .
\ee

\subsubsection{Black Hole Solution in the Near-Horizon Limit}

In this subsection, we consider the near-horizon limit of the asymptotically AdS$_7$ black hole metric. As mentioned in Subsection~\ref{sec:GeneralAdS7Review}, we should in principle take the Cardy limit of the near horizon solution \eqref{eq:GeneralAdS7MetricNew}. Applying the refined gravitational Cardy limit \eqref{eq:GeneralAdS7CardyLimitRefined}, we find that $B_1$ and $B_2$ are subleading compared to $B_3$. Therefore, in the near-horizon limit we obtain an AdS$_3$ geometry, just like the other cases. However, in practice the expressions of the coefficients are lengthy, so we consider an alternative near-horizon metric discussed in \cite{Chow:2008dp}. That is, the metric \eqref{eq:GeneralAdS7Metric} can be expressed in an equivalent form
\begin{align}
  ds^2 & = H^{2/5} \Bigg[ \frac{(r^2 + y^2) (r^2 + z^2)}{R} dr^2 + \frac{(r^2 + y^2) (y^2 - z^2)}{Y} dy^2 + \frac{(r^2 + z^2) (z^2 - y^2)}{Z} dz^2 \nonumber\\
  {} & \qquad\qquad + \frac{Y}{(r^2 + y^2) (y^2 - z^2)} \left(dt - \sum_{i=1}^3 \frac{(r^2 + a_i^2) \gamma_i}{a_i^2 - y^2} \frac{d\hat{\phi}_i}{\delta_i} - \frac{q \mathcal{A}}{H (r^2 + y^2) (r^2 + z^2)}  \right)^2 \nonumber\\
  {} & \qquad\qquad + \frac{Z}{(r^2 + z^2) (z^2 - y^2)} \left(dt - \sum_{i=1}^3 \frac{(r^2 + a_i^2) \gamma_i}{a_i^2 - z^2} \frac{d\hat{\phi}_i}{\delta_i} - \frac{q  \mathcal{A}}{H (r^2 + y^2) (r^2 + z^2)}  \right)^2 \nonumber\\
  {} & \qquad\qquad + \frac{a_1^2\, a_2^2\, a_3^2}{r^2 y^2 z^2} \Bigg(dt - \sum_{i=1}^3 \frac{(r^2 + a_i^2) \gamma_i}{a_i^2} \frac{d\hat{\phi}_i}{\delta_i} - \frac{q \mathcal{A}}{H (r^2 + y^2) (r^2 + z^2)} \left(1 + \frac{g y^2 z^2}{a_1 a_2 a_3} \right) \Bigg)^2 \Bigg]\, ,\label{eq:GeneralAdS7MetricNew 2}
\end{align}
where we have used the changes of coordinates \eqref{eq:GeneralAdS7ChangeCoord} and
\begin{align}
\begin{split}
  \hat{\phi}_i & \equiv \phi_i - a_i g^2 t\, ,\quad (i = 1,\, 2,\, 3)\, ,\\
  \gamma_i & \equiv a_i^2 (a_i^2 - y^2) (a_i^2 - z^2)\, ,\\
  \delta_i & \equiv a_i (1 - a_i^2 g^2) \prod_{j \neq i} (a_i^2 - a_j^2)\, .
\end{split}
\end{align}
Applying the following scaling to the new metric \eqref{eq:GeneralAdS7MetricNew 2}
\be
  r \to r_0 (1 + \lambda \rho)\, ,\quad \phi \to \widetilde{\phi}_i + \frac{\Omega_i^0}{2 \pi T'^0_H r_0 \lambda}\, \widetilde{t}\, ,\quad t \to \frac{\widetilde{t}}{2 \pi T'^0_H r_0 \lambda}\, ,
\ee
which is slightly different from the original Bardeen-Horowitz scaling \cite{Bardeen:1999px}, we obtain the near-horizon geometry in the limit $\lambda \to 0$
\begin{align}
  ds^2 & = H_0^{2/5} \Bigg[ \frac{U_0}{V} \left(- \rho^2 d\widetilde{t}^2 + \frac{d\rho^2}{\rho^2} \right) + \frac{(r_0^2 + y^2) (y^2 - z^2)}{Y} dy^2 + \frac{(r_0^2 + z^2) (z^2 - y^2)}{Z} dz^2 \nonumber\\
  {} & \qquad\qquad + \frac{Y}{(r_0^2 + y^2) (y^2 - z^2)} \left(\frac{2 r_0 (r_0^2 + z^2)}{V} \rho\, d\widetilde{t} + \sum_{i=1}^3 \frac{(r_0^2 + a_i^2) \gamma_i}{a_i^2 - y^2} \frac{d\widetilde{\phi}_i}{\delta_i} + \frac{q \widetilde{A}}{H_0 U_0} \right)^2 \nonumber\\
  {} & \qquad\qquad + \frac{Z}{(r_0^2 + z^2) (z^2 - y^2)} \left(\frac{2 r_0 (r_0^2 + y^2)}{V} \rho\, d\widetilde{t} + \sum_{i=1}^3 \frac{(r_0^2 + a_i^2) \gamma_i}{a_i^2 - z^2} \frac{d\widetilde{\phi}_i}{\delta_i} + \frac{q \widetilde{A}}{H_0 U_0} \right)^2 \nonumber\\
  {} & \qquad\qquad + \frac{a_1^2 a_2^2 a_3^2}{r_0^2 y^2 z^2} \Bigg(\frac{2}{V r_0} \left(U_0 - \frac{q g y^2 z^2}{a_1 a_2 a_3} \right) \rho\, d\widetilde{t} + \sum_{i=1}^3 \frac{(r_0^2 + a_i^2) \gamma_i}{a_i^2} \frac{d\widetilde{\phi}_i}{\delta_i} + \frac{q \widetilde{A}}{H_0 U_0} \left(1 + \frac{g y^2 z^2}{a_1 a_2 a_3} \right)  \Bigg)^2 \Bigg]\, ,\label{eq:GeneralAdS7NearHorizonMetric}
\end{align}
where
\begin{align}
\begin{split}
  U & \equiv (r^2 + y^2) (r^2 + z^2)\, ,\\
  \gamma_i & \equiv a_i^2 (a_i^2 - y^2) (a_i^2 - z^2)\, ,\\
  \delta_i & \equiv \Xi_i a_i \prod_{j \neq i} (a_i^2 - a_j^2)\, ,\\
  U_0 & \equiv U \Big|_{r = r_0} = (r_0^2 + y^2) (r_0^2 + z^2)\, ,\\
  H_0 & \equiv H \Big|_{r = r_0} = 1 + \frac{q}{(r_0^2 + y^2) (r_0^2 + z^2)}\, ,\\
  V & \equiv 6 r_0^2 + \sum_{i=1}^3 a_i^2 + \frac{3 (a_1 a_2 a_3 - q g)^2}{r_0^4} + g^2 \Bigg[15 r_0^4 + 6 r_0^2 \sum_{i=1}^3 a_i^2 + \sum_{1 \leq i < j \leq 3} a_i^2 a_j^2 + 2 q \Bigg]\, ,\\
  \widetilde{A} & \equiv - \frac{2 r_0 (2 r_0^2 + y^2 + z^2)}{V} \rho\, d\widetilde{t} - \sum_{i=1}^3 \gamma_i \frac{d\widetilde{\phi}_i}{\delta_i}\, .
\end{split}
\end{align}
Taking the refined gravitational Cardy limit \eqref{eq:GeneralAdS7CardyLimitRefined}, we can see that two of the three $U(1)$ circles in the near-horizon metric \eqref{eq:GeneralAdS7NearHorizonMetric} become degenerate. However, the remaining two $U(1)$ circles are still of the same order in the gravitational Cardy limit. This is expected, because we should take the gravitational Cardy limit of the near-horizon of the metric \eqref{eq:GeneralAdS7MetricNew} instead of \eqref{eq:GeneralAdS7MetricNew 2}, in order to have only one $U(1)$ circle fibered over AdS$_2$ in the near-horizon plus gravitational Cardy limit. Nevertheless, the gravitational Cardy limit reduces some redundant $U(1)$ circles, while keeping the essential information for the near-horizon Virasoro algebra.

\subsubsection{Black Hole Entropy from Cardy Formula}

We can apply the formalism described in Subsection~\ref{sec:Review Cardy}. The central charge and the extremal Frolov-Thorne temperature in the near-horizon region of the asymptotically AdS$_7$ BPS black holes \eqref{eq:GeneralAdS7NearHorizonMetric} were obtained in \cite{Chow:2008dp}. In the refined gravitational Cardy limit \eqref{eq:GeneralAdS7CardyLimitRefined}, the results are
\be
  c_L = \frac{48 \pi^2}{g^9 \epsilon^2 V}\, ,\quad T_L = \frac{g V}{32 \pi \epsilon}\, .
\ee
Hence, the black hole entropy from the Cardy formula in the gravitational Cardy limit is
\be\label{eq:GeneralAdS7CardyFormula}
  S_{BH} = \frac{\pi^2}{3} c_L T_L = \frac{\pi^3}{2 g^8 \epsilon^3}\, ,
\ee
which is exactly the same as the result from the gravity solution \eqref{eq:General AdS7 BPSthermo Cardy limit}.

\subsubsection{Comparison with Results from Boundary CFT}

As we discussed in Subsection~\ref{sec:General AdS7 S_BH from 6d (2,0)}, for the asymptotically AdS$_7$ black holes with general charges $(Q_1,\, Q_2)$ and angular momenta $(J_1,\, J_2,\, J_3)$, the entropy can be obtained from the boundary 6d $(2, 0)$ theory \cite{Choi:2018hmj, Nahmgoong:2019hko}, and the results are summarized in \eqref{eq:AdS7 S_BH from 6d} subject to the constraint \eqref{eq:AdS7 constraint from 6d}.

We have discussed a degenerate case in Subsection~\ref{sec:Special AdS7} with $Q_1 = Q_2$ and $J_1 = J_2 = J_3$. In this subsection, we have seen another degenerate case with $Q_1 = Q_2 = Q$ and $(J_1,\, J_2,\, J_3)$, which consequently has the black hole entropy
\be
  S_{BH} = 2 \pi \sqrt{\frac{6 Q^3 - N^3 (J_1 J_2 + J_2 J_3 + J_3 J_1)}{6 Q - N^3}}\, ,
\ee
subject to the constraint
\begin{align}
  {} & \frac{6 Q^3 - N^3 (J_1 J_2 + J_2 J_3 + J_3 J_1)}{6 Q - N^3}\nonumber\\
   & = \left[\frac{N^3}{3} (J_1 + J_2 + J_3) + 3 Q^2 \right]\cdot \left[1 - \sqrt{1 - \frac{\frac{2}{3} N^3 J_1 J_2 J_3 + Q^4}{\left(\frac{N^3}{3} (J_1 + J_2 + J_3) + 3 Q^2 \right)^2}} \right]\, ,
\end{align}
which are consistent with both the thermodynamic quantities on the gravity side \eqref{eq:GeneralAdS7BPSthermo} \eqref{eq:General AdS7 BPSthermo Cardy limit} and the black hole entropy in the gravitational Cardy limit from the Cardy formula \eqref{eq:GeneralAdS7CardyFormula} under the AdS$_7$/CFT$_6$ dictionary of parameters \cite{Choi:2018hmj, Nahmgoong:2019hko}
\be
  G_N = \frac{3 \pi^2}{16\, g^5 N^3}\, .
\ee

%%%%%%%%%%%%%%%%%%%%%%%%%%%%%%%%%%%%%%%%%%%
%%%%%%%%%%%%%%%%%%%%%%%%%%%%%%%%%%%%%%%%%%%
\section{Asymptotically AdS$_6$ Black Holes}\label{sec:AdS6}
%%%%%%%%%%%%%%%%%%%%%%%%%%%%%%%%%%%%%%%%%%%
%%%%%%%%%%%%%%%%%%%%%%%%%%%%%%%%%%%%%%%%%%%

In this section, we consider the asymptotically AdS$_6$ black holes and the corresponding gravitational Cardy limit. Similar to the other cases, we demonstrate that the AdS$_6$ black hole entropy can be computed in various ways as shown in Fig.~\ref{fig:M1}, and the other thermodynamic quantities scale correspondingly in the gravitational Cardy limit.

\subsection{AdS$_6$ Black Hole Solution}

In this subsection, we discuss the near-horizon plus Cardy limit of the non-extremal asymptotically AdS$_6$ black holes constructed in \cite{Chow:2008ip}, which are solutions to 6d $\mathcal{N}=4$ $SU(2)$ gauged supergravity.

The bosonic part of this class of solution is given by the metric, a scalar, a 1-form potential and a 2-form potential. The metric is
\begin{align}
  ds^2 & = H^{1/2} \Bigg[ \frac{(r^2 + y^2) (r^2 + z^2)}{R} dr^2 + \frac{(r^2 + y^2) (y^2 - z^2)}{Y} dy^2 + \frac{(r^2 + z^2) (z^2 - y^2)}{Z} dz^2 \nonumber\\
  {} & \qquad - \frac{R}{H^2 (r^2 + y^2) (r^2 + z^2)} \mathcal{A}^2 \nonumber\\
  {} & \qquad + \frac{Y}{(r^2 + y^2) (y^2 - z^2)} \left(dt' + (z^2 - r^2) d\psi_1 - r^2 z^2 d\psi_2 - \frac{q r \mathcal{A}}{H (r^2 + y^2) (r^2 + z^2)} \right)^2 \nonumber\\
  {} & \qquad + \frac{Z}{(r^2 + z^2) (z^2 - y^2)} \left(dt' + (y^2 - r^2) d\psi_1 - r^2 y^2 d\psi_2 - \frac{q r \mathcal{A}}{H (r^2 + y^2) (r^2 + z^2)} \right)^2 \Bigg]\, ,\label{eq:AdS6Metric}
\end{align}
while the 1-form potential, the 2-form potential and the scalar are
\begin{align}
\begin{split}
  A_{(1)} & = \frac{2 m r\, \textrm{sinh} (\delta)\, \textrm{cosh} (\delta)}{H (r^2 + y^2) (r^2 + z^2)} \mathcal{A} + \frac{\alpha_6}{dt}\, ,\\
  A_{(2)} & = \frac{q}{H (r^2 + y^2)^2 (r^2 + z^2)^2} \Bigg[ - \frac{y z \left(2 r (2 r^2 + y^2 + z^2) + q \right)}{H} dr \wedge \mathcal{A} \\
  {} & \qquad\qquad + z \left( (r^2 + z^2) (r^2 - y^2) + q r \right) dy \\
  {} & \qquad\qquad\qquad \wedge \left(dt' + (z^2 - r^2) d\psi_1 - r^2 z^2 d\psi_2 - \frac{q r \mathcal{A}}{H (r^2 + y^2) (r^2 + z^2)} \right) \\
  {} & \qquad\qquad + y \left( (r^2 + y^2) (r^2 - z^2) + q r \right) dz \\
  {} & \qquad\qquad\qquad \wedge \left(dt' + (y^2 - r^2) d\psi_1 - r^2 y^2 d\psi_2 - \frac{q r \mathcal{A}}{H (r^2 + y^2) (r^2 + z^2)} \right) \Bigg]\, ,\\
  X & = H^{-1/4}\, ,
\end{split}
\end{align}
where
\begin{align}
\begin{split}
  t' & \equiv \frac{t}{\Xi_a \Xi_b} - \frac{a^4 \phi_1}{\Xi_a\, a (a^2 - b^2)} - \frac{b^4 \phi_2}{\Xi_b\, b (b^2 - a^2)}\, ,\\
  \psi_1 & \equiv - \frac{g^2 t}{\Xi_a \Xi_b} + \frac{a^2 \phi_1}{\Xi_a\, a (a^2 - b^2)} + \frac{b^2 \phi_2}{\Xi_b\, b (b^2 - a^2)}\, ,\\
  \psi_2 & \equiv \frac{g^4 t}{\Xi_a \Xi_b} - \frac{\phi_1}{\Xi_a\, a (a^2 - b^2)} - \frac{\phi_2}{\Xi_b\, b (b^2 - a^2)}\, ,
\end{split}
\end{align}
and
\begin{align}
\begin{split}
  R & \equiv (r^2 + a^2) (r^2 + b^2) + g^2 \left[r (r^2 + a^2) + q \right] \left[r (r^2 + b^2) + q \right] - 2 m r\, ,\\
  Y & \equiv - (1 - g^2 y^2) (a^2 - y^2) (b^2 - y^2)\, ,\\
  Z & \equiv - (1 - g^2 z^2) (a^2 - z^2) (b^2 - z^2)\, ,\\
  H & \equiv 1 + \frac{q r}{(r^2 + y^2) (r^2 + z^2)}\, ,\\
  \mathcal{A} & \equiv dt' + (y^2 + z^2) d\psi_1 + y^2 z^2 d\psi_2\, ,\\
  q & \equiv 2 m\, \textrm{sinh}^2 (\delta)\, ,\qquad   \Xi_a \equiv 1 - a^2 g^2\, ,\qquad \Xi_b \equiv 1 - b^2 g^2\, .
\end{split}
\end{align}
Note that we have added a pure gauge term to the 1-form potential. It was shown in \cite{Chow:2008ip} that the metric \eqref{eq:AdS6Metric} can be written in an equivalent form
\begin{align}
  ds^2 & = H^{1/2} \Bigg[ \frac{(r^2 + y^2) (r^2 + z^2)}{R} dr^2 + \frac{(r^2 + y^2) (y^2 - z^2)}{Y} dy^2 + \frac{(r^2 + z^2) (z^2 - y^2)}{Z} dz^2 \nonumber\\
  {} & \quad \qquad + \frac{R Y Z}{H^2 \Xi_a^2\, \Xi_b^2\, a^2 b^2 (a^2 - b^2)^2 B_1 B_2} dt^2 + B_2 (d\phi_2 + v_{21}\, d\phi_1 + v_{20}\, dt)^2 \nonumber\\
  {} & \quad \qquad + B_1 (d\phi_1 + v_{10} dt)^2 \Bigg]\, ,\label{eq:AdS6MetricNew}
\end{align}
where $B_1$, $B_2$, $v_{10}$, $v_{20}$ and $v_{21}$ can be determined by comparing \eqref{eq:AdS6MetricNew} with \eqref{eq:AdS6Metric}. Because the explicit expressions of these coefficients are lengthy and not very illuminating, we do not list them here. Moreover, we notice a sign error in \cite{Chow:2008ip} for the term $\sim dt^2$ in \eqref{eq:AdS6MetricNew}.

The thermodynamic quantities can be expressed as
\begin{align}
\begin{split}\label{eq:AdS6thermo}
  E & = \frac{\pi}{3 \Xi_a \Xi_b} \left[2 m \left(\frac{1}{\Xi_a} + \frac{1}{\Xi_b} \right) + q \left(1 + \frac{\Xi_a}{\Xi_b} + \frac{\Xi_b}{\Xi_a} \right) \right]\, ,\\
  S & = \frac{2 \pi^2 \left[ (r_+^2 + a^2) (r_+^2 + b^2) + q r_+\right]}{3\, \Xi_a\, \Xi_b}\, ,\\
  T & = \frac{2 r_+^2 (1 + g^2 r_+^2) (2 r_+^2 + a^2 + b^2) - (1 - g^2 r_+^2) (r_+^2 + a^2) (r_+^2 + b^2) + 4 q g^2 r_+^3 - q^2 g^2}{4 \pi r_+ \left[ (r_+^2 + a^2) (r_+^2 + b^2) + q r_+ \right]}\, ,\\
  J_1 & = \frac{2 \pi m a (1 + \Xi_b\, \textrm{sinh}^2 (\delta))}{3\, \Xi_a^2\, \Xi_b}\, ,\qquad J_2 = \frac{2 \pi m b (1 + \Xi_a\, \textrm{sinh}^2 (\delta))}{3\, \Xi_b^2\, \Xi_a}\, ,\\
  \Omega_1 & = \frac{a \left[ (1 + g^2 r_+^2) (r_+^2 + b^2) + q g^2 r_+ \right]}{(r_+^2 + a^2) (r_+^2 + b^2) + q r_+}\, ,\qquad \Omega_2 = \frac{b \left[ (1 + g^2 r_+^2) (r_+^2 + a^2) + q g^2 r_+ \right]}{(r_+^2 + a^2) (r_+^2 + b^2) + q r_+}\, ,\\
  Q & = \frac{2 \pi m\, \textrm{sinh} (\delta)\, \textrm{cosh} (\delta)}{\Xi_a \Xi_b}\, ,\qquad \Phi = \frac{2 m r_+\, \textrm{sinh} (\delta)\, \textrm{cosh} (\delta)}{(r_+^2 + a^2) (r_+^2 + b^2) + q r_+}\, .
\end{split}
\end{align}
This class of asymptotically AdS$_6$ black hole solutions is characterized by four parameters $(m,\, \delta,\, a,\, b)$.
The BPS limit can be obtained by imposing the following condition
\be
  e^{2 \delta} = 1 + \frac{2}{(a + b) g}\, .
\ee
The absence of the naked closed timelike curves (CTCs) for these supersymmetric black holes requires an additional condition
\be
  q = \frac{\Xi_{a+} \Xi_{b+} (a + b) r_+}{(1 + a g + b g) g}\, ,
\ee
where in the BPS limit
\be\label{eq:AdS6 r_+}
  r_+ \equiv \sqrt{\frac{a b}{1 + a g + b g}}\, ,\quad 
  \Xi_{a+} \equiv 1 + a g\, ,\quad \Xi_{b+} \equiv 1 + b g\, .
\ee

\subsection{Gravitational Cardy Limit}

The Cardy-like limit for the 5d SCFT was defined in \cite{Choi:2019miv}
\be\label{eq:CardyLimitAdS6}
  |\omega_i| \ll 1\, ,\quad \Delta \sim \mathcal{O} (1)\, ,\qquad (\textrm{$i = 1,\, 2$})\, .
\ee
Using the following relations \cite{Choi:2018fdc}
\be
  \omega_i = - \lim_{T \to 0} \frac{\Omega_i - \Omega_i^*}{T}\, ,\quad \Delta = - \lim_{T \to 0} \frac{\Phi - \Phi^*}{T}\, ,
\ee
with $\Omega_i^* = g$ and $\Phi^* = 1$ denoting the BPS values of $\Omega_i$ and $\Phi$, we can find the gravitational counterpart of the Cardy-like limit \eqref{eq:CardyLimitAdS6}
\be
  \bigg| \left(\frac{\partial \Omega_i}{\partial T}\right)_{T=0} \bigg| \ll 1\, ,\quad \frac{\partial \Phi}{\partial T} \bigg|_{T=0} \sim \mathcal{O} (1)\, .
\ee
The equations $\left(\frac{\partial \Omega_i}{\partial T}\right)_* = 0$ have the roots
\be
  a = \frac{1}{g}\quad \textrm{and}\quad b = \frac{1}{g}\, .
\ee
Hence, the gravitational Cardy limit for the class of asymptotically AdS$_6$ black holes \eqref{eq:AdS6Metric} is
\be
  a \to \frac{1}{g}\quad \textrm{and}\quad b \to \frac{1}{g}\, .
\ee
Similar to the black hole solutions in the previous sections, we have
\begin{align}
	a_{i} g \to 1,
\end{align}
where $a_{i}=\{a,b\}$. We can introduce small parameters to denote the deviations from this limit, i.e.,
\be\label{eq:AdS6CardyLimit}
  a = \frac{1}{g} + \epsilon\, ,\quad b = \frac{1}{g} + \epsilon + \eta\, ,\quad \textrm{with } 0 \neq \eta \ll \epsilon\, .
\ee

Expanding in $\epsilon$ after expanding in $\eta$, we find the thermodynamic quantities \eqref{eq:AdS6thermo} in the BPS and gravitational Cardy limit \eqref{eq:AdS6CardyLimit} to the leading order
\begin{align}
\begin{split}\label{eq:AdS6BPSthermoCardy}
  S_* & = \frac{4 \pi^2}{9 g^6 \epsilon^2} + \mathcal{O} (\epsilon^{-1})\, ,\\
  J_1^* & = - \frac{8 \pi}{9 \sqrt{3} g^7 \epsilon^3} + \mathcal{O} (\epsilon^{-2})\, ,\\
  J_2^* & = - \frac{8 \pi}{9 \sqrt{3} g^7 \epsilon^3} + \mathcal{O} (\epsilon^{-2})\, ,\\
  Q^* & = \frac{2 \pi}{\sqrt{3} g^5 \epsilon^2} + \mathcal{O} (\epsilon^{-1})\, ,
\end{split}
\end{align}
which are consistent with \cite{Choi:2019miv, ZaffaroniSlides} and the Cardy-like limit on the field theory side \eqref{eq:CardyLimitAdS6}
\be
  \omega_i^* \sim \epsilon\, ,\quad \Delta_* \sim \mathcal{O} (1)\, .
\ee

\subsection{Black Hole Solution in the Near-Horizon + Gravitational Cardy Limit}

In the previous subsection, we have obtained the gravitational Cardy limit for the parameters on the gravity side. In this subsection, we discuss how the near-horizon metric changes when taking the gravitational Cardy limit. In Appendix~\ref{app:AdS6}, we verify explicitly that the resulting background is a solution of the 6d gauged supergravity equations of motion. In the following, we implement the gravitational Cardy limit in the space of parameters, which further simplifies the geometry.

We apply the following scaling near the horizon $r_+$ \eqref{eq:AdS6 r_+} to the asymptotically AdS$_6$ black hole metric \eqref{eq:AdS6MetricNew} in the BPS limit
\be
  r \to r_+ + \lambda \widetilde{r}\, ,\quad t \to \frac{\widetilde{t}}{\lambda}\, ,\quad \phi_1 \to \widetilde{\phi}_1 + g \frac{\widetilde{t}}{\lambda}\, ,\quad \phi_2 \to \widetilde{\phi}_2 + g \frac{\widetilde{t}}{\lambda}\, ,
\ee
with $\lambda \to 0$, and then take the AdS$_6$ gravitational Cardy limit \eqref{eq:AdS6CardyLimit}. To the leading order in $\epsilon$ and $\eta$, the metric becomes
\begin{align}
  ds^2 & = - \frac{\sqrt{3} g^2}{4} \sqrt{(1 + 3 g^2 y^2) (1 + 3 g^2 z^2) \left[3 + 3 g^4 y^2 z^2 + g^2 (y^2 + z^2) \right]}\, \widetilde{r}^2\, d\widetilde{t}^2 \nonumber\\
  {} & \quad \qquad + (1 + 3 g^2 y^2) (1 + 3 g^2 z^2) H_{*}(y,\, z) \frac{d\widetilde{r}^2}{144 g^2 \widetilde{r}^2} \nonumber\\
  {} & \quad \qquad + \frac{g^2 (1 + 3 g^2 y^2) (z^2 - y^2)}{3 (1 - g^2 y^2)^3} H_{*}(y,\, z)\, dy^2 \nonumber + \frac{g^2(1 + 3 g^2 z^2)(z^2 - y^2)}{3 (g^2 z^2 - 1)^3} H_{*}(y,\, z)\, dz^2 \nonumber\\
  {} & \quad \qquad + \frac{4 (1 - g^2 y^2)^2 (1 - g^2 z^2)^2 \left[z^2 + y^2 (1 + 2 g^2 z^2) \right]}{3 g^4 \left[3 + 3 g^4 y^2 z^2 + g^2 (y^2 + z^2) \right]^2 \epsilon^2 \eta^2} H_{*}(y,\, z) \left(d \widetilde{\phi}_1 - d \widetilde{\phi}_2 \right)^2 \nonumber\\
  {} & \quad \qquad + \frac{(1 - g^2 y^2) (g^2 z^2 - 1) (1 + 3 g^2 y^2) (1 + 3 g^2 z^2)}{12 g^6 (y^2 + z^2 + 2 g^2 y^2 z^2) \epsilon^2} H_{*}(y,\, z) \left(d\widetilde{\phi}_1 - \frac{\sqrt{3}}{2} g^3 \epsilon \widetilde{r} d\widetilde{t} \right)^2\, ,\label{eq:AdS6Metric in Cardy Limit}
\end{align}
where
\be
  H_* (y,\, z) \equiv \sqrt{1 + \frac{8}{(1 + 3 g^2 y^2) (1 + 3 g^2 z^2)}}.
\ee
Defining
\be
  \tau \equiv 6 g^2 \widetilde{t}\, , \qquad \chi \equiv \frac{\widetilde{\phi}_1 - \widetilde{\phi}_2}{g\, \eta}\, ,
\ee
we can rewrite the metric \eqref{eq:AdS6Metric in Cardy Limit} as
\begin{align}
  ds^2 & = H_{*}(y,\, z)\Bigg[\frac{(1 + 3 g^2 y^2) (1 + 3 g^2 z^2)}{144 g^2}\,\, \bigg(- \widetilde{r}^2\, d\tau^2 + \frac{d\widetilde{r}^2}{\widetilde{r}^2} \bigg) \nonumber\\
  {} & \quad + \frac{g^2 (1 + 3 g^2 y^2) (z^2 - y^2)}{3 (1 - g^2 y^2)^3}\, dy^2 \nonumber + \frac{g^2 (1 + 3 g^2 z^2)(z^2 - y^2) }{3 (g^2 z^2 - 1)^3} \, dz^2 \nonumber\\
  {} & \quad + \frac{4 (1 - g^2 y^2)^2 (1 - g^2 z^2)^2 \left[z^2 + y^2 (1 + 2 g^2 z^2) \right]}{3 g^2 \left[3 + 3 g^4 y^2 z^2 + g^2 (y^2 + z^2) \right]^2 \epsilon^2} d\chi^2 \nonumber\\
  {} & \quad + \frac{(1 - g^2 y^2) (g^2 z^2 - 1) (1 + 3 g^2 y^2) (1 + 3 g^2 z^2)}{12 g^6 (y^2 + z^2 + 2 g^2 y^2 z^2) \epsilon^2}  \left(d\widetilde{\phi}_1 - \frac{\sqrt{3}}{12} g \epsilon \widetilde{r} d\tau \right)^2\, \Bigg]\, .\label{eq:AdS6Metric in Cardy Limit New 2}
\end{align}

\subsection{Black Hole Entropy from Cardy Formula}

For the asymptotically AdS$_6$ black holes discussed in this section, we apply the Cardy formula to the near-horizon metric only after taking the gravitational Cardy limit. More explicitly, we first rewrite the metric \eqref{eq:AdS6Metric in Cardy Limit New 2} from Poincar\'e coordinates $(\widetilde{r},\, \tau)$ to global coordinates $(\hat{r},\, \hat{t})$ using the relations \eqref{eq:PoincareToGlobal 1} - \eqref{eq:PoincareToGlobal 3}. Consequently, the near-horizon metric in the gravitational Cardy limit \eqref{eq:AdS6Metric in Cardy Limit New 2} becomes
\begin{align}
  ds^2 & = H_{*}(y,\, z) \Bigg[\frac{(1 + 3 g^2 y^2) (1 + 3 g^2 z^2)}{144 g^2} \, \bigg( - (1 + \hat{r}^2)\, d\hat{t}^2 + \frac{d\hat{r}^2}{1 + \hat{r}^2} \bigg) \nonumber\\
  {} & \quad + \frac{g^2 (1 + 3 g^2 y^2) (z^2 - y^2)}{3 (1 - g^2 y^2)^3} \, dy^2 \nonumber + \frac{g^2 (z^2 - y^2) (1 + 3 g^2 z^2)}{3 (g^2 z^2 - 1)^3} \, dz^2 \nonumber\\
  {} & \quad + \frac{4 (1 - g^2 y^2)^2 (1 - g^2 z^2)^2 \left[z^2 + y^2 (1 + 2 g^2 z^2) \right]}{3 g^2 \left[3 + 3 g^4 y^2 z^2 + g^2 (y^2 + z^2) \right]^2 \epsilon^2}  d\chi^2 \nonumber\\
  {} & \quad + \frac{(1 - g^2 y^2) (g^2 z^2 - 1) (1 + 3 g^2 y^2) (1 + 3 g^2 z^2)}{12 g^6 (y^2 + z^2 + 2 g^2 y^2 z^2) \epsilon^2} \left(d\hat{\psi} - \frac{\sqrt{3}}{12} g \epsilon \hat{r} d\hat{t} \right)^2\,\Bigg]\, ,\label{eq:AdS6Metric in Cardy Limit New 3}
\end{align}
where
\be
  \hat{\psi} \equiv \widetilde{\phi}_1 - \frac{\sqrt{3}}{12} g \epsilon \gamma\, .
\ee

Applying the same formalism in Subsection~\ref{sec:Review Cardy} and choosing appropriate ranges of $y$ and $z$, we obtain the central charge and the extremal Frolov-Thorne temperature in the near-horizon region of the asymptotically AdS$_6$ BPS black holes as follows:
\be
  c_L = \frac{5 \pi}{3 \sqrt{3} g^5 \epsilon}\, ,\quad T_L = \frac{4 \sqrt{3}}{5 \pi g \epsilon}\, .
\ee
Using the Cardy formula, we can compute the black hole entropy of the asymptotically AdS$_6$ BPS black holes:
\be\label{eq:AdS6CardyFormula}
  S_{BH} = \frac{\pi^2}{3} c_L T_L = \frac{4 \pi^2}{9 g^6 \epsilon^2}\, ,
\ee
which is the same as the black hole entropy in the gravitational Cardy limit \eqref{eq:AdS6BPSthermoCardy} from the gravity side.

\subsection{Comparison with Results from Boundary CFT}

For the asymptotically AdS$_6$ BPS black holes, it was shown in \cite{Choi:2018fdc, Choi:2019miv, Crichigno:2020ouj} that the entropies of these black holes can be obtained from the boundary 5d $\mathcal{N} = 1$ superconformal field theories by extremizing an entropy function, which has also been studied in \cite{Hosseini:2019iad}. We can first compute the free energy in the large-$N$ limit using the 5d superconformal index. The entropy function is then defined as a Legendre transform of the free energy in the large-$N$ limit
\be\label{eq:AdS6EntropyFct}
  S (\Delta_I,\, \omega_i) = - \frac{i \pi}{81 g^4 G} \frac{\Delta^3}{\omega_1 \omega_2} + Q \Delta + \sum_{i=1}^2 J_i \omega_i + \Lambda \left(\Delta - \sum_{i=1}^2 \omega_i - 2 \pi i \right)\, .
\ee
In the Cardy-like limit \eqref{eq:CardyLimitAdS6}
\be
  \omega \sim \epsilon\, ,\quad \Delta_I \sim \mathcal{O} (1)\, ,
\ee
we can read off from the entropy function \eqref{eq:AdS6EntropyFct}
\be
  S \sim \frac{1}{\epsilon^2}\, ,\quad J \sim \frac{1}{\epsilon^3}\, ,\quad Q_I \sim \frac{1}{\epsilon^2}\, ,
\ee
which have been summarized in Table~\ref{Table:Intro}.

Similar to AdS$_{4, 5, 7}$, for AdS$_6$ the electric charge $Q$ and the angular momenta $J_i$ are real, while the chemical potential $\Delta$ and the angular velocities $\omega_i$ can be complex, and so can the entropy function $S$. By requiring that the black hole entropy $S_{BH}$ to be real after extremizing the entropy function $S$, we obtain one more constraint on $Q$ and $J_i$. More precisely, the asymptotically AdS$_6$ black hole entropy and the corresponding constraint are given implicitly by the following two relations \cite{Choi:2018fdc, Choi:2019miv}
\begin{align}
  S_{BH}^3 - \frac{2 \pi^2}{3 g^4 G_N} S_{BH}^2 - 12 \pi^2 \left(\frac{Q}{3 g} \right)^2\, S_{BH} + \frac{8 \pi^4}{3 g^4 G_N} J_1 J_2 & = 0\, ,\\
  \frac{Q}{3 g}\, S_{BH}^2 + \frac{2 \pi^2}{9 g^4 G_N} (J_1 + J_2) S_{BH} - \frac{4 \pi^2}{3} \left(\frac{Q}{3 g} \right)^3 & = 0\, ,
\end{align}
which are consistent with both the thermodynamic quantities on the gravity side \eqref{eq:AdS6thermo} and \eqref{eq:AdS6BPSthermoCardy} as well as the black hole entropy in the gravitational Cardy limit from the Cardy formula \eqref{eq:AdS6CardyFormula} under the AdS$_6$/CFT$_5$ dictionary of parameters \cite{Choi:2018fdc, Choi:2019miv}
\be
  \frac{1}{g^4 G_N} \sim N^{5/2}\, .
\ee

%%%%%%%%%%%%%%%%%%%%%%%%%%%%%%%%%%%%%%%%%%%
%%%%%%%%%%%%%%%%%%%%%%%%%%%%%%%%%%%%%%%%%%%
\section{Discussion}\label{sec:Discussions}
%%%%%%%%%%%%%%%%%%%%%%%%%%%%%%%%%%%%%%%%%%%
%%%%%%%%%%%%%%%%%%%%%%%%%%%%%%%%%%%%%%%%%%%

In this paper, we have discussed the near-horizon gravitational Cardy limit of asymptotically AdS$_{4, 5, 6, 7}$ black holes. The gravitational Cardy limit can be written universally as $|a_{i} g| \to 1$, where $a_i$ are parametrize angular momenta in units of the inverse AdS radius, $g$, for all the black hole solutions we analyzed. As we have explicitly shown in these examples, the gravitational Cardy limit leads to an AdS$_3$ geometry near the horizon  and is effectively an additional limit on the independent parameters of the black hole solutions. The macroscopic Bekenstein-Hawking entropy of asymptotically AdS black holes has recently been given a microscopic foundation using the dual boundary CFT$_{3,4,5,6}$. Our work relies on a near-horizon AdS$_3$ geometry  and we provide an effective microscopic description via the CFT$_2$ Cardy formula obtained from the algebra of asymptotic symmetries.

It is instructive to point out various analogies with the previous instance when string theory answered explicitly the problem of microstate counting for black hole entropy. In the mid 90's,  Strominger and Vafa  \cite{Strominger:1996sh}  used the full machinery of D-brane technology to provide a microscopic description of the Bekenstein-Hawking entropy of a class of asymptotically flat black holes. Viewing the D-brane description as the UV complete description of gravity, the analogy with the current developments is that the microscopic description of the entropy of AdS$_{d+1}$ black holes  in terms of field theory degrees of freedom in the dual CFT$_{d}$ boundary theory is the UV complete description.  After the UV complete description of the 90's, Strominger went on to provide a universal description \cite{Strominger:1997eq}, based only on the near-horizon symmetries exploiting  the AdS$_3$ near-horizon region and the asymptotic symmetry algebra computation of Brown and Henneaux \cite{Brown:1986nw}. Similar symmetry-based approaches were shown to apply to a wide variety of black holes  by Carlip \cite{Carlip:1998wz}. The results presented in \cite{Lu:2008jk, Chow:2008dp} and in this manuscript show that we can understand the entropy of asymptotically AdS black holes based only on near-horizon symmetries via the Kerr/CFT correspondence.

The satisfying aspect of this point of view resides in the separation-of-scales principle. Such a universal feature of gravity as the Bekenstein-Hawking entropy formula can certainly be explained using UV complete formulations of quantum gravity but must also be understood without recourse to the existence of such a  UV complete theory and could be determined strictly from low energy symmetry principles.

The point of view advocated in this manuscript leads to a number of interesting questions some of which we now describe.  It would be interesting to understand the field theory counterpart of the locally AdS$_3$ near-horizon region that arises from the Bardeen-Horowitz limit plus the gravitational Cardy limit. It clearly suggests the existence of an effective CFT$_2$  which we have used to microscopically compute the entropy but whose further details we do not know. Some  aspects of this effective CFT$_2$ were studied in  \cite{Nian:2020qsk, David:2020jhp} for the AdS$_5$ and the AdS$_4$ black holes, but it required going away from extremality. In the bigger picture described above, understanding how this effective CFT$_2$ embeds in the boundary CFT$_d$ is the dual to finding the UV complete description of the gravitational theory living near the horizon -- a worthy challenge. Along these lines, in this manuscript, we have only discussed the asymptotically AdS black holes in the BPS limit, hence at zero temperature.  It would be interesting to extend the discussion to near-extremal asymptotically AdS$_{4, 6, 7}$ black holes and to reproduce the Bekenstein-Hawking entropy formula from a near-horizon Cardy formula. When higher-derivative terms are included in the gravity theory, the black hole entropy does not obey the area law. It was shown in \cite{Azeyanagi:2009wf} that the central charge of the near-horizon asymptotic Virasoro symmetry also gets modified in the gravity with higher-derivative terms, while the Frolov-Thorne temperature and the Cardy formula still hold. Other higher-derivative aspects of AdS$_4$ black holes were recently considered \cite{Bobev:2020egg, Ghosh:2020rwf}. A tantalizing property of higher-derivative corrections in AdS$_5$ black holes was recently reported in \cite{Melo:2020amq}, which showed that the leading $\alpha'$-correction is absent in the BPS limit. This suggests that the central charge of the near-horizon asymptotic Virasoro symmetry remains the same in this case.

There is another line of attack that is worth sketching. Recall that the original setup for Cardy-like limits is 2d CFT. In this case, one simply has a formula for CFT$_2$ on $S^1\times S^1$ which effectively relates the high energy and low energy degrees of freedom. It is fair to think of this relation as a  UV/IR relation with the important characteristic of being controlled by the anomaly, $c$. Similar formulas have been developed in higher dimensions by Di Pietro and Komargodski in \cite{DiPietro:2014bca}  and further clarified in  \cite{Ardehali:2015bla, DiPietro:2016ond,Honda:2019cio}.  In particular, in four dimensions they found an effective description of theories in $S^1\times M^3$ whose effective action is controlled by anomaly coefficients. A similar analysis has been rigorously performed for a set of six-dimensional theories \cite{Choi:2018hmj, Nahmgoong:2019hko, Chang:2019uag}. More closely related to the questions we addressed in this paper is the recent work of Seok Kim and collaborators who have used an effective low energy action approach to find the leading term in the entropy function for the Cardy-like limit, first in ${\cal N}=4$ SYM as well as in the 6d ${\cal N} = (2, 0)$ SCFT living on $N$ M5-branes \cite{Choi:2018hmj}, and later for a more generic 4d ${\cal N}=1$ situation  \cite{Kim:2019yrz}.  These developments point to the possibility that the Cardy-like limit may be understood as the leading term in an effective field theory expansion.  Although for these cases in the BPS limit the Cardy-like free energy has been derived from the effective quantum field theory approach, higher order corrections as well as finite temperatures should be taken into account to go beyond the leading order in the BPS limit. It would be quite interesting to explore such possibilities on the field theory side and, ultimately, connect it with a more standard hydrodynamics approach on the gravity side  \cite{Banerjee:2012cr, Jensen:2013rga}.

Finally, it would be nice to develop what seems like a more natural  AdS$_2$ or SYK approach to the entropy of extremal  AdS black holes  as described in Fig.~\ref{fig:M1}. Some interesting work along this direction was performed in  \cite{Castro:2018ffi} for AdS$_5$ and more recently in  \cite{Moitra:2019bub} for AdS$_4$. Finding the connection between the AdS$_2$ and AdS$_3$ low energy descriptions in more details is an interesting problem.

%%%%%%%%%%%%%%%%%%%%%%%%%%%%%%%%%%%%%%%%%%%
%%%%%%%%%%%%%%%%%%%%%%%%%%%%%%%%%%%%%%%%%%%
\section*{Acknowledgements}
%%%%%%%%%%%%%%%%%%%%%%%%%%%%%%%%%%%%%%%%%%%
%%%%%%%%%%%%%%%%%%%%%%%%%%%%%%%%%%%%%%%%%%%

We would like to thank  Arash Arabi Ardehali, Sunjin Choi, Zhihao Duan, Martin Fluder, Alfredo Gonz\'alez Lezcano, Simeon Hellerman, Kentaro Hori, Seyed Morteza Hosseini, Cindy Keeler, Joonho Kim, Seok Kim, Sung-Soo Kim, Finn Larsen, Kimyeong Lee, Sungjay Lee, Yang Lei, June Nahmgoong, Ioannis Papadimitriou, Wei Song, Futoshi Yagi, Junya Yagi,  Masahito Yamazaki, Piljin Yi and Yang Zhou for many helpful discussions.  We also would like to thank David Chow for very useful communications. This  work was supported in part by the U.S. Department of Energy under grant DE-SC0007859. M.D. was supported by the NSF Graduate Research Fellowship Program under NSF Grant Number: DGE 1256260. J.N. was also supported by a Van Loo Postdoctoral Fellowship and he would like to thank Rutgers University, Arizona State University, UESTC, Southwest Jiaotong University, Sichuan University, KIAS and Kavli IPMU for warm hospitality during various stages of this work.

\appendix

%%%%%%%%%%%%%%%%%%%%%%%%%%%%%%%%%%%%%%%%%%%%%%%%%%%%%%%%
\section{Verifying the equations of motion for the near-horizon}\label{app:NearHorizonEOM}
Gravitational theories are nonlinear and, therefore, a truncated sector of a solution need not be a solution itself. For that reason, we explicitly verify that, in each instance, the near-horizon limit satisfies the equations of motion. This fact alone should inspire trust in the consistency of the resulting geometry and the potential existence and closure of a dual field theory sector.  Returning to the analogy with the BMN paradigm, this is equivalent to checking the equations of motion for the plane wave background  \cite{Berenstein:2002jq}.

\subsection{$AdS_5$}\label{app:AdS5}

We verify the equations of motion for the near-horizon geometry for $AdS_5$. The Lagrangian describing the solution in \cite{Chong:2005hr} is
\begin{align}
    \mathcal{L}=\left(R+12 g^{2}\right) * 1-\frac{1}{2} * F \wedge F+\frac{1}{3 \sqrt{3}} F \wedge F \wedge A,
\end{align}
and the equations of motion are
\begin{align}
R_{ab}-\frac{1}{2} F_{ac}\tensor{F}{_b^c}+\frac{1}{3} g_{ab}\left(\frac{1}{4}F^{2}+12g^2\right)=0,
\qquad
d \star F-\frac{1}{\sqrt{3}} F \wedge F=0.
\end{align}
In order to facilitate the computation, we turn to a veilbein description for the near-horizon geometry,
\begin{align}
\begin{split}
    e_{0}&=%\text{pm}
    \sqrt{\frac{a}{10 a g^2+2 g}} \tilde{r}d\tau,
    \\
    e_{1}&=\sqrt{\frac{a}{10 a g^2+2 g}}\frac{d\tilde{r}}{\tilde{r}},
    \\
    e_{2}&= \sqrt{\frac{2a}{g-a g^2}} d\theta,
    \\
    e_{3}&= p_1\left(p_2 \left((-3 a g \cos 2 \theta +a g-4)d\tilde{\psi}-6 a g \sin ^2\theta d\tilde{\phi} \right)+3 a \left(a^2 g^2+a g-2\right) \tilde{r}d\tau\right),
    \\
    e_{4}&=p_3\left(3 a(1- ag) \tilde{r}d\tau+2 p_2 d\tilde{\phi}\right),
    \end{split}
\end{align}
where
\begin{align}
\begin{split}
    p_1&=-\frac{\cos\theta}{(1-ag) (5 a g+1) \sqrt{2(a g+2) (3 a g \cos (2 \theta )-a g+4)}},
    \\
    p_2&= \sqrt{\frac{a (a g+2)}{g}} (5 a g+1),
    \\
    p_3&=\frac{\sin\theta}{(5 a g+1)\sqrt{(1- ag)(3 a g \cos (2 \theta )-a g+4)}}.
    \end{split}
\end{align}
Note that this coframe describes the near-horizon, which is computed using \cite{Bardeen:1999px}. After applying the near-horizon geometry and gauge fixing, the gauge potential is
\begin{align}
    A_{(1),\text{near}}&=-\frac{\sqrt{6} (1-ag)}{\sqrt{a g+2} \sqrt{5 a g+1}}e_1-\frac{\sqrt{6ag} \cos \theta}{\sqrt{3 a g \cos 2 \theta-a g+4}}e_4 \nonumber
    \\&-\frac{2\sqrt{3ag(1-ag)} \sin \theta }{(\sqrt{ag+2} \sqrt{3 a g \cos 2 \theta -a g+4}}e_{5}.
    %\frac{\sqrt{3}}{2} \left(\frac{2 a \left(d\tilde{\psi} \cos ^2\theta +d\tilde{\phi} \sin ^2\theta\right)}{a h-1}-\frac{\pm \tilde{r} \sqrt{a (a h+2)} d\tilde{t}}{\sqrt{h} (5 a h+1)}\right)+O\left(\lambda ^1\right)
\end{align}
Note that the exterior derivative and the near-horizon geometry limit commute to give an equivalent expression for the gauge field,
\begin{align}
\begin{split}
    F_{(2),\text{near}}=dA_{(1),\text{near}}&=
    \frac{\sqrt{3a g (a g+2)}}{a}e_1\wedge e_2 +2 g \sin\theta\sqrt{\frac{3(1-a g)}{3 a g \cos 2\theta-a g+4}} e_3\wedge e_4
    \\&+ g  \cos \theta\sqrt{\frac{6(a g+2)}{(3 a g \cos 2\theta-ag+4)}} e_3\wedge e_5.
    %\frac{\sqrt{3}}{2}\left(\frac{2a \sin 2\theta}{ah-1}(d\theta\wedge d\tilde{\phi}-d\theta\wedge d\tilde{\psi})-\frac{\pm \sqrt{a (a h+2)}}{\sqrt{h} (5 a h+1)}d\tilde{r}\wedge d\tilde{t}\right).
    \end{split}
\end{align}
Then,
\begin{align}
\begin{aligned} \label{fwedgef}
    F_{(2),\text{near}}\wedge F_{(2),\text{near}}=6 g^{3/2} \frac{\sqrt{a g+2}}{\sqrt{a (3 a g \cos 2 \theta-a g+4)}}&\left(2 \sqrt{1-a g}\sin\theta e_1\wedge e_2\wedge e_3\wedge e_4 \right.
    \\&\left.-\sqrt{2 a g+4}\cos\theta e_1\wedge e_2\wedge e_3\wedge e_5 \right).
    %\frac{\pm 3\sqrt{a (a h+2)}}{\sqrt{h} (5 a h+1)}\frac{a \sin 2\theta}{ah-1}\left(d\tilde{t}\wedge d\tilde{r} \wedge d\theta \wedge d\tilde{\phi}-d\tilde{t}\wedge d\tilde{r} \wedge d\theta \wedge d\tilde{\psi}\right)
\end{aligned}
\end{align}
The other term gives
\begin{align}
\begin{split} \label{dstaroff}
    d\star F_{(2),\text{near}}=&
    d\left[\sqrt{6} g \cos \theta \sqrt{\frac{a g+2}{3 a g \cos 2 \theta-a g+4}}e_1\wedge e_2\wedge e_4
    -\frac{\sqrt{3a g (a g+2)} }{a}e_3\wedge e_4\wedge e_5\right.
    \\&\left.
    +2 \sqrt{3} g \sin \theta  \sqrt{\frac{1-a g}{3 a g \cos 2 \theta-a g+4}}e_1\wedge e_2\wedge e_5
    \right]
    \\=&
    \frac{2 \sqrt{3(ag+2)} g^{3/2}}{\sqrt{a (3 a g \cos 2 \theta-a g+4)}}
        \left(2 \sqrt{1-a g} \sin \theta e_1\wedge e_2\wedge e_3\wedge e_4 \right.
        \\& \left.-\sqrt{2ag+4}\cos \theta e_1\wedge e_2\wedge e_3\wedge e_5 \right).
    %\sqrt{g}\frac{3}{2(3!)}\frac{4a\cos 2\theta}{ah-1}(d\theta \wedge d\tilde{r}\wedge d\tilde{t}\wedge d\tilde{\psi}-d\theta\wedge d\tilde{r}\wedge d\tilde{t} \wedge d\tilde{\phi},
    \end{split}
\end{align}
Comparing equations \eqref{fwedgef} and \eqref{dstaroff}, we can see that the equation of motion for the gauge potential is satisfied. For the Einstein equations, the geometric data we need are the nonzero components of the Ricci tensor
\begin{align}
\begin{split}
    R_{00,\text{near}}&=-R_{11,\text{near}}=\frac{g (11 a g+4)}{2 a},
    \\
    R_{22,\text{near}}&=-\frac{g (5 a g-2)}{2 a},
    \\
    R_{33,\text{near}}&=-\frac{g \left(9 a^2 g^2 \cos 2 \theta-a^2 g^2+14 a g-4\right)}{a (3 a g \cos 2 \theta-a g+4)},
    \\
    R_{34,\text{near}}&=-\frac{3 \sqrt{2} g^2 \sqrt{2-a g-a^2 g^2} \sin \theta \cos \theta }{3 a g \cos 2 \theta -a g+4},
    \\
    R_{44,\text{near}}&=-\frac{g \left(21 a^2 g^2 \cos 2 \theta-11 a^2 g^2-12 a g \cos 2 \theta+28 a g-8\right)}{2 a (3 a g \cos 2\theta-a g+4)}.
    \end{split}
\end{align}
We also need the explicit nonzero contractions of the gauge field $F_{ac,\text{near}}\tensor{F}{_b^c_{,\text{near}}} \equiv \mathcal{F}_{ab,\text{near}}$
\begin{align}
\begin{split}
    \mathcal{F}_{00,\text{near}}&=\frac{3 g (a g+2)}{a}, 
    \qquad
     \mathcal{F}_{11,\text{near}}=-\mathcal{F}_{00,\text{near}},
    \qquad  \mathcal{F}_{22,\text{near}}=3 g^2,
    \qquad \\
    \mathcal{F}_{33,\text{near}}&=\frac{12 g^2 (1-a g) \sin ^2\theta}{3 a g \cos 2 \theta -a g+4},
    \qquad
    \mathcal{F}_{34,\text{near}}=
    -\frac{3g^2 \sin 2\theta \sqrt{2(1-ag)(2+ag)}}{3ag\cos 2\theta - ag +4},
    \\
    \mathcal{F}_{44,\text{near}}&=\frac{6 g^2 (a g+2) \cos ^2\theta}{3 a g \cos 2\theta-a g+4},
    \qquad
    F_{ab,\text{near}}F^{ab,\text{near}}=-\frac{12 g}{a}.
    \end{split}
\end{align}
The equations of motion are then verified once we impose these expressions.

%%%%%%%%%%%%%%%%%%%%%%%%%%%%%%%%%%%%%%%%%%%%%%%%%%%%%%%%%%%

\subsection{$AdS_4$}\label{app:AdS4}

The 4d $\mathcal{N}=4$ gauged supergravity can be obtained by the truncation of the 11d supergravity \cite{Chong:2004na}
\begin{align}
    \begin{split}
        \mathcal{L}_{4}=& R * 1-\frac{1}{2} * \mathrm{d} \varphi \wedge \mathrm{d} \varphi-\frac{1}{2} e^{2 \varphi} * \mathrm{d} \chi \wedge \mathrm{d} \chi-\frac{1}{2} e^{-\varphi} * F_{(2) 2} \wedge F_{(2) 2}-\frac{1}{2} \chi F_{(2) 2} \wedge F_{(2) 2} \\
        &-\frac{1}{2\left(1+\chi^{2} e^{2 \varphi}\right)}\left(e^{\varphi} * F_{(2)1} \wedge F_{(2) 1}-e^{2 \varphi} \chi F_{(2) 1} \wedge F_{(2) 1}\right) \\
        &-g^{2}\left(4+2 \cosh \varphi+e^{\varphi} \chi^{2}\right) * 1,
    \end{split}
\end{align}
where $\varphi$ and $\chi$ are the dilaton and axion. The subscript in parenthesis denotes the degree of the form. The solution has two pairwise equal charges and therefore two gauge potential $A_{(1)1}$ and $A_{(1)2}$. The equations of motion are
\begin{align}
\begin{split}
   0=&\mathrm{d}\left(\frac{1}{1+\chi^2 e^{2\varphi}}\left(e^{\varphi} \star F_{(2)1}- e^{2\varphi} \chi F_{(2)1}\right)\right),
   \\
   0=&\mathrm{d}\left(e^{-\varphi}\star F_{(2)2}+\chi F_{(2)2}\right),
   \\
   \begin{split}
    0=\, & - \mathrm{d}\star \mathrm{d}\varphi - e^{2\varphi}\star \mathrm{d}\chi \wedge \mathrm{d}\chi+\frac{1}{2}e^{-\varphi}\star \mathrm{d}A_{(1)2}\wedge \mathrm{d}A_{(1)2}-g^2(2\sinh\varphi+e^{\varphi}\chi^2)\star 1
    \\&+\frac{e^{\varphi } \left(e^{2 \varphi } \chi ^2-1\right)}{2 \left(e^{2 \varphi } \chi ^2+1\right)^2} \star \mathrm{d}A_{(1)1}\wedge \mathrm{d}A_{(1)1}+\frac{e^{2 \varphi } \chi }{\left(e^{2 \varphi } \chi ^2+1\right)^2} \mathrm{d}A_{(1)1}\wedge \mathrm{d}A_{(1)1},
    \end{split}
   \\
    0=&-\mathrm{d}(e^{2\varphi}\star d\chi) -\frac{1}{2}\mathrm{d}A_{(1)2}\wedge \mathrm{d}A_{(1)2}+
        \frac{e^{3 \varphi } \chi }{\left(e^{2 \varphi } \chi ^2+1\right)^2}\star \mathrm{d}A_{(1)1}\wedge \mathrm{d}A_{(1)1} 
        -2g^2 e^{\varphi}\chi\star 1
        \\&+\frac{e^{2 \varphi }-e^{4 \varphi } \chi ^2}{2 \left(e^{2 \varphi } \chi ^2+1\right)^2}\mathrm{d}A_{(1)1}\wedge \mathrm{d}A_{(1)1},
   \\
    0=&R_{ab}-\frac{1}{2}g_{ab}R-\frac{1}{2}\left(\nabla_a \varphi \nabla_b \varphi -\frac{1}{2}\nabla^c \varphi \nabla_c \varphi g_{ab}\right)-\frac{1}{2}e^{-\varphi} \left(F_{ac,2}F_{b,2}^{c}-\frac{1}{4}F_{cd,2}F^{cd}_{2} g_{ab}\right)
        \\&-\frac{1}{2}e^{2\varphi}\left(\nabla_a \chi \nabla_b \chi -\frac{1}{2}\nabla^c \chi \nabla_c \chi g_{ab}\right)
        -\frac{e^{\varphi}}{2(1+\chi^2 e^{2\varphi})} \left(F_{ac,1}F_{b,1}^{c}-\frac{1}{4}F_{cd,1}F^{cd}_{1} g_{ab}\right) \\&+\frac{1}{2}g^2 (4+2\cosh \varphi +e^{\varphi}\chi^2) g_{ab}.
    \end{split}
\end{align}
The convenient veilbein for this black hole solution is
\begin{align}
    \begin{split}
    e_{0}&=G_1\sqrt{\cos2 \theta+x^2 y^2}\tilde{r} d\tau,
    \\
    e_{1}&=G_1\sqrt{\cos 2 \theta+x^2 y^2}\frac{d\tilde{r}}{\tilde{r}},
    \\
    e_{2}&=\sqrt{2}\sqrt{\frac{\cos 2 \theta +x^2 y^2}{h^2 \left(-2 \cos 2 \theta+x^4 y^4-2 x^2 y^2-1\right)}}d\theta,
    \\
    e_{3}&=G_4 \sin\theta \sqrt{\frac{-2 \cos2 \theta+x^4 y^4-2 x^2 y^2-1}{\left(x^2 y^2-1\right)^2 \left(\cos 2 \theta +x^2 y^2\right)}} (G_5 \tilde{r}d\tau+d\tilde{\phi}),
    \end{split}
\end{align}
where $G_{1},G_{4},G_{5}$ are constants
\begin{align}
\begin{split}
    G_{1}&=\frac{xy\sqrt{2\left(x^4-1\right) \left(y^4-1\right)}}{g}\left(x^{10} \left[y^{10}+y^6\right)+x^8 \left(6 y^8-8 y^4\right)+x^6 y^2 \left(y^8-10 y^4+5\right)\right.
    \\& \left.-2 x^4 \left(4 y^8-7 y^4+1\right)+x^2 y^2 \left(5 y^4-3\right)-2 y^4 \right]^{-1/2},
    \\
    G_{4}&=\frac{\sqrt{2} \left(x^2 y^2-1\right)}{g \left| x^2 y^2-3\right| },
    \\
    G_{5}&=G_1^2 \frac{g^2 \left(x^2+y^2\right) \left(x^2 y^2-3\right) \left(x^2 y^2-1\right)^{3/2}}{\sqrt{2} x y \sqrt{\left(x^4-1\right) \left(y^4-1\right)}},
    \end{split}
\end{align}
that depend on the parameters $\delta_1,\delta_2$ of the solution, which we have redefined as
\begin{align}
    \delta_1 \equiv \ln x, \qquad \delta_2 \equiv \ln y.
\end{align}
The scalar fields in the near-horizon are
\begin{align}
\begin{split}
    \chi_{\text{near}}&=-\frac{x \cos (\theta ) \left(x^2-y^2\right) \left(x^2 y^2+1\right)\sqrt{2\left(y^4-1\right) \left(x^2 y^2-1\right)}}{y \left(x^2 \left(y^4-1\right) \cos (2 \theta )+x^6 y^4-x^4 y^2-x^2+y^2\right)\sqrt{x^4-1}},
    \\
    \left(e^{\varphi}\right)_{\text{near}}&=\frac{x^2 \left(y^4-1\right) \cos (2 \theta )+x^6 y^4-x^4 y^2-x^2+y^2}{x^2 \left(y^4-1\right) \left(\cos (2 \theta )+x^2 y^2\right)},
\end{split}
\end{align}
and for the gauge potentials after adding  a pure gauge term for convenience,
\begin{align}
\begin{split}
    A_{(1)1,\text{near}}=&\frac{M_1}{\cos (2 \theta )+x^2 y^2}\left[
    \frac{M_2 \left(-\left(x^4-1\right) y^2 \cos (2 \theta )+x^4 y^2 \left(y^4-2\right)-x^2 \left(y^4-1\right)+y^2\right)}{G_1 \sqrt{\cos (2 \theta )+x^2 y^2}}e_1 \right.
    \\&\left. +M_4 \sin^2 \theta\left(\frac{ \csc (\theta ) \left(x^2 y^2-1\right)\sqrt{\cos (2 \theta )+x^2 y^2}}{G_4 \sqrt{-2 \cos (2 \theta )+x^4 y^4-2 x^2 y^2-1}}e_4-\frac{G_5}{G_1 \sqrt{\cos (2 \theta )+x^2 y^2}}e_1\right)
    \right],
    \\
    A_{(1)2,\text{near}}=&\frac{M_1}{\cos (2 \theta )+x^2 y^2}\left[\frac{M_2 \left(-x^2 \left(y^4-1\right) \cos (2 \theta )+x^6 y^4-x^4 y^2+x^2 \left(1-2 y^4\right)+y^2\right)}{G_1 \sqrt{\cos (2 \theta )+x^2 y^2}}e_1
    \right.\\&\left.
    +
    M_4 \sin^2 \theta \left( \frac{ \csc (\theta ) \left(x^2 y^2-1\right)\sqrt{\cos (2 \theta )+x^2 y^2}}{G_4 \sqrt{-2 \cos (2 \theta )+x^4 y^4-2 x^2 y^2-1}}e_4-\frac{G_5}{G_1\sqrt{\cos (2 \theta )+x^2 y^2}}\right) e_1\right],
    \end{split}
\end{align}
where
\begin{align}
\begin{split}
    M_1 &= \frac{x^2 y^2-1}{\sqrt{2}},
    \\
    M_2 &= -G_5 \frac{2 \left(x^2 y^2+1\right)}{g \left(x^2+y^2\right) \left(x^2 y^2-3\right) \left(x^2 y^2-1\right)^2},
    \\
    M_4 &= \frac{4}{3 g-g x^2 y^2}.
\end{split}
\end{align}

%%%%%%%%%%%%%%%%%%%%%%%%%%%%%%%%%%%%%%%%%%%%%%%%%%%%%%%%%%%

\subsection{$AdS_7$}\label{app:AdS7}

In this section, we are interested in charged, rotating $AdS_7$ black hole solutions as studied in \cite{Chow:2007ts,Chong:2004dy}. The Lagrangian is
\begin{align}
    \begin{aligned}
        \mathcal{L}_{7}=& R \star 1-\frac{1}{2} \sum_{i=1}^{2} \star \mathrm{d} \varphi_{i} \wedge \mathrm{d} \varphi_{i}-\frac{1}{2} \sum_{I=1}^{2} X_{I}^{-2} \star F_{(2)}^{I} \wedge F_{(2)}^{I}-\frac{1}{2} X_{1}^{2} X_{2}^{2} \star F_{(4)} \wedge F_{(4)} \\
        &+2 g^{2}\left(8 X_{1} X_{2}+4 X_{1}^{-1} X_{2}^{-2}+4 X_{1}^{-2} X_{2}^{-1}-X_{1}^{-4} X_{2}^{-4}\right) \star 1 \\
        &+g F_{(4)} \wedge A_{(3)}+F_{(2)}^{1} \wedge F_{(2)}^{2} \wedge A_{(3)},
    \end{aligned}
\end{align}
where
\begin{align}
    X_{1}=\mathrm{e}^{-\varphi_{1} / \sqrt{10}-\varphi_{2} / \sqrt{2}}, \quad X_{2}=\mathrm{e}^{-\varphi_{1} / \sqrt{10}+\varphi_{2} / \sqrt{2}}, \quad F_{(2)}^{I}=\mathrm{d} A_{(1)}^{I}, \quad F_{(4)}=\mathrm{d} A_{(3)},
\end{align}
where we have fixed a typographical error corresponding to a minus sign in one of the terms in the Lagrangian. The bosonic fields include two scalars $\varphi_1$ and $\varphi_2$, the graviton, a 3-form potential $A_{(3)}$, and two U(1) gauge potentials $A_{(1)}^{I}, I=1,2$. We study two different solutions to this Lagrangian. The first solution is more general with two charges set equal but different angular momenta. The equations of motion corresponding to the scalars and gauge fields are
\begin{align}
    \begin{split}
        \Box \varphi_{1}&= \frac{8}{\sqrt{10}} g^{2}\left(4 X_{1} X_{2}-3 X_{1}^{-1} X_{2}^{-2}-3 X_{1}^{-2} X_{2}^{-1}+2 X_{1}^{-4} X_{2}^{-4}\right) + \frac{1}{2 \sqrt{10}} \sum_{I=1}^{2} X_{I}^{-2} F^{I a b} F_{a b}^{I}
        \\&-\frac{1}{12 \sqrt{10}} X_{1}^{2} X_{2}^{2} F^{a b c d} F_{a b c d}, \\
        \Box \varphi_{2}&= \frac{1}{2 \sqrt{2}}\left(X_{1}^{-2} F^{1 a b} F_{a b}^{1}-X_{2}^{-2} F^{2 a b} F_{a b}^{2}\right)+4 \sqrt{2} g^{2}\left(X_{1}^{-1} X_{2}^{-2}-X_{1}^{-2} X_{2}^{-1}\right), \\
        0&=\mathrm{d}\left(X_{1}^{-2} \star F_{(2)}^{1}\right) - F_{(2)}^{2} \wedge F_{(4)}, \\
        0&=\mathrm{d}\left(X_{2}^{-2} \star F_{(2)}^{2}\right) -F_{(2)}^{1} \wedge F_{(4)}, \\
        0&=\mathrm{d}\left(X_{1}^{2} X_{2}^{2} \star F_{(4)}\right) -2 g F_{(4)}-F_{(2)}^{1} \wedge F_{(2)}^{2},
    \end{split}
\end{align}
and for the graviton, we have 
\begin{align}
    \begin{aligned}
        0&=R_{ab}-\frac{1}{2}R \,g_{ab}-g^{2}\left(8 X_{1} X_{2}+4 X_{1}^{-1} X_{2}^{-2}+4 X_{1}^{-2} X_{2}^{-1}-X_{1}^{-4} X_{2}^{-4}\right) g_{a b}
        \\&
        -\sum_{i=1}^{2}\left(\frac{1}{2} \nabla_{a} \varphi_{i} \nabla_{b} \varphi_{i}-\frac{1}{4} \nabla^{c} \varphi_{i} \nabla_{c} \varphi_{i} g_{a b}\right)-\sum_{I=1}^{2} X_{I}^{-2}\left(\frac{1}{2} \tensor{F}{_{a}^{I c}} F_{b c}^{I}-\frac{1}{8} F^{I c d} F_{c d}^{I} \, g_{a b}\right) \\
        &-X_{1}^{2} X_{2}^{2}\left(\frac{1}{12} \tensor{F}{_{a}^{c d e}} F_{b c d e}-\frac{1}{96} F^{c d e f} F_{c d e f} \,g_{a b}\right).
    \end{aligned}
\end{align}
We can truncate this solution as constructed in \cite{Chong:2004dy}, where the two charges and angular momenta are set equal. This truncation can be done by letting $X=X_1=X_2=e^{-\varphi/\sqrt{10}}, \varphi_2=0$ and $A_{(1)}=A_{(1)}^1=A_{(1)}^2$ and the Lagrangian of interest becomes
\begin{align}
    \begin{split}
        \mathcal{L}_{7}=& R \star 1-\frac{1}{2} \star \mathrm{d} \varphi_{1} \wedge \mathrm{d} \varphi_{1}-X^{-2} \star F_{(2)} \wedge F_{(2)}-\frac{1}{2} X^{4} \star F_{(4)} \wedge F_{(4)} \\
        &+2 g^{2}\left(8 X^{2}+8 X^{-3}-X^{-8}\right) \star 1+F_{(2)} \wedge F_{(2)} \wedge A_{(3)}+g F_{(4)} \wedge A_{(3)},
    \end{split}
\end{align}
and the equations of motion are
\begin{align}
    \begin{split}
    0&=\mathrm{d} \star \mathrm{d} \varphi -\frac{2X^{-2}}{\sqrt{10}} \star F_{(2)} \wedge F_{(2)}+\frac{2X^{4}}{\sqrt{10}} \star F_{(4)} \wedge F_{(4)}-\frac{16g^{2}}{\sqrt{10}}\left(2 X^{2}-3X^{-3}+X^{-8}\right) \star 1,
    \\
    0&=d\left(X^{-2} \star F_{(2)}\right)-F_{(2)}\wedge F_{(4)},
    \\
    0&=\mathrm{d}\left(X_{1}^{4} \star F_{(4)}\right) - 2 g F_{(4)}-F_{(2)}^{1} \wedge F_{(2)}^{2},
    \end{split}
\end{align}
and for the graviton
\begin{align}
    \begin{split}
        0=&R_{ab}-\frac{1}{2}R \, g_{ab}-\frac{1}{2}\left(\nabla_a\varphi\nabla_b\varphi-\frac{1}{2}\nabla^c\varphi\nabla_c\varphi g_{ab}\right)
        -X^{-2}\left(\tensor{F}{_{a}^{c}}\tensor{F}{_{bc}}-\frac{1}{4}F^{cd}F_{cd}\,g_{ab}\right)
        \\&-g^2\left(8 X^{2}+8 X^{-3}-X^{-8}\right)-\frac{1}{12}X^{4}\left(\tensor{F}{_a^{cde}}\tensor{F}{_{bcde}}-\frac{1}{8}F^{cdef}F_{cdef}\,g_{ab}\right).
    \end{split}
\end{align}
The fields corresponding to the solution in \cite{Chong:2004dy} are
\begin{align}
    \begin{split}
    X&=H^{-1/5},
    \\
    A_{(1)}&=\frac{2 m \sinh (\delta) \cosh (\delta)}{  \rho ^4\Xi H}(dt - a \sigma) + \frac{\alpha_{70}\, dt}{\Xi_{-}},
    \\
    A_{(3)}&=\frac{ \left( a m \sinh^2 (\delta) \right) \sigma \wedge d\sigma}{\rho ^2\Xi  \Xi_- } + \alpha_{71} \,dt\wedge d\theta \wedge d\psi +\alpha_{72} \,dt\wedge d\xi \wedge d\phi +\alpha_{73} \,dt\wedge d\xi \wedge d\psi,
    \end{split}
\end{align}
where we have  added pure gauge terms to  both potentials $A_{(1)}$ and $A_{(3)}$ for convenience. More precisely, after taking the near-horizon geometry, we have
\begin{align}
    \alpha_{70} =-1,
    \qquad
    \alpha_{71} = -\beta \sin \theta \sin ^2\xi,
    \qquad
    \alpha_{72} = \beta \sin 2 \xi,
    \qquad
    \alpha_{73} = \beta \cos \theta \sin 2 \xi,
\end{align}
where
\begin{align}
    \beta &= -\frac{4 \left(e^{2 \delta }-1\right)}{\left(-13 e^{2 \delta }-9 e^{4 \delta }+9 e^{6 \delta }+5\right) g^2}.
\end{align}
A convenient veilbein for the near-horizon is
\begin{align}
    \begin{split}
    e_{1}&= p_1 \tilde{r}d\tau,
    \\
    e_{2}&= p_1 \frac{d\tilde{r}}{\tilde{r}},
    \\
    e_{3}&= p_2 \left(p_3 \tilde{r}d\tau + p_4 \left( \sin ^2\xi (d\phi+\cos \theta d\psi)+2 d\tilde{\chi} \right) \right),
    \\
    e_{4}&= p_5 d\xi,
    \\
    e_{5}&=p_5 \sin \xi d\theta, 
    \\
    e_{6}&=p_5 \sin \theta \sin \xi d\psi,
    \\
    e_{7}&=p_5 \sin \xi \cos \xi (d\phi+\cos \theta d\psi),
    \end{split}
\end{align}
where
\begin{align}
    \begin{split}
    p_1 &= \frac{1}{g}\frac{2^{3/5}3^{1/5}(e^{2 \delta }+1)^{1/5}}{\sqrt{6 e^{2 \delta }+27 e^{4 \delta }+43}},
    \\
    p_2 &= \frac{1}{2^{2/5} 3^{3/10}g}\frac{1}{\left(1-3 e^{2 \delta }\right)\left(5-3 e^{2 \delta }\right)}\frac{1}{\sqrt{\left(e^{2 \delta }+1\right)^{3/5} \left(9 e^{2 \delta }-7\right)}},
    \\
    p_3 &= -\frac{16  \left(3 e^{2 \delta }-5\right)^{3/2}  \left(2 e^{2 \delta }+3 e^{4 \delta }-1\right)}{6 e^{2 \delta }+27 e^{4 \delta }+43} \sqrt{\frac{3(9 e^{2 \delta }-7)}{\left(-2 e^{2 \delta }+3 e^{4 \delta }-5\right)}},
    \\
    p_4 &= -30 e^{2 \delta }+27 e^{4 \delta }+7,
    \\
    p_5 &= \frac{1}{g}\frac{2^{8/5}}{3^{3/10}}\frac{(e^{2 \delta }+1)^{1/5}}{\sqrt{ \left(-2 e^{2 \delta }+3 e^{4 \delta }-5\right)}}.
    \end{split}
\end{align}
In the near-horizon limit, the fields in the veilbein basis become
\begin{align}
    \begin{split}
    X_{\text{near}}&=\frac{2^{2/5}}{3^{1/5} (e^{2 \delta }+1)^{1/5}},
    \\
    A_{(1),\text{near}}&=\frac{2^{2/5} 3^{3/10} \left(e^{2 \delta }+1\right)^{4/5} \left(\left(15-9 e^{2 \delta }\right) e_1+\sqrt{6 e^{2 \delta }+27 e^{4 \delta }+43} e_3\right)}{\sqrt{44 e^{2 \delta }+210 e^{4 \delta }+108 e^{6 \delta }+243 e^{8 \delta }-301}},
    \\
    A_{(3),\text{near}}&=\frac{(54 e^{2 \delta }+27 e^{4 \delta }-101)(e_1\wedge e_5\wedge e_6-e_1\wedge e_4\wedge e_7)}{2^{9/5} 3^{1/10} (e^{2 \delta }+1)^{1/10} \sqrt{9 e^{2 \delta }-7} \sqrt{6 e^{2 \delta }+27 e^{4 \delta }+43}}\\&-\frac{\left(99 e^{2 \delta }-117 e^{4 \delta }+81 e^{6 \delta }-215\right)(e_3\wedge e_4\wedge e_7-e_3\wedge e_5\wedge e_6)}{2^{4/5} 3^{1/10} (e^{2 \delta }+1)^{1/10} \sqrt{9 e^{2 \delta }-7} \left(6 e^{2 \delta }+27 e^{4 \delta }+43\right)}.
    \end{split}
\end{align}

%%%%%%%%%%%%%%%%%%%%%%%%%%%%%%%%%%%%%%%%%%%%%%%%%%%%%%%%%%%

\subsection{$AdS_6$}\label{app:AdS6}

The field content consists of the graviton, a 2-form $A_{(2)}$, the scalar $\varphi$ and one U(1) gauge potential $A_{(1)}$ after truncation, as shown in \cite{Chow:2008ip}. After appropriate rescaling and gauge transformations, the 6d Lagrangian is given by
\begin{align}\begin{aligned}
\mathcal{L}_{6}=& R \star 1-\frac{1}{2} \star \mathrm{d} \varphi \wedge \mathrm{d} \varphi-X^{-2}\left(\star F_{(2)} \wedge F_{(2)}+g^{2} \star A_{(2)} \wedge A_{(2)}\right)-\frac{1}{2} X^{4} \star F_{(3)} \wedge F_{(3)} \\
&+g^{2}\left(9 X^{2}+12 X^{-2}-X^{-6}\right) \star 1-F_{(2)} \wedge F_{(2)} \wedge A_{(2)}-\frac{g^{2}}{3} A_{(2)} \wedge A_{(2)} \wedge A_{(2)},
\end{aligned}\end{align}
where
\begin{align}
    X=e^{-\varphi/\sqrt{8}}.
\end{align}
The equations of motion are\begin{align}\begin{aligned}
G_{a b}=& \frac{1}{2} \nabla_{a} \varphi \nabla_{b} \varphi-\frac{1}{4} \nabla^{c} \varphi \nabla_{c} \varphi g_{a b}+X^{-2}\left(F_{a}^{c} F_{b c}-\frac{1}{4} F^{c d} F_{c d} g_{a b}\right) \\
&+X^{-2}g^2\left( A_{a}^{c} A_{b c}-\frac{1}{4} A^{c d} A_{c d} g_{a b}\right)+X^{4}\left(\frac{1}{4} F_{a}^{c d} F_{b c d}-\frac{1}{24} F^{c d e} F_{c d e} g_{a b}\right) \\
&+\frac{g^{2}}{2}\left(9 X^{2}+12 X^{-2}-X^{-6}\right) g_{a b},
\end{aligned}\end{align}
\begin{align}\begin{aligned}
&\Box \varphi=\frac{1}{\sqrt{8}} X^{-2}\left(F^{a b} F_{a b}+g^2A^{ a b} A_{a b}\right)-\frac{1}{3 \sqrt{8}} X^{4} F^{a b c} F_{a b c}+\frac{3}{\sqrt{2}} g^{2}\left(3 X^{2}-4 X^{-2}+X^{-6}\right),\\
&\mathrm{d}\left(X^{-2} \star F_{(2)}\right)=-F_{(2)} \wedge F_{(3)},\\
&\begin{array}{l}
%\mathrm{d}\left(X^{-2} \star F_{(2)}^{I}\right)+\frac{3 g}{\sqrt{2}} \epsilon_{I J K} X^{-2} A_{(1)}^{J} \wedge \star F_{(2)}^{K}=-F_{(2)}^{I} \wedge F_{(3)} \\
\mathrm{d}\left(X^{4} \star F_{(3)}\right)=-F_{(2)} \wedge F_{(2)}-g^2 A_{(2)} \wedge A_{(2)}- 2 g^2 X^{-2} \star F_{(2)}.
\end{array}
\end{aligned}\end{align}
We omit the veilbein and several other details for this black hole as the expressions are quite long. The scalar and the $U(1)$ gauge field in the near-horizon limit take the form
%The veilbein for this black hole is r solution is
%\begin{align}
%    e_0 &=
%    \\
%    e_1 &=
%    \\
%    e_2 &=
%    \\
%    e_3 &=
%    \\
%    e_4 &=
%    \\
%    e_5 &=
%\end{align}
%where
%\begin{align}
%    \dots
%\end{align}
%The fields in the near horizon geometry are given by
\begin{align}
\begin{split}
    \chi_{\text{near}}^4&=\frac{g \left(a \left(b+g y^2\right)+y^2 (b g+1)\right) \left(a \left(b+g z^2\right)+z^2 (b g+1)\right)}{(a g+b g+1) \left(a^2 b (b g+1)+a \left(b^2+b g \left(y^2+z^2\right)+g^2 y^2 z^2\right)+g y^2 z^2 (b g+1)\right)},
    \\
    A_{(1),\text{near}}&=W_1\left(W_2\tilde{r}d\tilde{t}+W_3 d\tilde{\phi_1}+W_{4}d\tilde{\phi_2} \right),
    %\\
    %A_{(2),\text{near}}&=
    \end{split}
\end{align}
where
\begin{align}
\begin{split}
    W_1=&\frac{\sqrt{a b}}{\sqrt{a g+b g+1} \left(a^2 b (b g+1)+a \left(b^2+b g \left(y^2+z^2\right)+g^2 y^2 z^2\right)+g y^2 z^2 (b g+1)\right)},
    \\
    W_2=&\left[\left(b^2-a^2\right) \Xi_a \Xi_b (a g+b g+1) \left(a^2 \left(3 b^2+b g \left(y^2+z^2\right)-g^2 y^2 z^2\right)\right.\right.
    \\ & \left.+a \left(b^2 g \left(y^2+z^2\right)+b \left(-2 g^2 y^2 z^2+y^2+z^2\right)-2 g y^2 z^2\right)-y^2 z^2 (b g+1)^2\right]
    \\ &\left[a^4 g^2 \left(b^2 g^2+6 b g+1\right)+2 a^3 g \left(b^3 g^3+7 b^2 g^2+7 b g+1\right)\right.
    \\&\left.+a^2 \left(b^4 g^4+14 b^3 g^3+30 b^2 g^2+14 b g+1\right)+2 a b \left(3 b^3 g^3+7 b^2 g^2+7 b g+3\right)\right.
    \\& \left.+b^2 (b g+1)^2 \right]^{-1},
    \\
    W_3=&\frac{b \left(a^2-y^2\right) \left(a^2-z^2\right) \left(b^2 g^2-1\right)}{\sqrt{\frac{a b}{a g+b g+1}}},
    \\
    W_4=&-\frac{a \left(a^2 g^2-1\right) \left(b^2-y^2\right) \left(b^2-z^2\right)}{\sqrt{\frac{a b}{a g+b g+1}}}.
    \end{split}
\end{align}
Note that we have added a pure gauge term to the 1-form $\alpha_{6} dt$, where $\alpha_{6}=-1$.

\bibliography{GravitationalCardy}
\bibliographystyle{utphys}

\end{document}